\documentclass[iop,apj,tighten,twocolappendix,numberedappendix]{emulateapj}
\usepackage{apjfonts}
\usepackage{graphicx}
\usepackage{amsmath,amssymb}
\usepackage{color}
\usepackage{bm}
\usepackage{url}
\usepackage[breaklinks,colorlinks,citecolor=blue,linkcolor=red]{hyperref} 
\usepackage[all]{hypcap} 
\usepackage{enumitem}

\def\LaTeX{L\kern-.36em\raise.3ex\hbox{a}\kern-.15em 
    T\kern-.1667em\lower.7ex\hbox{E}\kern-.125emX}

\newcommand{\Mpc}{\mathrm{Mpc}}
\newcommand{\Gpc}{\mathrm{Gpc}}
\newcommand{\Msun}{M_{\odot}}
\newcommand{\Hz}{\mathrm{Hz}}

\newcommand{\yr}{\mathrm{yr}}

\begin{document}

\title{Measurement Accuracy of Inspiraling Eccentric Neutron Star and Black Hole Binaries Using Gravitational Waves}

\author{L\'aszl\'o Gond\'an\altaffilmark{1} and Bence Kocsis\altaffilmark{1}}
  \affil{$^1$E\"otv\"os University, Institute of Physics, P\'azm\'any P. s. 1/A, Budapest, 1117, Hungary} 
 
\label{firstpage}

\begin{abstract} 
 In a recent paper, we determined the measurement accuracy of physical parameters for eccentric, precessing, non-spinning, inspiraling, stellar-mass black hole - black hole (\mbox{BH--BH}) binaries for the upcoming second-generation LIGO/VIRGO/KAGRA detector network at design sensitivity using the Fisher matrix method. Here we extend that study to a wide range of binary masses including neutron star - neutron star (\mbox{NS--NS}), \mbox{NS--BH}, and \mbox{BH--BH} binaries with BH masses up to $110 \, \Msun$. The measurement error of eccentricity $e_{10 \,\rm Hz}$ at a gravitational-wave (GW) frequency of $10 \, {\rm Hz}$ is in the range $(10^{-4}-10^{-3}) \times (D_L/ 100\,\rm Mpc)$ for \mbox{NS--NS}, \mbox{NS--BH}, and \mbox{BH--BH} binaries at a luminosity distance of $D_L$ if $e_{10 \,\rm Hz} \gtrsim 0.1 $. For events with masses and distances similar to the detected $10$ GW transients, we show that nonzero orbital eccentricities may be detected if $0.081 \lesssim e_{10 \,\rm Hz}$. Consequently, the LIGO/VIRGO/KAGRA detector network at design sensitivity will have the capability to distinguish between eccentric waveforms and circular waveforms. In comparison to circular inspirals, we find that the chirp mass measurement precision can improve by up to a factor of $\sim 20$ and $\sim 50-100$ for \mbox{NS--NS} and \mbox{NS--BH} binaries with BH companion masses \mbox{$\lesssim 40 \, \Msun$}, respectively. The identification of eccentric sources may give information on their astrophysical origin; it would indicate merging binaries in triple or higher multiplicity systems or dynamically formed binaries in dense stellar systems such as globular clusters or galactic nuclei. 
\end{abstract}


\maketitle

\section{Introduction} 
\label{sec:Intro}

 The Advanced Laser Interferometer Gravitational-wave Observatory\footnote{\url{http://www.ligo.caltech.edu/}} (aLIGO) detectors \citep{Aasietal2015} and Advanced Virgo\footnote{\url{http://www.ego-gw.it/}} (AdV, \citealt{Acerneseetal2015}) have made the first $10$ detections of gravitational waves (GWs) during the first and second observing runs \citep{Abbottetal2016c,Abbottetal2016b,Abbottetal2017,LIGOColl2017,Abbottetal2017_2,Abbottetal2017_3,Abbottetal2018_b}, and herald the beginning of a new branch of observational astronomy. With the two upcoming instruments KAGRA\footnote{\url{http://gwcenter.icrr.u-tokyo.ac.jp/en/}} \citep{Somiya2012} and LIGO-India\footnote{\url{http://www.gw-indigo.org/}} \citep{Iyeretal2011,Unnikrishnan2018}, the network of advanced GW detectors is expected to continue making detections of GWs and answering fundamental questions about their astrophysical sources \citep{Abbottetal2018}.   
 
 In \citet{Gondanetal2018}, we determined the expected accuracy with which the aLIGO--AdV--KAGRA detector network may determine the eccentricity and other physical parameters of \mbox{$10 \, \Msun - 10 \, \Msun$} and \mbox{$30 \, \Msun - 30 \, \Msun$} precessing eccentric black hole - black hole (\mbox{BH--BH}) binaries. We found that measurement errors for the orbital eccentricity at formation, $e_0$, and eccentricity at the last stable orbit (LSO), $e_{\rm LSO}$, can be as low as about $10^{-3} \times D_L/ (1 \, \Gpc)$ and $10^{-4} \times D_L/(1 \, \Gpc)$, respectively, where $D_L$ is the luminosity distance. Such low-eccentricity errors may give the \mbox{aLIGO--AdV--KAGRA} detector network the capability to distinguish among quasi-circular and low-eccentricity binaries and among different astrophysical formation channels. 
 
 In this paper, we extend \citet{Gondanetal2018} to determine the expected measurement errors of $e_0$, $e_{\rm LSO}$, and other physical parameters for precessing eccentric neutron star-neutron star (\mbox{NS--NS}), neutron star-black hole (\mbox{NS--BH}), and \mbox{BH--BH} binaries. We also determine the measurement error of the residual eccentricity $e_{10{\rm Hz}}$ at which the peak GW frequency ($f_\mathrm{GW}$; \citealt{Wen2003}) of binaries first enters the $10 \, \mathrm{Hz}$ frequency band of advanced GW detectors. Further, we extend our investigations to a wide mass range of eccentric \mbox{BH--BH} binaries, and examine the measurement errors of binary parameters as a function of $e_{10{\rm Hz}}$.
 
 The majority of \mbox{NS--NS} and \mbox{NS--BH} binaries may originate from isolated binaries in galactic fields (see \citet{Kruckowetal2018,Chruslinskaetal2018,Vignaetal2018} and references therein), and these binaries are expected to be well circularized by the time their GW signals enter the advanced detectors' frequency band \citep{Kowalskaetal2011}. However, the eccentricity of binaries in the advanced detectors' frequency band may be non-negligible for many other astrophysical channels. These include tidal capture and collision of two NSs in globular clusters (GCs) that have undergone core collapse \citep{Leeetal2010}, binary--single interactions in GCs \citep{Samsingetal2014}, Kozai resonance in GCs \citep{Thompson2011} and galactic nuclei (GNs) \citep{AntoniniPerets2012,PetrovichAntonini2017,Giacomoetal2018}, non-hierarchical triple systems in GCs and nuclear star clusters \citep{ArcaSedaetal2018}, and close fly-bys between NSs and/or BHs in GNs can lead to the formation of eccentric \mbox{NS--NS} and \mbox{NS--BH} binaries due to GW emission \citep{OLearyetal2009,KocsisLevin2012,Tsang2013}. These channels typically produce binaries with orbital eccentricities beyond $e_0 \gtrsim 0.9$ at formation, and these binaries may remain eccentric by the time they enter the advanced GW detectors' frequency band.
 
 Previous and recent parameter estimation studies of \mbox{NS--NS} and \mbox{NS--BH} binaries mostly focused on quasi-circular binaries \citep{CutlerFlanagan1994,PoissonWill1995,NicholsonVecchio1998,Nissankeetal2010,Choetal2013,Chatziioannouetal2014,Favata2014,OShaughnessyetal2014,Rodriguezetal2014,Berryetal2015,Canizaresetal2015,Milleretal2015,Veitchetal2015,Farretal2016,Vinciguerraetal2017,Biweretal2018,Coughlinetal2018,Dudietal2018,Fanetal2018,ZhaoWen2018} because their predicted merger rate densities are high. The recent detection of the first $10$ GW transients constrain the merger rate density of NS binaries in the universe to $110-3840 \, \Gpc^{-3} \yr^{-1}$ \citep{Abbottetal2018_b}, which is consistent with previous rate predictions \citep{Kalogeraetal2004}. With the non-detection of \mbox{NS--BH} binaries during the first and second observing runs, an upper limit of $610 \, \Gpc^{-3} \yr^{-1}$ is drawn for the merger rate density of such systems \citep{Abbottetal2018_b}, while recent estimates predict the merger rate density to be in the range $1-1500 \, \Gpc^{-3} \yr^{-1}$ (see \citealt{Cowardetal2012,Petrilloetal2013,Bausweinetal2014,Berger2014,Siellezetal2014,deMinkBelcz2015,Dominiketal2015,Fongetal2015,Jinetal2015,Kimetal2015,Abbottetal2016g,Belczynskietal2016,Vangionietal2016,Chruslinskaetal2018} for recent rate predictions and \citealt{Abadieetal2010} for a review of past estimates).
 
 However, several theoretical studies have shown that the detection rates of eccentric \mbox{NS--NS} and \mbox{NS--BH} binaries may be non-negligible. \citet{Leeetal2010} showed that the merger rate density of tidal capture and collision of two NSs in GCs that have undergone core collapse \citep{Fabianetal1975,Pooleyetal2003,Grindlayetal2006} peak around $z \simeq 0.7$ at values of \mbox{$\sim 55 \, \Gpc^{-3} \yr^{-1}$} that drops to $\sim 30 \, \Gpc^{-3} \yr^{-1}$ by $z = 0$. However, the rates are reduced for less optimistic NS retention fractions \citep{Tsang2013}. Eccentric \mbox{NS--NS} and \mbox{NS--BH} binaries could result from hierarchical triples through the Kozai mechanism. This has been suggested to occur in  mergers around supermassive BHs in GNs for \mbox{NS--BH} binaries with a merger rate density of order $0.1 \, \Gpc^{-3} \yr^{-1}$ \citep{AntoniniPerets2012,PetrovichAntonini2017}. Recently, \citet{Giacomoetal2018} revisited these results and found an \mbox{NS--BH} and \mbox{NS--NS} merger rate density of $0.06 - 0.1 \, \Gpc^{-3} \yr^{-1}$ and $0.04 - 0.16 \, \Gpc^{-3} \yr^{-1}$, respectively. Merger rate densities of $\sim 0.7 \, \Gpc^{-3} \yr^{-1}$ can be expected from binary--single interactions from GCs \citep{Samsingetal2014}. Close fly-bys between NSs and/or BHs in GNs may result in eccentric \mbox{NS--NS} and \mbox{NS--BH} binaries with estimated merger rate densities of $0.04- 6 \, \Gpc^{-3} \yr^{-1}$ and $0.05 - 0.6 \, \Gpc^{-3} \yr^{-1}$, respectively \citep{Tsang2013}.
 
 While eccentric \mbox{NS--NS} and \mbox{NS--BH} binaries are expected to be rare compared to quasi-circular binaries, they are interesting for at least three reasons. First, significantly larger amounts of mass are ejected during eccentric \mbox{NS--NS} binary mergers than in quasi-circular \mbox{NS--NS} mergers \citep{EastPretorius2012,Rosswogetal2013,Radiceetal2016,Papenfortetal2018,Chaurasiaetal2018,Vivekanandjietal2018}. Thus, despite being rare, they may nevertheless contribute a significant fraction of the overall \textit{r}--process element abundances, and may give birth to bright electromagnetic emission in the infrared and radio bands relative to quasi-circular binaries. Additionally, eccentric \mbox{NS--NS} binaries produced by dynamical captures in dense stellar environments can potentially produce \textit{r}--process nuclei at very low metallicity and account for the \textit{r}--process enhancements seen in some carbon--enhanced metal-poor stars \citep{RamirezRuizetal2015}. 
 
 A second reason eccentric \mbox{NS--NS} and \mbox{NS--BH} binaries are of interest is that they may improve the test of general relativity in the strong-field dynamics limit \citep{Abbottetal2016i,Abbottetal2016c,Abbottetal2016b,Abbottetal2016h,Abbottetal2017,Abbottetal2017_2,TheLIGOVirgo2018}. For eccentric binaries, a larger fraction of the energy is emitted in GWs at relativistic-velocities during pericenter passages than in quasi-circular inspirals \citep{Loutreletal2014}. Thus, the GWs from eccentric binaries could capture effects from the highly nonlinear dynamical regime during pericenter passages. 
 
 Third, detections of many eccentric and quasi-circular \mbox{NS--NS}, \mbox{NS--BH}, and \mbox{BH--BH} binaries have the potential to constrain the astrophysical origin of GW source populations. The different formation channels leading to \mbox{NS--NS},  \mbox{NS--BH}, and \mbox{BH--BH} mergers may be distinguished using the measurement of the correlations between the orbital eccentricity and the chirp mass \citep{Breiviketal2016}, the correlation between the total mass and spins \citep{PostnovKuranov2017,ArcaSedaBenacquista2018}, the eccentricity in the LIGO/Virgo \citep{Cholisetal2016,SamsingRamirezRuiz2017,ArcaSedaetal2018,Gondanetal2017,RandallXianyu2018,RandallXianyu2017,Samsingetal2018_2,Zevinetal2018} and Laser Interfermeter Space Antenna\footnote{\url{https://www.elisascience.org/}} (LISA) bands \citep{Nishizawaetal2016,Nishizawaetal2017,DOrazioSamsing2018,SamsingDorazio2018}, the spin orientations \citep{Rodriguezetal2016c,Farretal2017_a,LiuLai2017,Stevensonetal2017,TalbotThrane2017,Vitaleetal2017_1,Farretal2018,Gerosaetal2018,LiuLai2018,Lopezetal2018}, the effective spin parameter \citep{Antoninietal2018,Ngetal2018,Schoderetal2018,Zaldarriagaetal2018}, the mass distribution \citep{Stevensonetal2015,OLearyetal2016,Mandeletal2017,Zevinetal2017,Fishbachetal2017}, the spin distribution \citep{Fishbachetal2017}, and other waveform features \citep{Meironetal2017,Inayoshietal2017,Kremeretal2018,Samsingetal2018}. Further, kilonova signatures could provide information on the orbital properties of dynamically captured \mbox{NS--NS} binaries \citep{Papenfortetal2018}. See Section 5.4 in \citet{Baracketal2018} for an overview of ways to distinguish between different \mbox{BH--BH} merger source populations, including mergers in GNs and GCs.  
 
 Current search algorithms based on matched filtering are highly optimized for the detection of quasi-circular compact binary sources (see \citet{Abbottetal2016c,Abbottetal2016b,Abbottetal2017,LIGOColl2017,Abbottetal2017_2,Abbottetal2017_3} for a review of search algorithms) and burst-like GW signals \citep{Abbottetal2016KK}. Several algorithms have been developed to recover the physical parameters of quasi-circular and burst-like sources \citep{Cannonetal2012,Priviteraetal2014,CornishLittenberg2015,LittenbergCornish2015,Veitchetal2015,Klimenkoetal2016,SingerPrice2016,Millhouseetal2018}. However, GW sources that do not fall into these two categories may be missed by these search algorithms \citep{BrownZimmerman2010,HuertaBrown2013,Huertaetal2014,Tiwarietal2016,Huertaetal2018}. The maximum eccentricity for a quasi-circular source to be detected with current techniques based on circular templates is $\sim 0.1$ \citep{BrownZimmerman2010,Eastetal2013,HuertaBrown2013,Huertaetal2014,Huertaetal2017,Huertaetal2018,Loweretal2018}. Recently, several search methods have been developed to detect the waveforms of eccentric compact binaries \citep{Taietal2014,Coughlinetal2015,Tiwarietal2016}, and the development of algorithms recovering the parameters of compact binaries on eccentric orbits is underway.
 
 The paper is organized as follows. In Section \ref{sec:Methodology}, we summarize the method and describe the numerical setup we use to obtain the measurement errors for the physical parameters of precessing eccentric \mbox{NS--NS} and \mbox{NS--BH} binaries. We present our results in Section \ref{sec:Results}. Finally, we discuss our conclusions in Section \ref{sec:DiscAndConcl}. Details on the derivation of parameter measurement errors are presented in Appendices \ref{sec:ParamestErrore10Hz} and \ref{sec:DL_Dep}.
 
 In this paper, we use geometric units $c=G=1$. We work in the observer frame assuming a binary at cosmological redshift $z$, and neglect peculiar velocity and weak--lensing effects \citep{Kocsisetal2006}. In this frame, all of the formulae have redshifted mass parameters \mbox{$m_z = (1 + z) m$}.

\section{Methodology} 
\label{sec:Methodology}
 
 In this section, we summarize the method we used to obtain the parameter measurement accuracy of GWs generated by precessing eccentric \mbox{NS--NS} and \mbox{NS--BH} binaries. We refer the reader for more details to \citet{Gondanetal2018}.

\subsection{Waveform Model in Time Domain}
\label{subsec:TimeDomWaveform}
 
 We use the waveform model of \citet{MorenoGarridoetal1994} and \citet{MorenoGarridoetal1995}, which describes the quadrupole waveform emitted by a non-spinning binary on a Keplerian orbit undergoing slow pericenter precession. We ignore the radiation of higher multipole orders, which are typically subdominant at least in cases where the initial pericenter distance is not close to a grazing or zoom--whirl configuration and the initial velocity is much lower than the speed of light \citep{Davisetal1972,Bertietal2010,Healyetal2016}. For the time evolution of the orbital parameters semimajor axis and eccentricity, we adopt \citet{Peters1964}, who assumes quadrupole radiation and adiabatic evolution of orbital parameters. We include pericenter precession in the leading-order post--Newtonian (1PN) limit, which leads to a time-dependent angle of the pericenter. The time-domain waveform under these approximations is given explicitly in \citet{Mikoczietal2012} and \citet{Gondanetal2018}. For simplicity, we restrict our attention to the repeated burst and eccentric inspiral phase and neglect the contribution of the tidal interaction and disruption, merger, and ringdown phases to the signal.\footnote{Fully general-relativistic hydrodynamical simulations of highly eccentric \mbox{NS--NS} and \mbox{NS--BH} mergers \citep{Stephensetal2011,EastPretorius2012,Eastetal2012,Goldetal2012,Eastetal2015,Dietrichetal2015,Paschalidisetal2015,Eastetal2016,Radiceetal2016,Vivekanandjietal2018} and hydrodynamical simulations with Newtonian gravity of dynamical-capture \mbox{NS--NS} and \mbox{NS--BH} mergers \citep{Leeetal2010,Piranetal2013,Rosswogetal2013} revealed that the outcome of such mergers significantly depends on the impact parameter and the intrinsic parameters of the binary, i.e., the component masses, the initial orbital eccentricity, the initial spins, and the NS equation of state.}

\subsection{Phases of Binary Evolution}
\label{subsec:PhaseEvolution}
 
 We follow \citet{OLearyetal2009} to characterize the binary orbit with its orbital eccentricity $e$ and its dimensionless pericenter distance $\rho_{\rm p}= r_{p}/M_{\rm tot}$, which shrink strictly monotonically in time \citep{Peters1964}. Here, $ r_{p}$ is the orbital separation between the two objects at pericenter, and $M_{\rm tot}$ is the total mass of the binary. When the orbital eccentricity is high $e \gtrsim 0.7$, the time-domain waveform consists of repeated bursts, which are emitted during pericenter passages \citep{KocsisLevin2012}. As the orbital eccentricity decreases due to GW emission, the GW signal changes into an eccentric inspiral waveform. 
 
 For precessing eccentric \mbox{BH--BH} sources, we truncate the waveform  when the binary reaches the LSO, which we calculate in the leading-order approximation in the test-mass geodesic zero-spin limit for which \mbox{$\rho_\mathrm{p,LSO} = (6 + 2e_\mathrm{LSO})/(1 + e_\mathrm{LSO})$} \citep{Cutleretal1994}. Thus, we consider the waveform model between \mbox{$0 < e_\mathrm{LSO} < e < e_0 < 1$} or equivalently between \mbox{$0 < \rho_\mathrm{p,LSO} < \rho_\mathrm{p} < \rho_{\mathrm{p}0}$}. Here we note that \mbox{$4 < \rho_\mathrm{p,LSO} < 6$} for $e_\mathrm{LSO} \in ]0,1[$.  
 
 For precessing eccentric \mbox{NS--NS} and \mbox{NS--BH} sources, we account for the repeated burst and eccentric inspiral phase ignoring tidal effects and matter exchange among the components \citep{Blanchet2014}. This approximation breaks down abruptly in the late inspiral phase, but they are reasonable in the earlier parts of the GW signal because of the steep separation dependence of these effects. Simulations of dynamical-capture \mbox{NS--NS} mergers show that tidal deformation and matter exchange become significant when the NSs first come into contact \citep{Goldetal2012,Eastetal2012,Eastetal2016,Radiceetal2016}, which happens when $\rho_\mathrm{p} \lesssim 8.5$  for non-spinning binaries \citep{Eastetal2012}. Simulations of dynamical-capture \mbox{NS--BH} mergers with an NS mass of $1.35 \, \Msun$ and a mass ratio of $q = 1/4$ (where \mbox{$0 < q \leqslant 1$}) in \citet{Eastetal2015} showed that binaries with $\rho_\mathrm{p} \lesssim 6.5$ merge on the initial encounter, while those with $\rho_\mathrm{p} \gtrsim 7.5$ go back out on a quasi-elliptic orbit after flyby. These limits shift down for higher-mass BHs because tidal effects on the NSs scale with $q^{-1/3}$; see Equation (3) in \citet{Leeetal2010}. For precessing eccentric \mbox{NS--NS} and \mbox{NS--BH} binaries, we terminate the signal at $\rho_\mathrm{p}= 8.5$ and $7.5$, respectively. Note that this is a conservative estimate for \mbox{NS--BH} binaries with mass ratios well below $1/4$. 
 
 In the following, we refer to $\rho_\mathrm{p,LSO}$ as the dimensionless pericenter distance at which the signal is terminated in our calculations as specified above and denote the corresponding eccentricity by $e_\mathrm{LSO}$.

\subsection{GW Detectors Used in the Analysis}
\label{subsec:GWdetectors}
 
 Similar to \citet{Gondanetal2018}, we use the \mbox{aLIGO--AdV--KAGRA} detector network and adopt the design sensitivities for the two aLIGO \citep{Abbottetal2016a}, the AdV \citep{Abbottetal2016a}, and KAGRA \citep{Somiya2012} detectors.\footnote{Note, further, that the lower and upper limits of the advanced detector's band are typically $10 \, \mathrm{Hz}$ and $10^4 \, \mathrm{Hz}$, respectively.} The locations and orientations of these detectors are summarized in Table \ref{tab:DetCord}, and we define the orientation angle $\psi$ for each detector as the angle measured clockwise from north between the $x$-arm of the detector and the meridian that passes through the position of the detector (see Appendix B in \citet{Gondanetal2018} for the geometric conventions of detectors). We assume that the noise in each detector is stationary colored Gaussian with zero mean, and that it is uncorrelated between the different detectors.

\subsection{Parameters of Precessing Eccentric Binaries}
\label{subsec:ParamsPrecBinaries} 
 
 The adopted waveform model involves $12$ free parameters\footnote{The logarithms are introduced for convenience in order to measure the relative errors, e.g., \mbox{$ (\Delta D_L) / D_L = \Delta (\ln D_L)$}.}, which are $t_c$, $\Phi_c$, $\gamma_c$, $\theta_N$, $\phi_N$, $\theta_L$, $\phi_L$, $\ln D_L$, $\ln \mathcal{M}_z$, $\ln M_{{\rm tot},z}$, $e_0$ and $e_\mathrm{LSO}$. Here, $t_c$ is the coalescence time, $\Phi_c$ is the orbital phase at $t_c$, $\gamma_c$ is the angle of periapsis at $t_c$, $\theta_N$ and $\phi_N$ are polar angles describing the sky position angles of the source in a coordinate system fixed relative to the center of the Earth such that $\theta_N = 0$ is along the geographic north pole and $\phi_N$ is along the prime meridian, $\theta_L$ and $\phi_L$ are the angular momentum vector direction angles in spherical coordinates defined similar to $\theta_N$ and $\phi_N$, $D_L$ is the luminosity distance between the center of the Earth and the GW source, $\mathcal{M}_z$ is the redshifted chirp mass at cosmological redshift $z$, $M_{{\rm tot},z}$ is the redshifted total mass at cosmological redshift $z$, $e_0$ is the orbital eccentricity at formation, and $e_\mathrm{LSO}$ is the eccentricity at the LSO. 
 
 The first $10$ parameters ($t_c$, $\Phi_c$, $\gamma_c$, $\theta_N$, $\phi_N$, $\theta_L$, $\phi_L$, $\ln D_L$, $\ln \mathcal{M}_z$, and $\ln M_{{\rm tot},z}$) are measurable independently for precessing nearly circular binaries, where $\gamma_c$ and $M_{{\rm tot},z}$ dependence of the waveform is due to the first post-Newtonian effect of pericenter precession. Further, we showed in \citet{Gondanetal2018} that $e_0$ and $e_\mathrm{LSO}$ can also be extracted from the waveform model if it leaves an imprint at the beginning and the end of the waveform, respectively (i.e., if the binary forms and the inspiral terminates in the advanced GW detectors' frequency band).
 
 For the polar angles ($\theta_N, \phi_N$), we calculate and present the minor and major axes ($a_N, b_N$) of the corresponding 2D sky location error ellipse introduced in \citet{LangHughes2006}. We do the same for the polar angles ($\theta_L, \phi_L$) by constructing the corresponding minor and major axes of the binary orientation error ellipse ($a_L, b_L$). We measure the errors for the rest of the parameters, where $\Delta D_L / D_L = \Delta (\ln D_L)$, and similarly for $\mathcal{M}_z$ and $M_{{\rm tot},z}$. 
 
 Note that the marginalized measurement errors of other parameters describing the source can be determined by linear combinations of the covariance matrix elements of the adopted parameters. For instance, the measurement error of $\rho_{\mathrm{p}0}$ is determined as 
\begin{align}  \label{eq:deltarhop0}
  (\Delta \rho_{\mathrm{p}0})^2 
 = & \left( \frac{\partial \rho_\mathrm{p}(e_0,e_\mathrm{LSO})}{\partial e_0} \right)^2 (\Delta e_0)^2
   \nonumber\\ 
   & + \left( \frac{\partial \rho_\mathrm{p}(e_0,e_\mathrm{LSO})}{\partial e_\mathrm{LSO}} \right)^2 
  (\Delta e_\mathrm{LSO})^2
  \nonumber\\
  & + 2 \frac{\partial \rho_\mathrm{p} (e_0,e_\mathrm{LSO})}{\partial e_0} \frac{\partial \rho_\mathrm{p}(e_0,
  e_\mathrm{LSO})}{\partial e_\mathrm{LSO}} \left\langle \Delta e_0 \Delta e_\mathrm{LSO} \right\rangle\,
\end{align}
 because $\rho_\mathrm{p}$ can be expressed as a function of $e_0$ and $e_\mathrm{LSO}$ using the adopted adiabatic evolution equations and the equation defining the $e_\mathrm{LSO}$ \citep{Gondanetal2018}. Here and in Equations (\ref{eq:deltae10Hz}) and (\ref{eq:errdeltav}) below, $\left\langle \Delta \lambda_i \Delta \lambda_j \right\rangle$ denotes the cross-correlation coefficient between the inferred parameter values $\lambda_i$ and $\lambda_j$ relative to their fiducial values.
 
 Similarly, we express $e_{10{\rm Hz}}$ as a function of $e_{\rm LSO}$ and  $M_{{\rm tot},z}$; see Appendix \ref{sec:ParamestErrore10Hz} for details. Its measurement error for binaries forming outside of the detector frequency bands is formally
\begin{align}  \label{eq:deltae10Hz}
 (\Delta e_{10 {\rm Hz}})^2 
 = & \left( \frac{\partial e_{10 {\rm Hz}}(e_\mathrm{LSO}, \ln M_{{\rm tot},z})}{\partial e_\mathrm{LSO}} \right)^2  (\Delta e_\mathrm{LSO} )^2
   \nonumber\\ 
   & + \left( \frac{\partial e_{10 {\rm Hz}}(e_\mathrm{LSO}, \ln M_{{\rm tot},z})}{\partial \ln M_{{\rm tot},z}} \right)^2  (\Delta \ln M_{{\rm tot},z} )^2
  \nonumber\\
  & + 2 \frac{\partial e_{10 {\rm Hz}} (e_\mathrm{LSO}, \ln M_{{\rm tot},z} )}{\partial e_\mathrm{LSO}} 
  \nonumber\\
  & \times \frac{\partial e_{10 {\rm Hz}}( e_\mathrm{LSO}, \ln M_{{\rm tot},z})}{\partial \ln M_{{\rm tot},z} } \left\langle \Delta e_\mathrm{LSO} \Delta \ln M_{{\rm tot},z} \right\rangle \, .
\end{align} 
 
 The NSs' equation of state leaves its imprint on the GW signal during the late stages of the inspiral when the two bodies are tidally distorted  \citep{Blanchet2014}, and during the merger and post-merger phases \citep{FaberRasio2012,BaiottiRezzolla2017,PaschalidisStergioulas2017}. Thus, the tidal deformability information can be efficiently extracted from the late inspiral, merger, and post-merger phases of the waveform \citep{HarryHinderer2018}. However, as already mentioned, for this first study of precessing eccentric \mbox{NS--NS} and \mbox{NS--BH} sources, we ignore the tidal deformabilities in this study for simplicity (Section \ref{subsec:PhaseEvolution}).
 
 Following \citet{Kocsisetal2007}, we group the parameters as
\begin{align} \label{eq:fast}
  \bm{\lambda}_\mathrm{fast} & = \lbrace \Phi_c,  t_c, \ln \mathcal{M}_z, \ln M_{{\rm tot},z}, \gamma_c, e_{\rm LSO} \rbrace 
  \, , \\
  \bm{\lambda}_\mathrm{slow} & = \lbrace \ln D_L, \theta_N, \phi_N, \theta_L, 
  \phi_L \rbrace \, , \label{eq:slow}
\end{align} 
 where the $\bm{\lambda} _\mathrm{fast}$ fast parameters are related to the high-frequency GW phase, while the $\bm{\lambda} _\mathrm{slow}$ slow parameters appear only in the slowly varying amplitude of the GW signal. Slow parameters are mostly determined from a comparison  of the GW signals measured by the different detectors in the network.

\begin{table}
\centering  
   \begin{tabular}{@{}lrrr}
     \hline
       Detector  &  East Long.  &  North Lat. &  Orientation $\psi$  \\
     \hline\hline
       LIGO H  & $ -119.4^{\circ}$ & $46.5^{\circ}$ & $-36^{\circ}$ \\
       LIGO L  &   $ -90.8^{\circ}$   &   $ 30.6^{\circ} $ & $-108^{\circ}$  \\
       VIRGO &    $ 10.5^{\circ}$  &  $ 43.6^{\circ}$ & $ 20^{\circ}$  \\
       KAGRA &    $ 137.3^{\circ}$ &   $ 36.4^{\circ}$ & $ 65^{\circ}$  \\
     \hline
   \end{tabular} 
   \caption{ \label{tab:DetCord} Locations and orientations of the considered GW detectors in the coordinate system defined in Appendix B in \citet{Gondanetal2018}. LIGO H marks the Advanced LIGO detector in Hanford, WA, and LIGO L marks the Advanced LIGO detector in Livingston, LA. }
\end{table}

\subsection{Mass Ranges}
\label{subsec:CompMassRange}
 
 We set the lower bound of the BH mass range of interest to $5 \, M_{\odot}$ according to the observations of X-ray binaries \citep{Bailynetal1998,Ozeletal2010,Farretal2011,Belczynskietal2012}. In order to set the upper bound, we consider the following arguments: (i) supernova theory \citep{Belczynskietal2014,Belczynskietal2016b,Marchantetal2016,Gilmeretal2017,Mapellietal2017,Woosleyetal2017} predicts a mass gap between $\sim 50 \, M_{\odot}$ and $\sim 135 \, M_{\odot}$ in the BH initial mass function, (ii) studies based on the current GW observations of BH mergers and population synthesis find that the gap may start between $\sim 40 \, M_{\odot}$ and $\sim 76 \, M_{\odot}$ \citep{FishbachHolz2017,Baietal2018,Giacobboetal2018}, and (iii) the evolution of the BH initial mass function may fill the mass gap due to mergers of BHs but drops off quickly beyond $\sim 100 \, M_{\odot}$ \citep{Christianetal2018}. Since the mass function of stellar-mass BHs is still uncertain and systems more massive than a few times $100 \, M_{\odot}$ will not lie in the most sensitive band of aLIGO-type instruments during the eccentric inspiral, we do not consider systems beyond $\sim 110 \, M_{\odot}$ in this paper. 
 
 For NSs, statistical analyses based on observations of millisecond pulsars, double NS systems, and X-ray binaries predict a bimodal mass distribution with a sharp and dominant peak at $\sim 1.35 \, M_{\odot}$ \citep{Valentimetal2011,Ozeletal2012,Kiziltanetal2013,Antoniadisetal2016,OzelFreier2016,Alsingetal2018}. We set the NS mass to $1.35 \, M_{\odot}$ in this study.
 
 We note that we validated the codes in a larger mass range between $1 \, \Msun$ and $110 \, \Msun$ (Sections \ref{subsec:FreqDomWaveform} and \ref{subsec:CodeValidation}).\footnote{Here, the lower bound corresponds to the estimated minimum mass of NSs \citep{Valentimetal2011,Ozeletal2012,Kiziltanetal2013,Antoniadisetal2016,OzelFreier2016,Alsingetal2018}, the upper bound corresponds to the above considered upper bound of the BH mass range of interest, and we ignore the mass gap between NSs and BHs, i.e., between $2.17 \, \Msun$ (see \citet{MargalitMetzger2017} and references therein) and $5 \, \Msun$.}

\newpage 
 
\subsection{Waveform Model in Frequency Domain} 
\label{subsec:FreqDomWaveform}
 
 The expressions defining the signal-to-noise ratio ($S/N$) and the Fisher matrix components are both given in Fourier space, hence we calculate the Fourier transform of the adopted time-domain measured signal. The time-dependent GW signal measured by the $k$th detector can be given as
\begin{align}  \label{eq:hkt}
  h_k(t) & = h_+(t) F_{+,k} \left[\theta_N(t), \phi_N, \theta_L, \phi_L \right] 
  \nonumber  \\ 
 & + h_{\times}(t) F_{\times,k} \left[\theta_N(t), \phi_N, \theta_L, \phi_L \right] \, ,
\end{align}
 where $h_+(t)$ and $h_{\times}(t)$ are the two polarization states of the adopted waveform model defined in Section \ref{subsec:TimeDomWaveform}, and $F_{+,k}$ and $ F_{\times,k}$ are the antenna factors of the $k$th detector quantified in Appendix B in \citet{Gondanetal2018}. Because we ignore spins in this study, the angular momentum vector direction $(\theta_L, \phi_L)$ is conserved during the eccentric inspiral \citep{CutlerFlanagan1994}. 
 
 The polar angle of the source $\theta_N$ relative to the detector depends on the rotation phase of the Earth during the day, hence, Earth's rotation may in principle leave an imprint on the signal. To see if this may be significant, we determine the total time the GW signal spends in the advanced GW detectors' frequency band for different harmonics as a function of $\rho_{\mathrm{p}0}$ following \citet{Gondanetal2018}. In comparison with the analysis presented in Appendix C in \citet{Gondanetal2018}, we extend the component mass range to $[1 \, M_{\odot}, 110 \, M_{\odot}]$ (Section \ref{subsec:CompMassRange}), and we draw $e_0$ between $[0.9, \,1[$ in order to investigate binaries that enter the advanced GW detectors' frequency band with moderate or high eccentricities. We find that for high $\rho_{\mathrm{p}0}$, the signal effectively circularizes by the time it enters the advanced GW detectors' frequency band, and it spends seconds to several minutes in the advanced detectors' frequency band \citep[see also][]{KocsisLevin2012}. For lower $\rho_{\mathrm{p}0}$ (i.e., when binaries enter the advanced GW detectors' frequency band with moderate or high eccentricities), the signal may spend up to several hours in the detector bands.\footnote{ For instance, higher orbital harmonics of the signal for an eccentric \mbox{NS--NS} binary with $e \sim 0.9$ and $\rho_{p0}=150$, which forms inside the advanced GW detectors' frequency band (Section \ref{subsec:NumSims_2}), spend up to $2\times 10^4\,\rm s$ in the bands; see Appendix C in \citet{Gondanetal2018} for details of these calculations. The maximum time in a band scales with $M_{\rm tot}$ \citep{KocsisLevin2012,Gondanetal2018}. } However, more than $\sim 99 \%$ of the S/N accumulates in only a few seconds to a few $\times 10$ of minutes; see Appendix C in  \citet{Gondanetal2018} for details. Thus, the Earth's rotation and orbit around the Sun may help to decrease the parameter measurement errors only for rare very high S/N events, but in most cases, this can be neglected for precessing eccentric \mbox{NS--NS}, \mbox{NS--BH}, and \mbox{BH--BH} binaries. In this paper, we neglect these low S/N modulation effects.

\subsection{Parameter Values for Calculations as a Function of $\rho_{{\rm p}0}$}
\label{subsec:NumSetup}
 
 Here we list the assumed parameter values for calculations used to measure the S/N and parameter measurement errors as a function of $\rho_{{\rm p}0}$ and $e_{10\,\Hz}$, respectively.
 
 First, we carry out simulations for precessing eccentric \mbox{NS--NS}, \mbox{NS--BH}, and \mbox{BH--BH} binaries in which we examine the $e_0$, $\rho_{\mathrm{p}0}$, and component mass dependent properties of the total network S/N ($\mathrm{S/N} _\mathrm{tot}$, see Section 4 in \citealt{Gondanetal2018}) and the measurement errors of binary parameters. To do so, we determine these quantities as a function of $\rho_{\mathrm{p}0}$ for arbitrarily fixed component masses, luminosity distance, binary direction and orientation angles, and orbital eccentricity at formation as follows.
 
 We present the results of our analysis for an arbitrary choice of angular parameters $\theta_N = \pi/2$, $\phi_N = \pi/3$, $\theta_L = \pi/4$, and $\phi_L  = \pi/5$. Further, to explore the range of parameter errors, we repeated our calculations for $100$ other randomly chosen angles by drawing $(\cos \theta_N, \cos \theta_L)$ and ($\phi_N,\phi_L$) from isotropic distributions between $[-1, 1]$ and $[0, 2\pi]$, respectively. We preformed the calculations for $D_L = 100 \, \mathrm{Mpc}$, and scaled the results for an arbitrary choice of $D_L$ up to $1 \, \Gpc$.\footnote{The parameter measurement errors and the $\mathrm{S/N}_\mathrm{tot}$ roughly scale as $\propto D_L$ and $\propto 1/D_L$, respectively, between $100 \, \Mpc$ and $1 \, {\rm Gpc}$; see Appendix \ref{sec:DL_Dep} for details.} We present results for $e_0 = 0.9$ at any given $\rho_{{\rm p}0}$ and also examine higher $e_0$ values ($0.9 < e_0 < 1$) to show that the results are not very sensitive to this assumption \citep{Gondanetal2018}. By increasing $\rho_{p0}$ beyond the value where the binary forms in the band, we cover the full possible range of eccentricities at the point when the binary enters the detector frequency bands (Section~\ref{subsec:NumSims_2}).We assume the fiducial value \mbox{$t_c = \Phi_c = \gamma_c = 0$} because these parameters are responsible for an overall phase shift of the waveform and thus did not randomize their values \citep{Gondanetal2018}. 
 
 After fixing the component masses, $m_A$ and $m_B$, and the luminosity distance, $D_L$, the redshifted mass parameters $\mathcal{M}_z$ and $M_{{\rm tot},z}$ can be given as $\mathcal{M}_z = (1+z) (m_A m_B)^{3/5}(m_A + m_B)^{-1/5}$ and $M_{{\rm tot},z} = (m_A + m_B) (1 + z)$, respectively. By assuming a flat $\Lambda$CDM cosmology, $z$ is calculated numerically as
\begin{equation}   \label{eq:CovDLz} 
  D_L = \frac{ (1+z)c }{ H_0 } \int_0 ^z \frac{dz'}{\sqrt{ \Omega_\mathrm{M} 
  (1+z')^3 + \Omega_{\Lambda} }} \, ,  
\end{equation}
 where  $H_0 = 68 \mathrm{km} \, \mathrm{s}^{-1} \mathrm{Mpc}^{-1}$ is the Hubble constant, and \mbox{$\Omega_\mathrm{M} = 0.304$} and \mbox{$\Omega_{\Lambda} = 0.696$} are the density parameters for matter and dark energy, respectively \citep{PlanckCollaboration2014a,PlanckCollaboration2014b}.
 
 Following \citet{Gondanetal2018}, we set the lower bound of the $\rho_{\mathrm{p}0}$ range of interest to $5$ for \mbox{BH--BH} binaries. Further, for \mbox{NS--NS} and \mbox{NS--BH} binaries, the lower bound of the $\rho_{\mathrm{p}0}$ range is set to $9$ and $8$, respectively, according to the $\rho_\mathrm{p}$ cutoffs; see Section \ref{subsec:PhaseEvolution} for details. These low values are likely for eccentric \mbox{BH--BH} binaries (see \citealt{Gondanetal2018} for a summary of channels), and also likely for eccentric \mbox{NS--NS} and \mbox{NS--BH} sources that form through the GW capture mechanism in high-velocity dispersion environments such as GNs and GCs \citep{OLearyetal2009}, tidal capture during close encounters in GCs that underwent core collapse \citep{Leeetal2010}, binary--single interaction in GCs \citep{Samsingetal2014}, and dynamical multi-body interaction in the core-collapsed regions of star clusters without a central massive BH \citep{AntoniniRasio2016}. We refer to the circular limit by setting $e_0 \rightarrow 0$ for arbitrary fixed $\rho_{\mathrm{p}0}$ or $\rho_{\mathrm{p}0} \rightarrow \infty$ for arbitrary fixed $e_0 \leqslant 1$; see Figure 5 in \citet{OLearyetal2009}. In practice, we set $\rho_{\mathrm{p}0} = 800$. We set the upper bound of the $\rho_{\mathrm{p}0}$ range to this value because in this case, binaries are expected to be well circularized by the time their peak GW frequency enters the advanced GW detectors' frequency band over the considered ranges of component mass and $e_0$.

\newpage 
 
\subsection{Calculations as a Function of $e_{\rm 10 Hz}$} 
\label{subsec:NumSims_2} 
 
 The orbital parameters $\rho_\mathrm{p}$ and $e$ both shrink due to GW emission \citep{Peters1964}, while $f_\mathrm{GW}$ continuously increases with shrinking $\rho_\mathrm{p}$ and $e$ \citep{Wen2003}. As a consequence, binaries in the above introduced analysis that have high $\rho_{\mathrm{p}0}$ enter the advanced GW detectors' frequency band with relatively low residual eccentricities, while those with relatively low $\rho_{\mathrm{p}0}$ form in the advanced GW detectors' frequency band. Recent studies \citep{BrownZimmerman2010,Eastetal2013,HuertaBrown2013,Huertaetal2014,Huertaetal2017,Huertaetal2018,Loweretal2018} have shown that the classification of binaries as quasi-circular or eccentric binaries depends on this eccentricity, i.e., an eccentric binary can be misclassified as quasi-circular binary if the residual eccentricity exceeds $\sim 0.1$. Therefore, similar to the analysis presented in Section \ref{subsec:NumSetup}, we carry out simulations for precessing eccentric \mbox{NS--NS}, \mbox{NS--BH}, and \mbox{BH--BH} binaries in which we examine the $e_{10\,\rm Hz}$ and component mass dependent properties of the $\mathrm{S/N} _\mathrm{tot}$ and the measurement errors of binary parameters. 
 
 For binaries that form outside of the advanced GW detectors' frequency band, it is customary to define the measurement errors of binary parameters as a function of $e_{10\,\rm Hz}$ when the peak GW frequency first enters the advanced GW detectors' frequency band, i.e., \mbox{$f_{\rm GW} = f_{\rm det} = 10\,\rm Hz$}. Here, we adopt \citet{Wen2003} to define $f_{\rm GW}$ as 
 \begin{equation}  \label{eq:fGW}
  f_{\rm GW} = \frac{2 (1+e)^{1.1954}}{(1-e^2)^{3/2}} \frac{M_{\mathrm{tot},z}^{1/2}}{2 \pi a^{3/2}} \, .
\end{equation}
For each component mass and $e_0$ defined in Section \ref{subsec:NumSetup}, we determine the initial dimensionless pericenter distance $\rho_{\rm p0,det}$ below which binaries form within the advanced GW detectors' frequency band. Using $a (1-e)= M_{\mathrm{tot},z} \rho_\mathrm{p}$ and Equation (\ref{eq:fGW}), we find that  
\begin{equation}  \label{eq:rhoDet_Gen}
  \rho_{\rm p0,det} =  \left( (1+e_0)^{0.3046} \pi M_{\mathrm{tot},z} f_{\rm det}\right)^{-2/3} \, ,
\end{equation} 
which reduces in the $e_0 \rightarrow 1$ limit to \citep{Gondanetal2017}
\begin{equation}  \label{eq:rhoaLIGO}
  \rho_{\rm p0,det} = 40.9 \left(\frac{M_{\mathrm{tot},z}}{20\,\Msun}\right)^{-2/3} \left( \frac{f_{\rm det}}{10\,\mathrm{Hz}}  \right)^{-2/3}  \, .
\end{equation} 
 
 Next, we calculate $\rho_{\mathrm{p}0}$ as a function of $e_{\rm 10 Hz}$ at fixed component masses and $e_0$ in order to carry out simulations similar to those presented in Section \ref{subsec:NumSetup}, but by initializing them with $e_{\rm 10 Hz}$ through $\rho_{\mathrm{p}0}$. We evolve systems from their initial orbital parameters using the evolution equations of \citet{Peters1964}, therefore the function $\rho_{\rm p} \equiv \rho_{\rm p}(e_0,\rho_{\mathrm{p}0},e)$ is known for any initial condition; see Equation (\ref{eq:rhoe}). After combining Equation (\ref{eq:fGW}) and $a (1-e) = M_{\mathrm{tot},z} \rho_\mathrm{p}$ together with $\rho_{\rm p}(e_0,\rho_{\mathrm{p}0},e)$, $\rho_{\mathrm{p}0}$ can be given for binaries forming outside of the detector frequency bands (i.e., $\rho_{\mathrm{p}0} \geqslant \rho_{\rm p0,det}$) as a function of $e_{\rm 10 Hz}$ at fixed $e_0$ and $M_{\mathrm{tot},z}$ by solving numerically the equation $f_{\rm det} = f_{\rm GW}$ for $\rho_{\mathrm{p}0}$, which can be written as
\begin{equation}  \label{eq:erho_fGW10Hz}
  f_{\rm det} = \left[(1 + e_{\rm 10 Hz})^{0.3046} M_{\mathrm{tot},z} \pi [\rho_{\rm p}( e_{\rm 10 Hz} , e_0,  \rho_{\rm p0})]^{3/2}\right]^{-1} \, .
\end{equation}
 
 For these calculations involving $e_{\rm 10 Hz}$, we focus on the range of $e_{\rm 10 Hz}$ between $[8 \times 10^{-4}, 0.9]$. For each choice of component masses and initial orbital eccentricity, we calculate the corresponding $\rho_{{\rm p}0}$ using Equation (\ref{eq:erho_fGW10Hz}), and use these values to evaluate the $\mathrm{S/N} _\mathrm{tot}$ and the measurement errors of binary parameters. Other parameters are chosen as in Section \ref{subsec:NumSetup}. 
 
\begin{figure*}
    \centering
    \includegraphics[width=80mm]{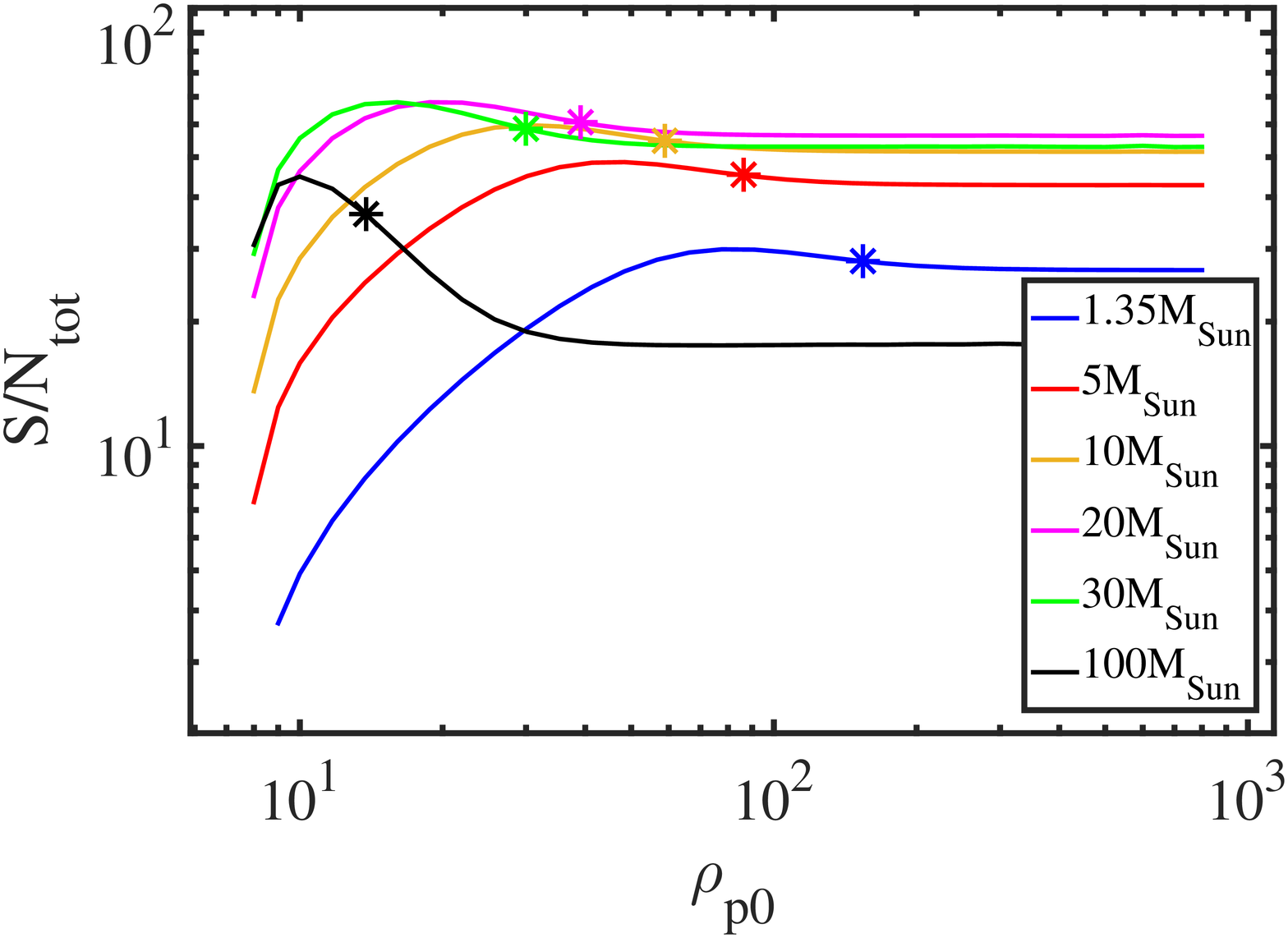}
    \includegraphics[width=80mm]{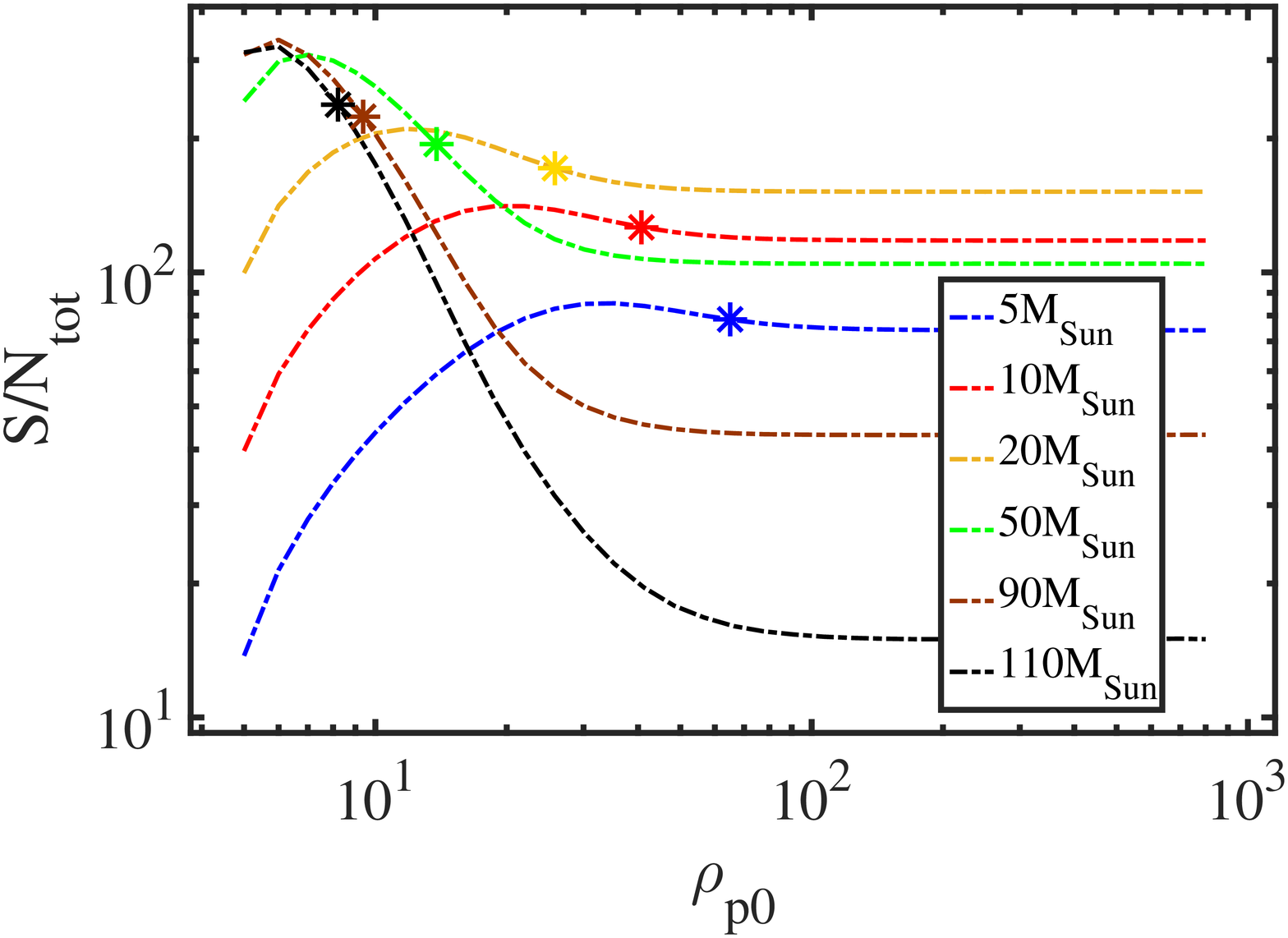}  \\
    \includegraphics[width=80mm]{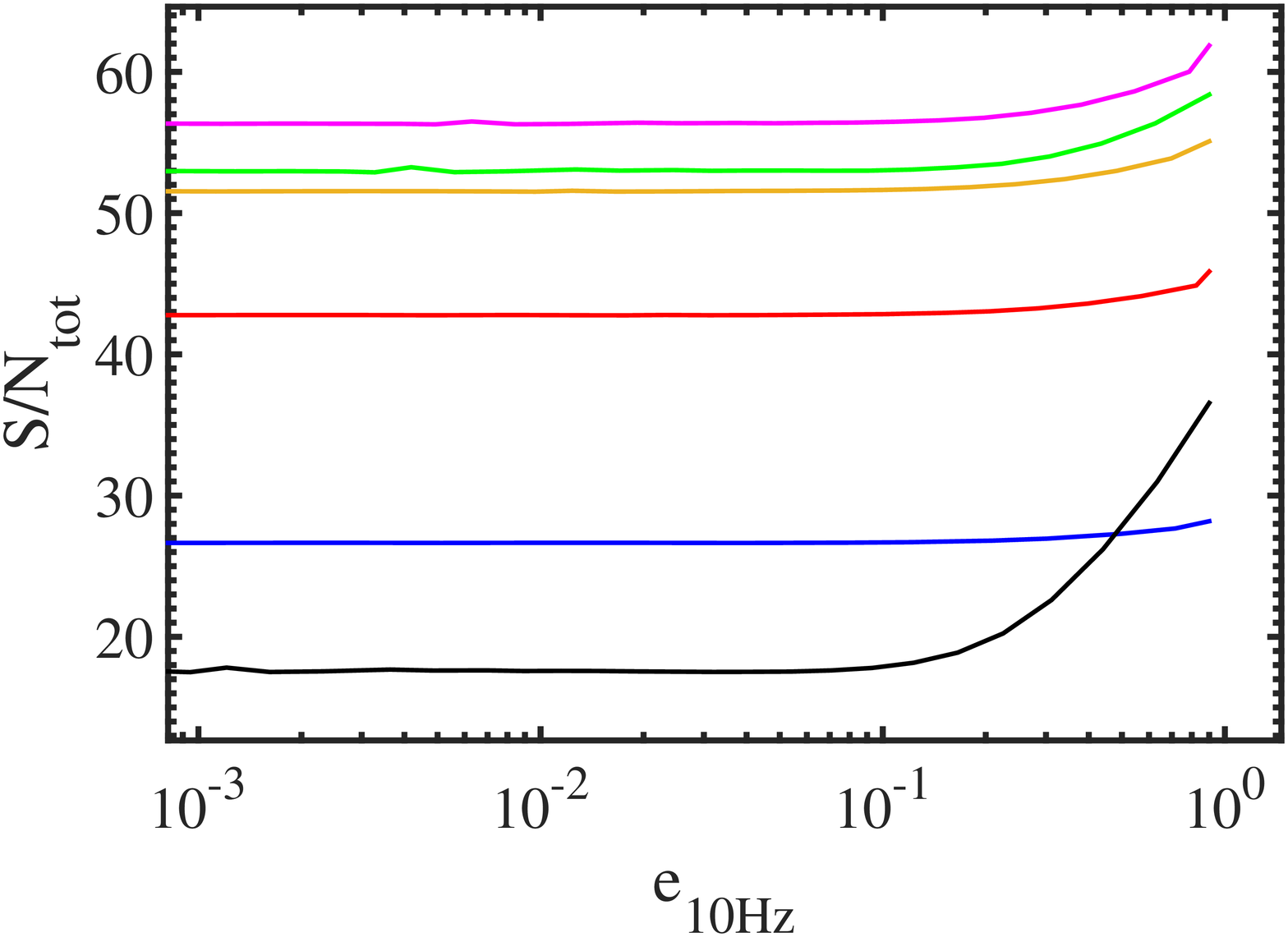}
    \includegraphics[width=80mm]{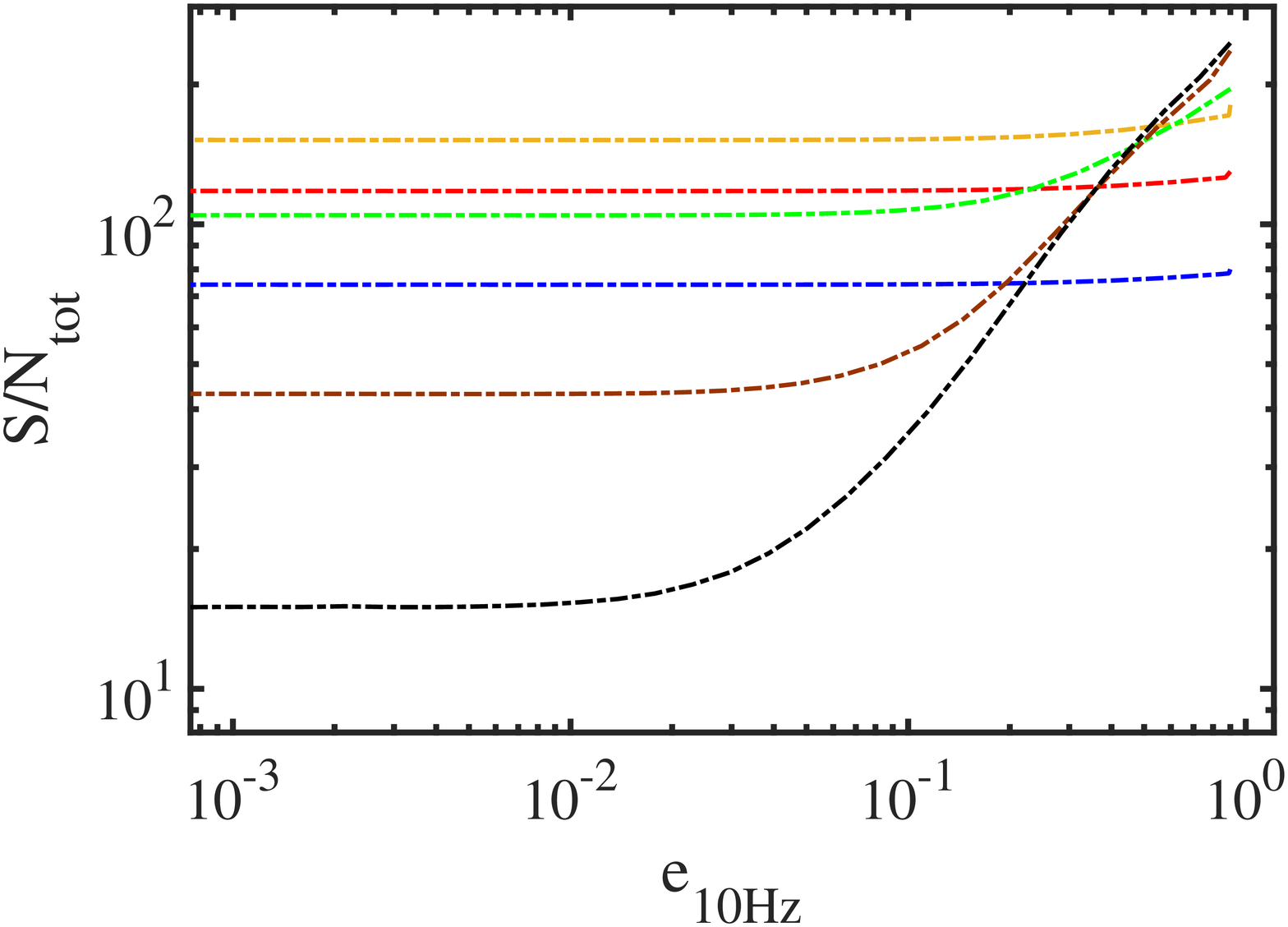}  
  \caption{Total GW network S/N ($\mathrm{S/N}_\mathrm{tot}$) for precessing eccentric \mbox{NS--NS} and \mbox{NS--BH} binaries (left panels) and precessing, eccentric, equal-mass \mbox{BH--BH} binaries (right panels) as a function of their initial dimensionless pericenter distance $\rho_{\mathrm{p}0}$ for initial eccentricity $e_0 = 0.9$ (top panels) and as a function of their eccentricity $e_{\rm 10 Hz}$ at which the peak GW frequency ($f_\mathrm{GW}$; \citet{Wen2003}) $f_\mathrm{GW} = 10 \, \mathrm{Hz}$ (bottom panels). In the left panels, the NS mass is $1.35 \, M_{\odot}$ and the mass of the companion is labeled in the legend. For each curve in the top panels, a star indicates $\rho_{\rm p0,det}$ defined as the $\rho_{\mathrm{p}0}$ at which $f_\mathrm{GW}$ first enters the advanced GW detectors' frequency band at $10 \, \mathrm{Hz}$; see Section \ref{subsec:NumSims_2} for details. Binaries with $\rho_{\mathrm{p}0} \leqslant \rho_{\rm p0,det}$ have $f_\mathrm{GW} \geqslant 10 \, \mathrm{Hz}$ and form inside the advanced GW detectors' frequency band. The bottom panels show binaries where the binaries form outside of the advanced GW detectors' frequency band with $\rho_{\mathrm{p}0} \geqslant \rho _{\rm p0,det} $. The source direction and orientation angular parameters are fixed arbitrarily at $\theta_N=\pi/2$, $\phi_N = \pi/3$, $\theta_L = \pi/4$, and $\phi_L = \pi/5$ (see Table \ref{tab:DetCord} for the parameters of the detector network). Results are displayed for $D_L=100 \, \Mpc$, which may be calculated for other source distances as $\mathrm{S/N}_\mathrm{tot} \times (100 \, \Mpc / D_L)$. The accuracy of this scaling is $5-12 \%$ ($7-17 \%$) for $500 \, \Mpc$ ($1 \, \Gpc$); see Appendix \ref{sec:DL_Dep} for details. We find similar trends with $\rho_{\mathrm{p}0}$ for other random choices of binary direction and orientation (not shown). The maximum of  $\mathrm{S/N} _\mathrm{tot}$ is sensitive to the binary mass and converges asymptotically to the circular limit for high $\rho_{\mathrm{p}0}$ and for low $e_{\rm 10 Hz}$.}   \label{fig:SNRtot_rho}
\end{figure*}

\subsection{Validation of Codes for \mbox{NS--NS}, \mbox{NS--BH}, and \mbox{BH--BH} Binaries}
\label{subsec:CodeValidation}
 
 We adopt the same computationally efficient formulae to calculate the S/N and the Fisher matrix components that were derived and validated in Appendices D and E in \citet{Gondanetal2018}. It was shown that the results of these codes for the $\mathrm{S/N} _\mathrm{tot}$ and the measurement errors of binary parameters are qualitatively consistent with those presented in previous studies \citep{Arunetal2005,OLearyetal2009,Yunesetal2009,KyutokuSeto2014,Sunetal2015,Maetal2017}; see Section 6.3 in \citet{Gondanetal2018} for details. We repeated the validation tests of the source codes for precessing eccentric \mbox{NS--NS} and \mbox{NS--BH} sources with the corresponding $\rho_\mathrm{p}$ cutoffs (Section \ref{subsec:TimeDomWaveform}) by drawing the component masses between $1 \, \Msun$ and $ 110 \, \Msun$ (Section \ref{subsec:CompMassRange}) with all other parameter ranges and distributions fixed as in Appendix E in \citet{Gondanetal2018}. We found similar trends with those presented in \citet{Gondanetal2018}. This justifies using the same codes to calculate the S/N and the Fisher matrix components for precessing \mbox{NS--NS} and \mbox{NS--BH} binaries on eccentric inspiraling orbits given the approximations listed above.

\section{Results} 
\label{sec:Results}
 
 We start with a brief summary of results on the parameter errors as a function of system parameters in Section \ref{subsec:overview} and present details in the following subsections. In Section \ref{subsec:MeasErrEccRhoP0} we present the measurement errors of eccentricity at different stages of the binary evolution and the initial dimensionless pericenter distance. In Section \ref{subsec:MeasErrSlowParamChirpMass} we show the measurement errors for source distance, sky location, binary orientation, and redshifted chirp mass. Next, in Section \ref{subsec:MeasErrCharVelDispOfHost} we present results for the measurement accuracy of the characteristic relative velocity of the sources. In Sections \ref{subsec:MeasErrorsInitialOrbParams} and \ref{subsec:MassDepOfErrors} we discuss the initial orbital parameter and mass dependence of measurement errors, respectively. Finally, in Section \ref{subsec:MeasErrFixSNR} we present measurement errors of various source parameters for fixed $\mathrm{S/N}_\mathrm{tot}$.

\subsection{Overview of Results}
\label{subsec:overview}
 
\begin{figure*}
    \centering
    \includegraphics[width=80mm]{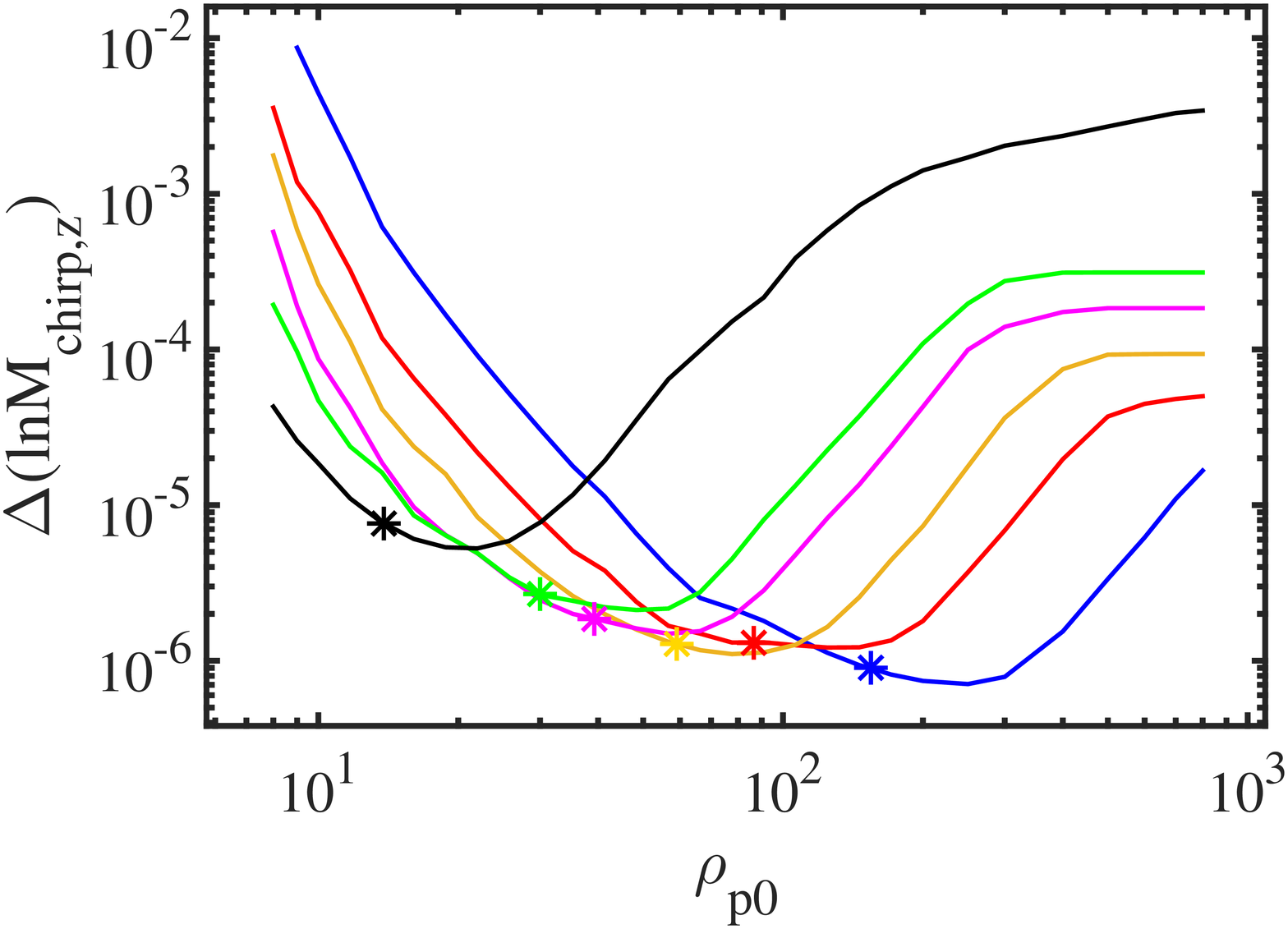}
    \includegraphics[width=80mm]{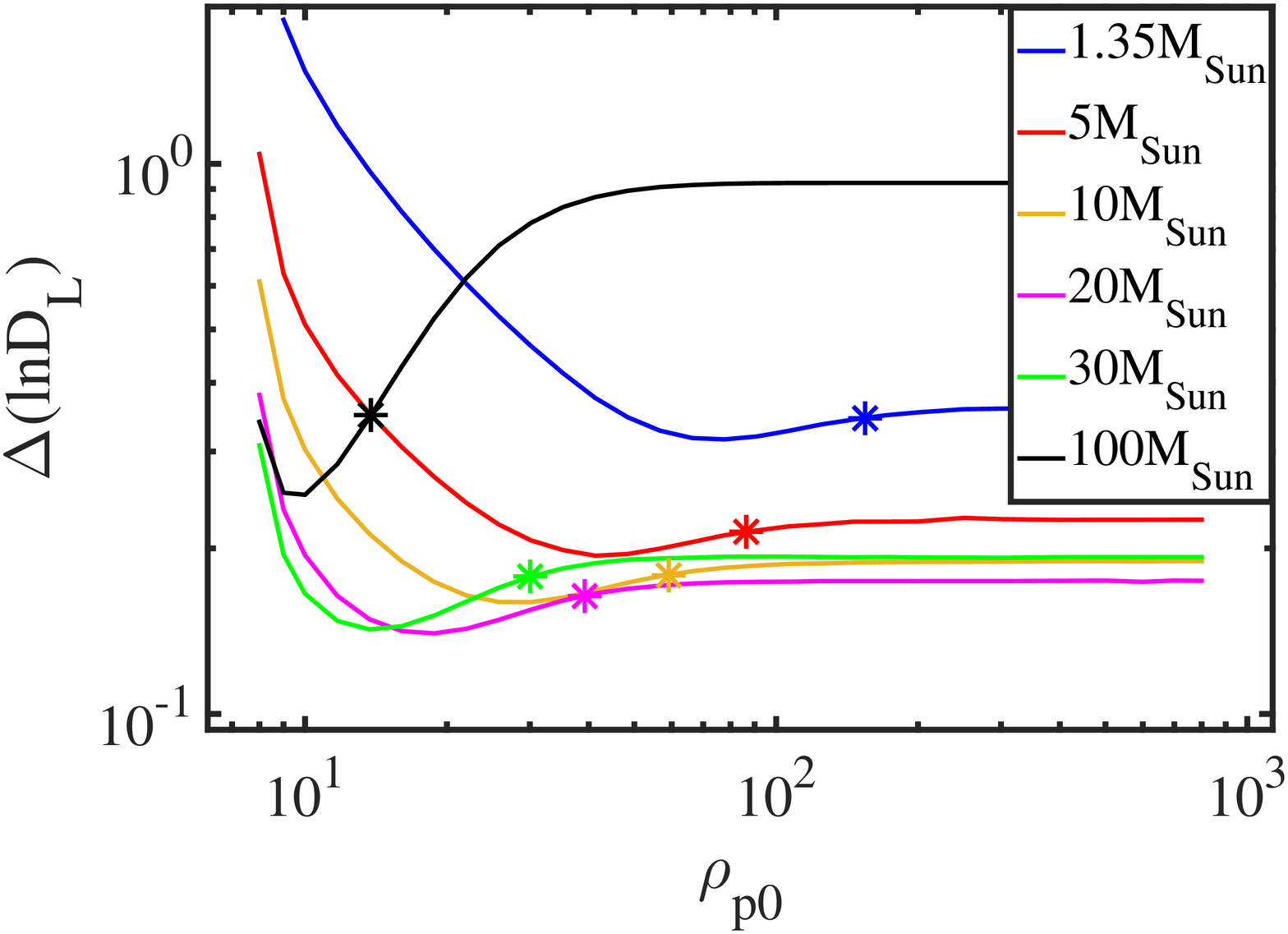}
\\
    \includegraphics[width=80mm]{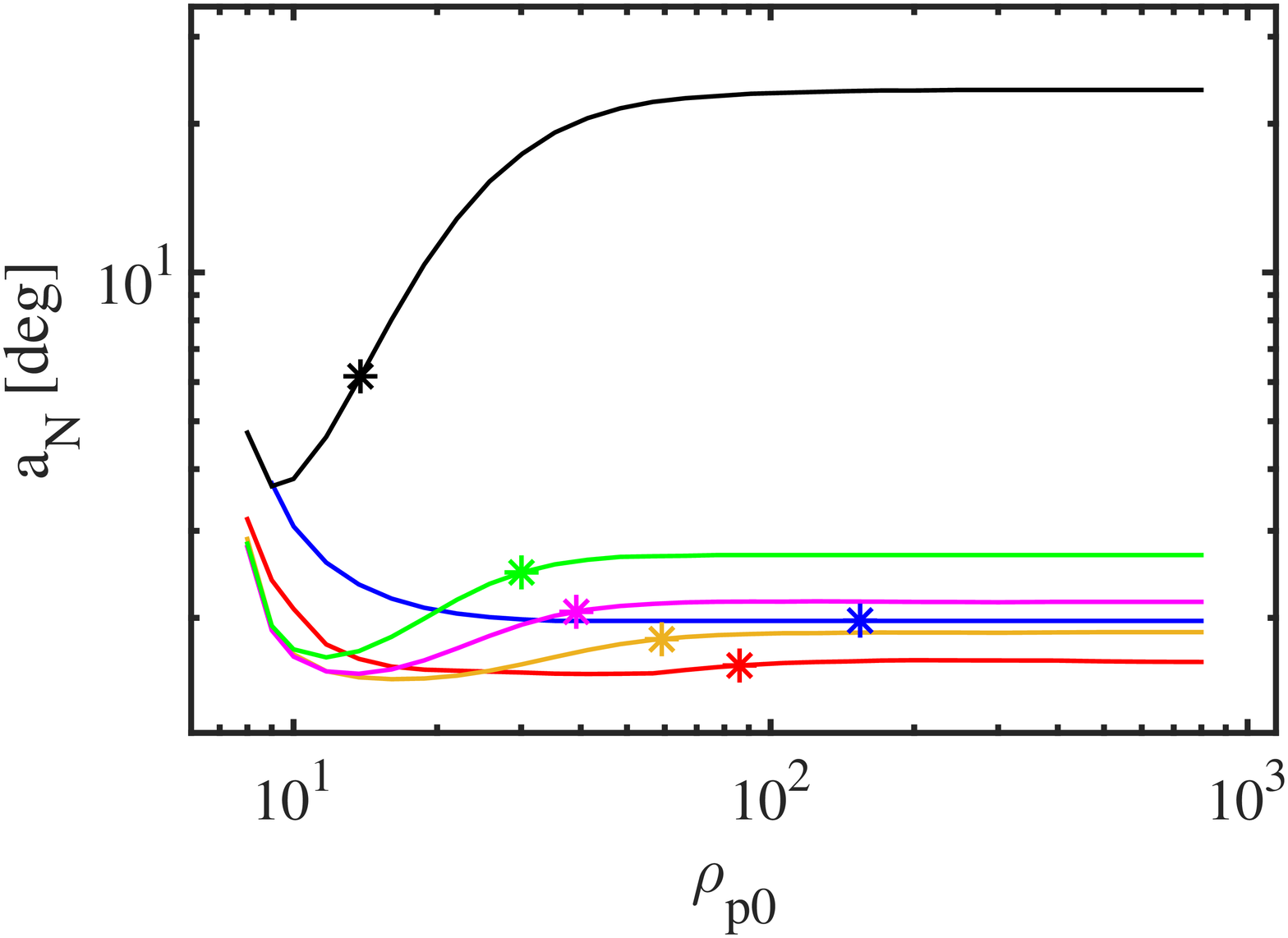}
    \includegraphics[width=80mm]{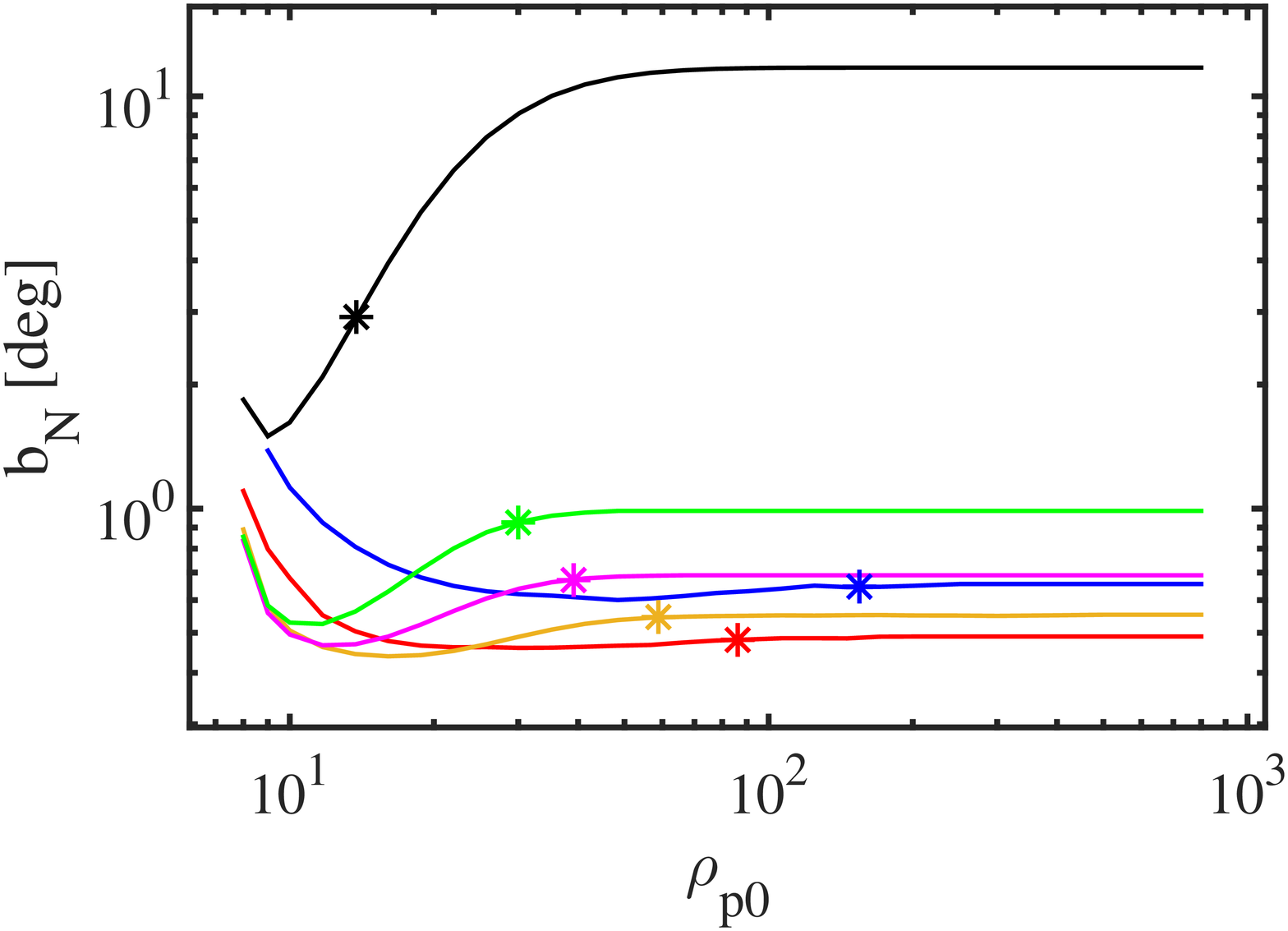}
\\
    \includegraphics[width=80mm]{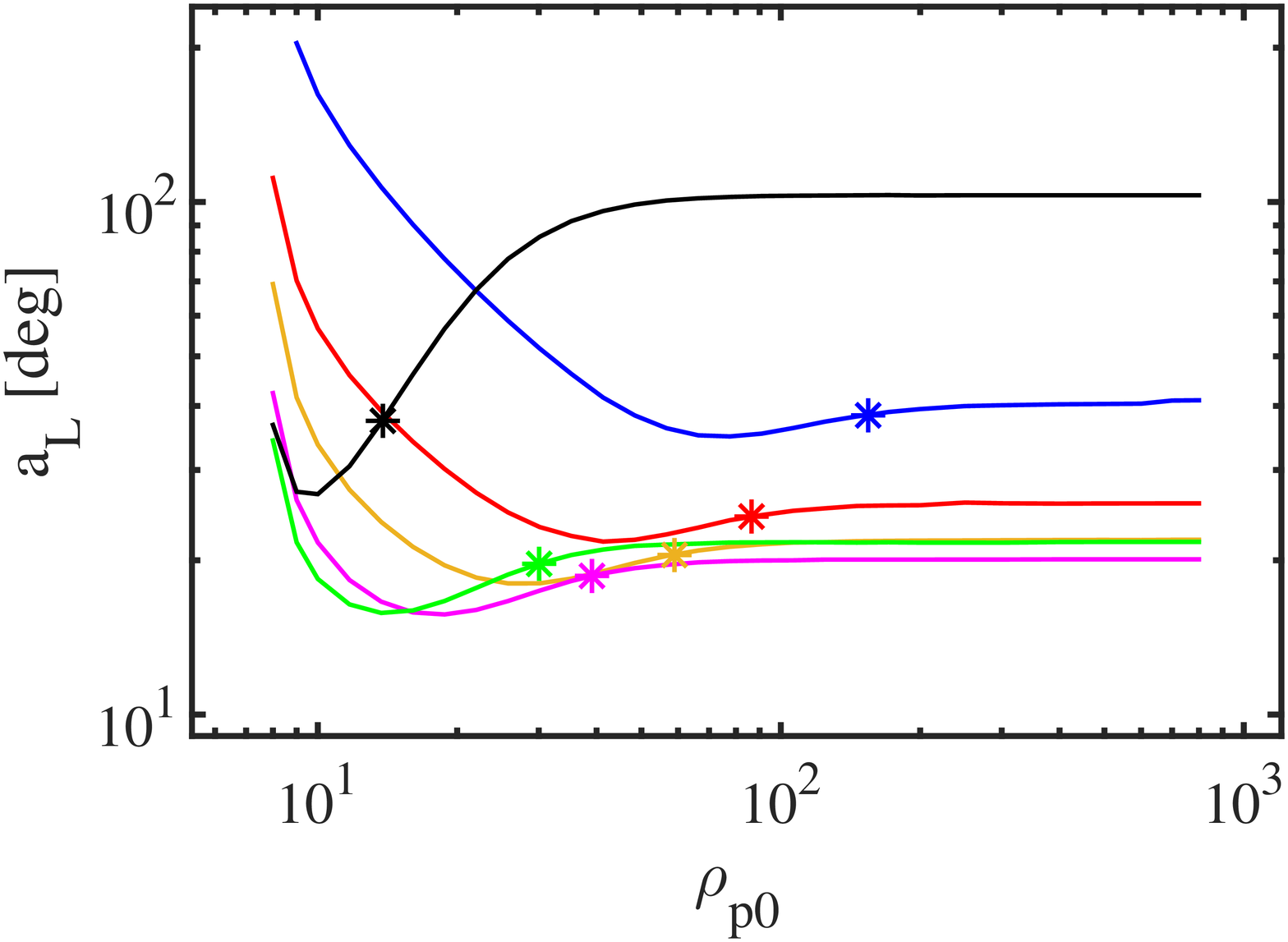}
    \includegraphics[width=80mm]{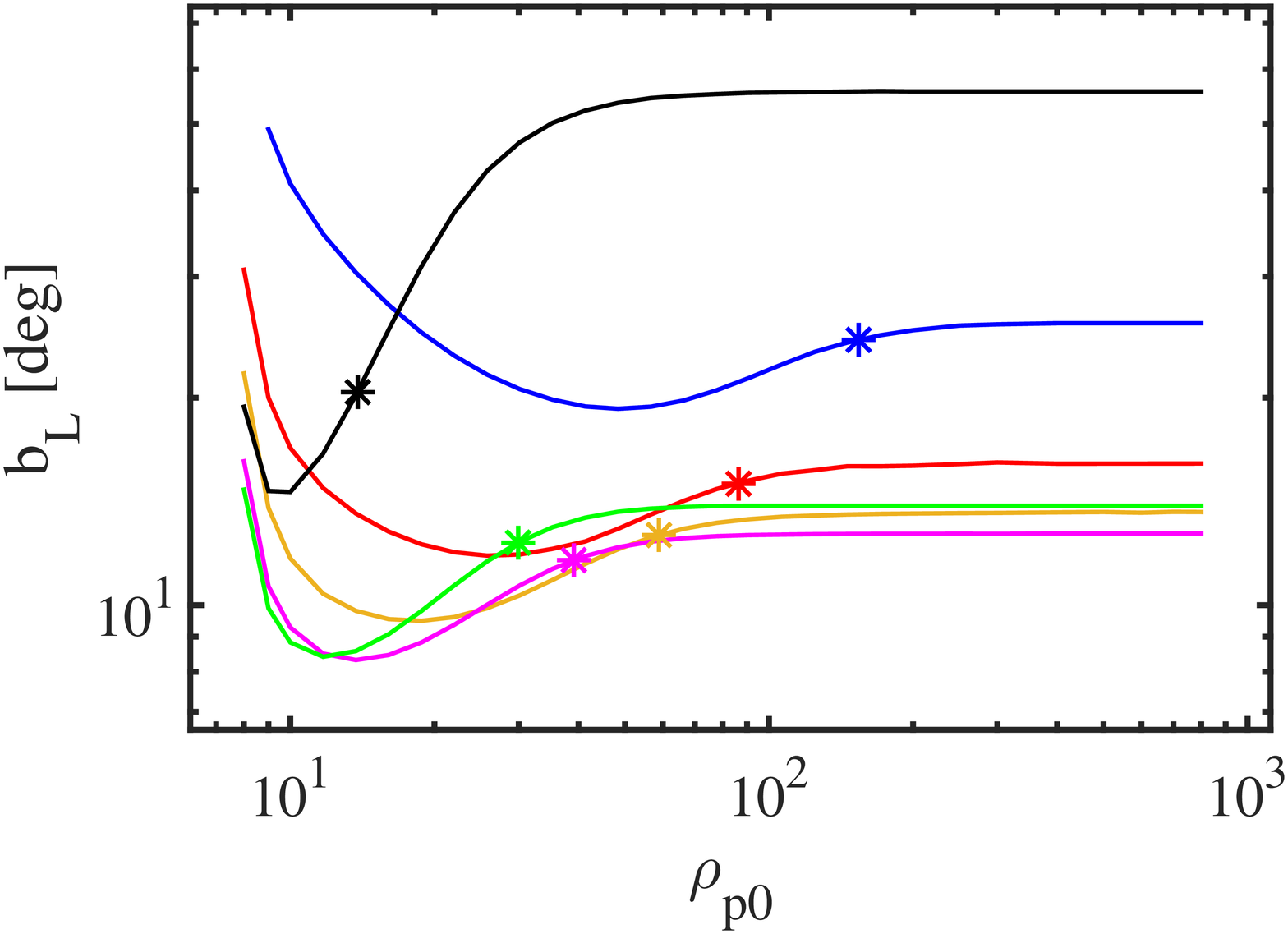}
  \caption{Measurement errors of various source parameters as a function of $\rho_{\mathrm{p}0}$ for precessing eccentric \mbox{NS--NS} and \mbox{NS--BH} binaries with different companion masses as shown in the legend with all other binary parameters fixed as in the left panel of Figure \ref{fig:SNRtot_rho}. Stars indicate the case where the binary forms at $f_{\rm GW} = 10\,\rm Hz$ as in Figure \ref{fig:SNRtot_rho}. The binary parameter measurement errors are displayed for $D_L = 100 \, \Mpc$, but can be scaled to other $D_L$ values by multiplying the displayed results by a factor of $D_L / 100 \, \Mpc$. This scaling is accurate to within $3-15 \%$ ($5-21 \%$) for $500 \, \Mpc$ ($1 \, \Gpc$); see Appendix \ref{sec:DL_Dep} for details. First row left: Redshifted chirp mass, $\Delta \mathcal{M}_z / \mathcal{M}_z= \Delta (\mathrm{ln} \mathcal{M}_z)$. First row right: Luminosity distance $\Delta D_L / D_L = \Delta (\mathrm{ln} D_L)$. Second row: Semi-major (left) and semi-minor (right) axes of the sky position error ellipse, $a_N$ and $b_N$. Third row: Semimajor (left) and semiminor (right) axes of the error ellipse for the binary orbital plane normal vector direction, $a_L$ and $b_L$. The measurement errors converge asymptotically to the circular limit for high $\rho_\mathrm{p0}$. We find similar trends with $\rho_{\mathrm{p}0}$ for other random choices of binary direction and orientation (not shown).  }  \label{fig:ParamEst_BHNS_Circular}
\end{figure*}

\begin{figure*}
   \centering
  \includegraphics[width=80mm]{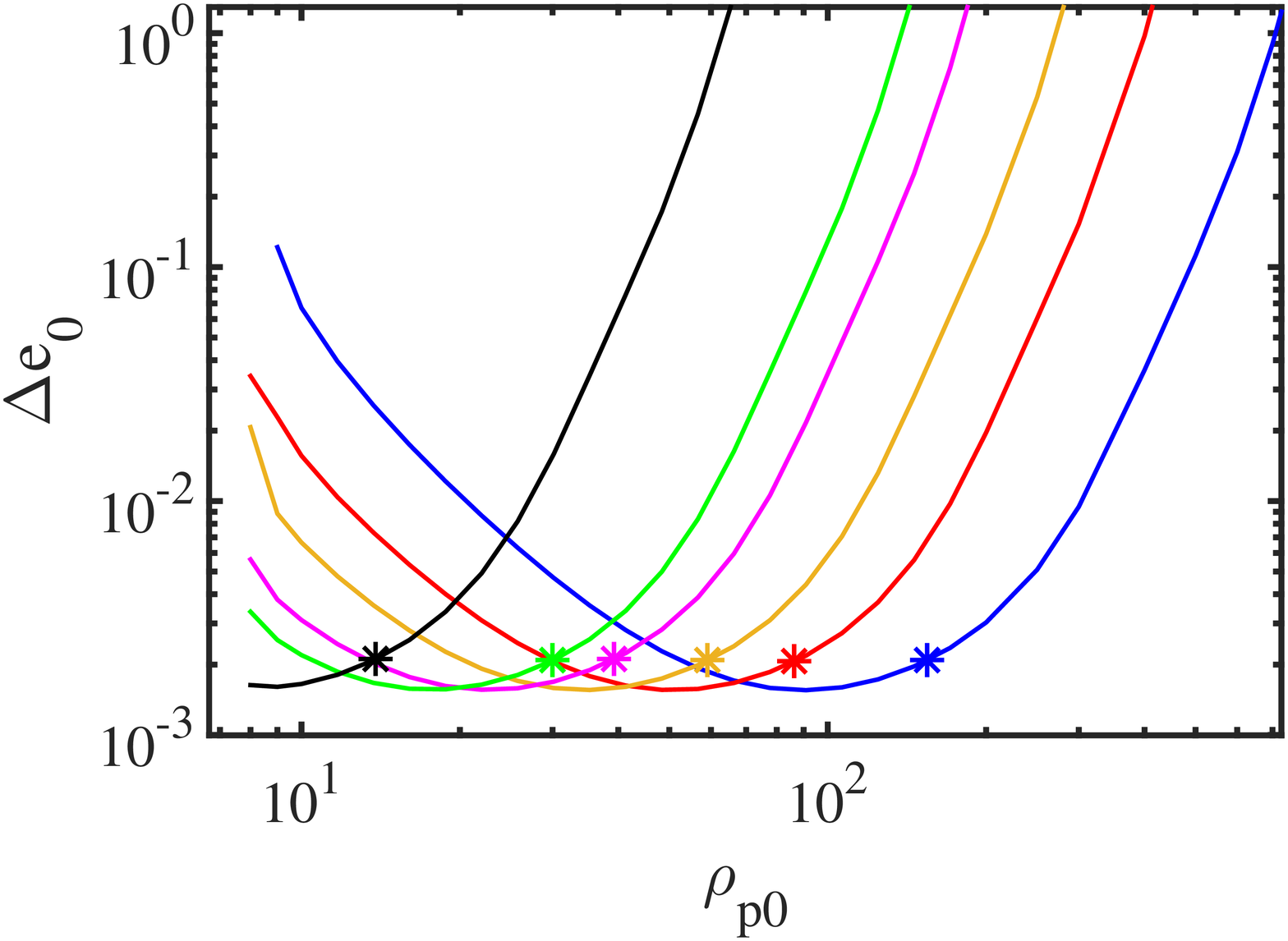}
  \includegraphics[width=80mm]{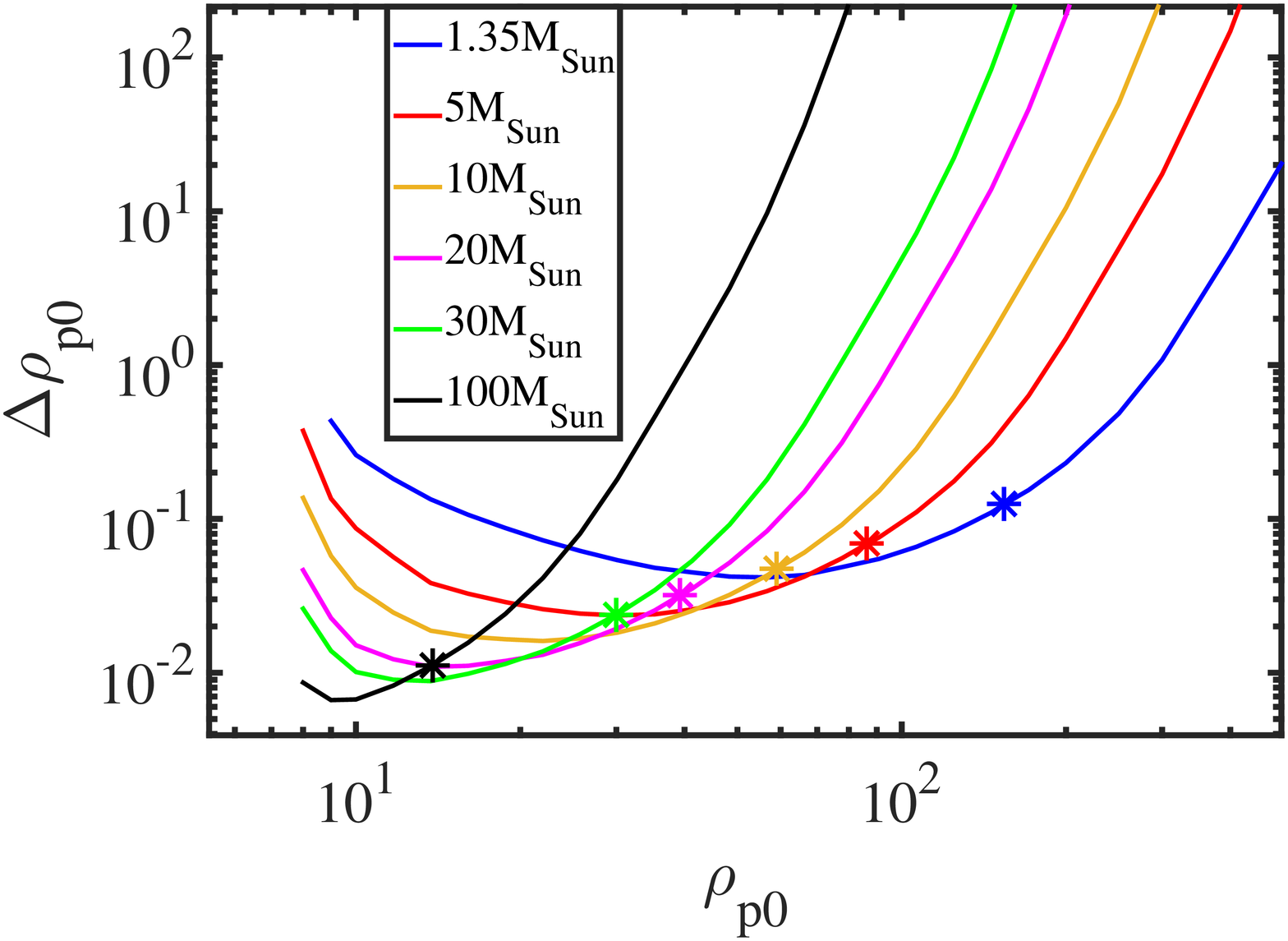}   
\\
  \includegraphics[width=80mm]{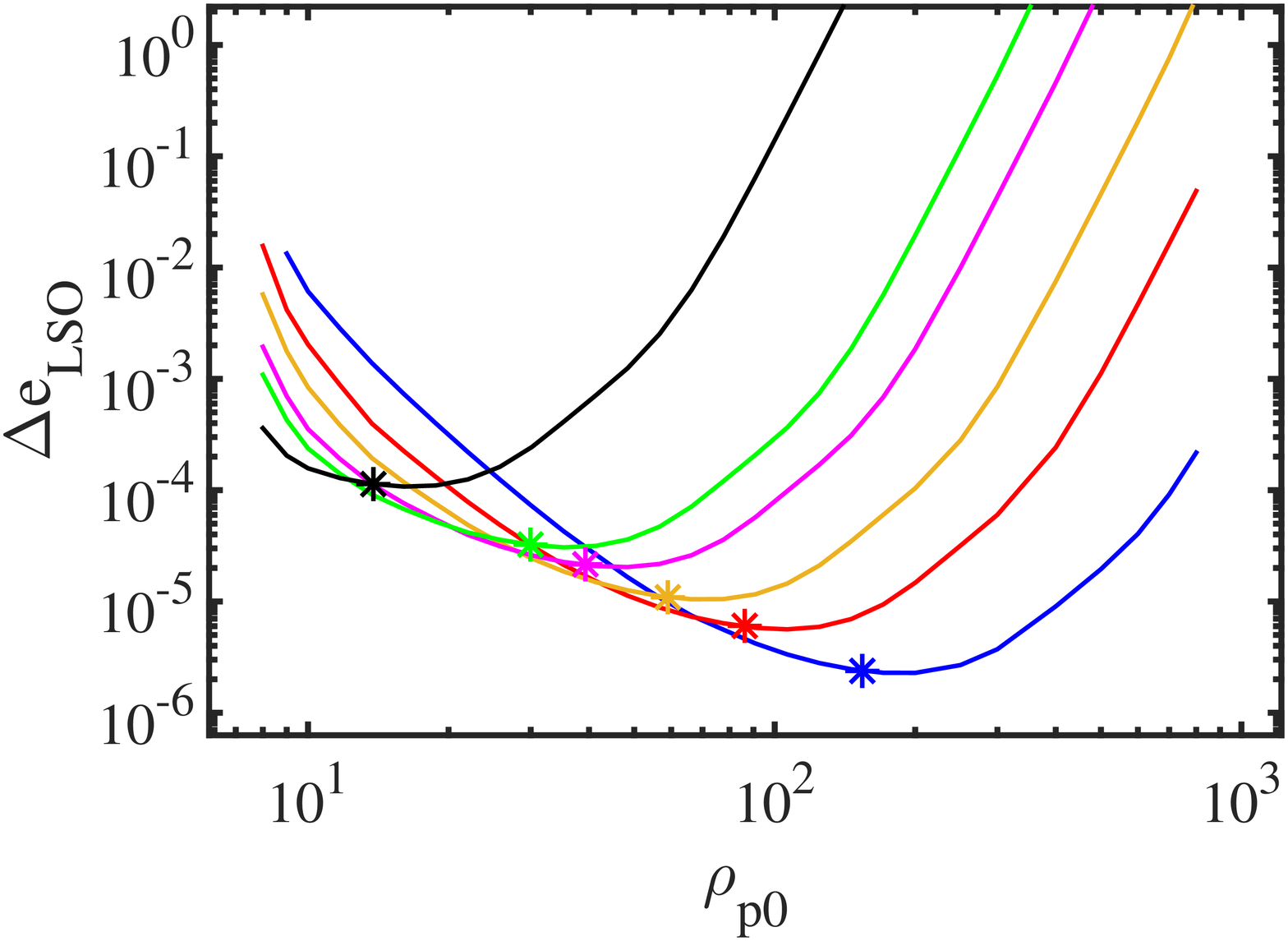}
  \caption{Same as Figure \ref{fig:ParamEst_BHNS_Circular} but showing the \mbox{NS--NS} and \mbox{NS--BH} binary measurement errors  for source parameters specific to eccentric binaries: initial eccentricity $\Delta e_0$ (top left panel), initial dimensionless pericenter distance $\Delta \rho_{\mathrm{p}0}$ (top right panel), and eccentricity at the last stable orbit, $\Delta e_\mathrm{LSO}$ (bottom panel). The measurement error of these parameters increases rapidly for high $\rho_{\mathrm{p}0}$. The measurement error of $\Delta e_{\rm LSO}$ is inaccurate for high $\rho_{\mathrm{p}0}$ because Fisher matrix algorithm becomes invalid in this regime \citep{Gondanetal2018}. Similar to Figure \ref{fig:ParamEst_BHNS_Circular}, results are displayed for $D_L = 100 \, \Mpc$, but may be scaled to other $D_L$ values by multiplying the displayed results by a factor of $D_L / 100 \, \Mpc$. We find similar trends with $\rho_{\mathrm{p}0}$ for other random choices of binary direction and orientation (not shown). }  \label{fig:ParamEst_BHNS_Eccentric}
\end{figure*}

\begin{figure*}
  \centering
  \includegraphics[width=80mm]{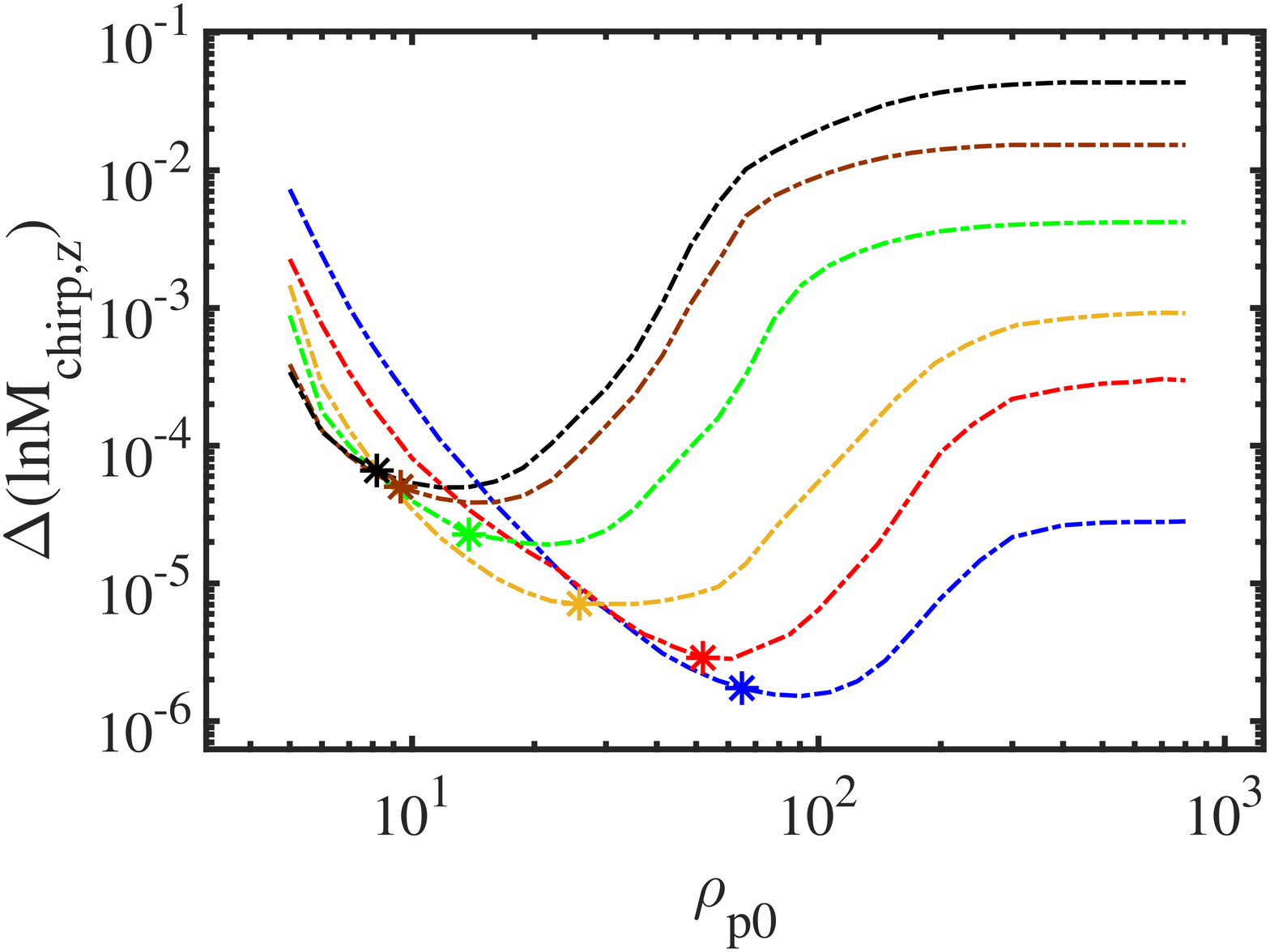}
  \includegraphics[width=80mm]{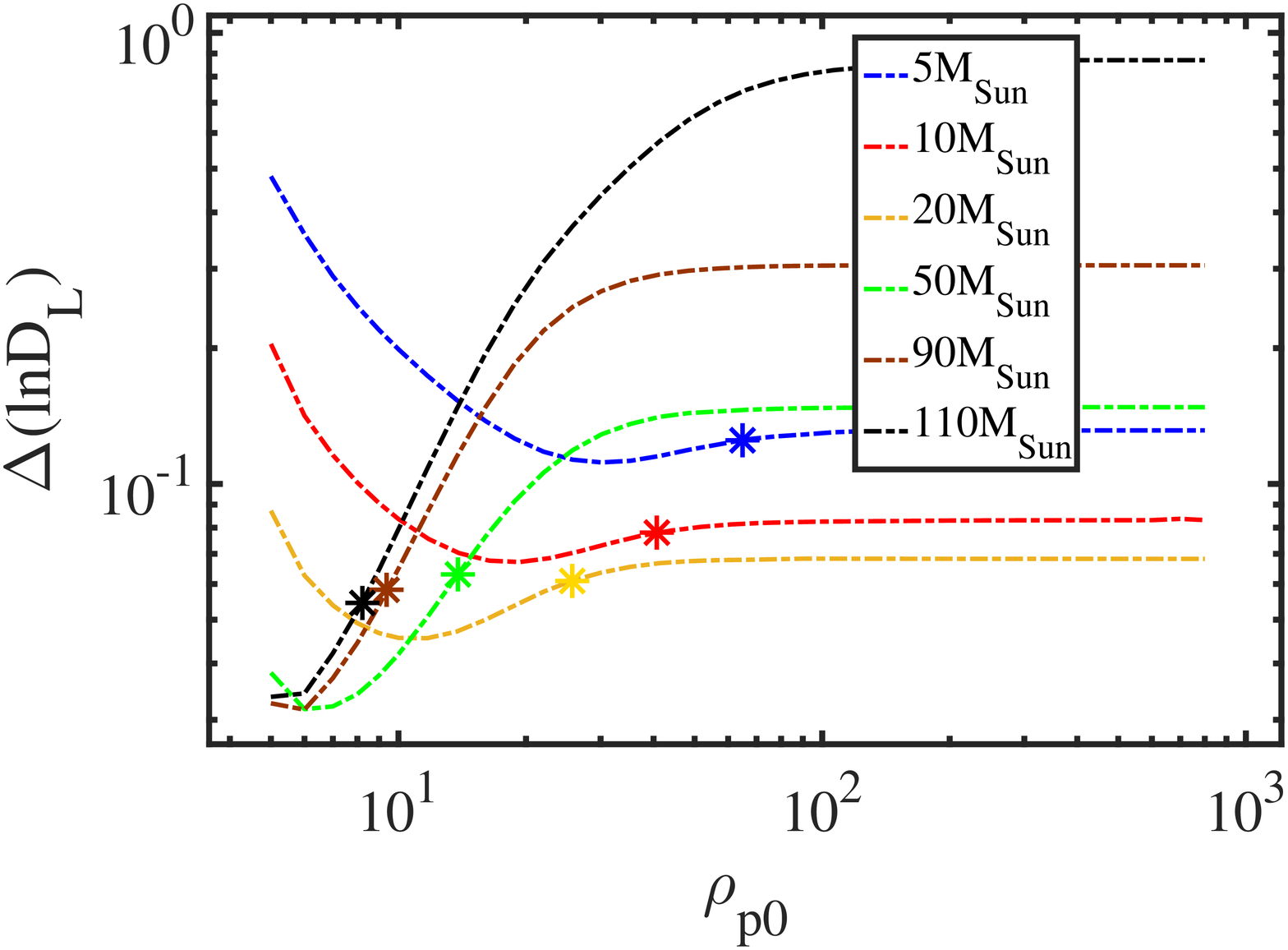}
\\
  \includegraphics[width=80mm]{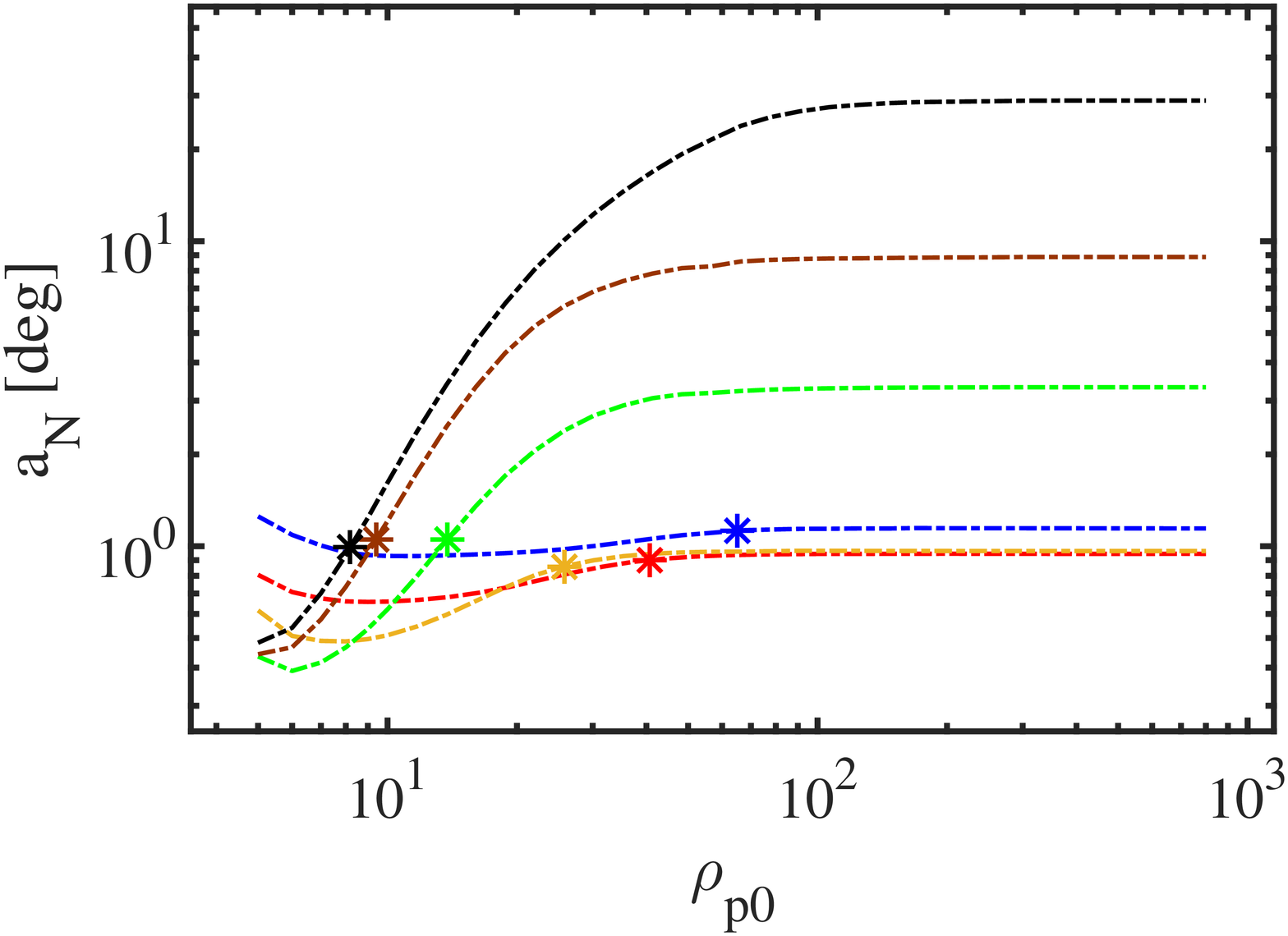}
  \includegraphics[width=80mm]{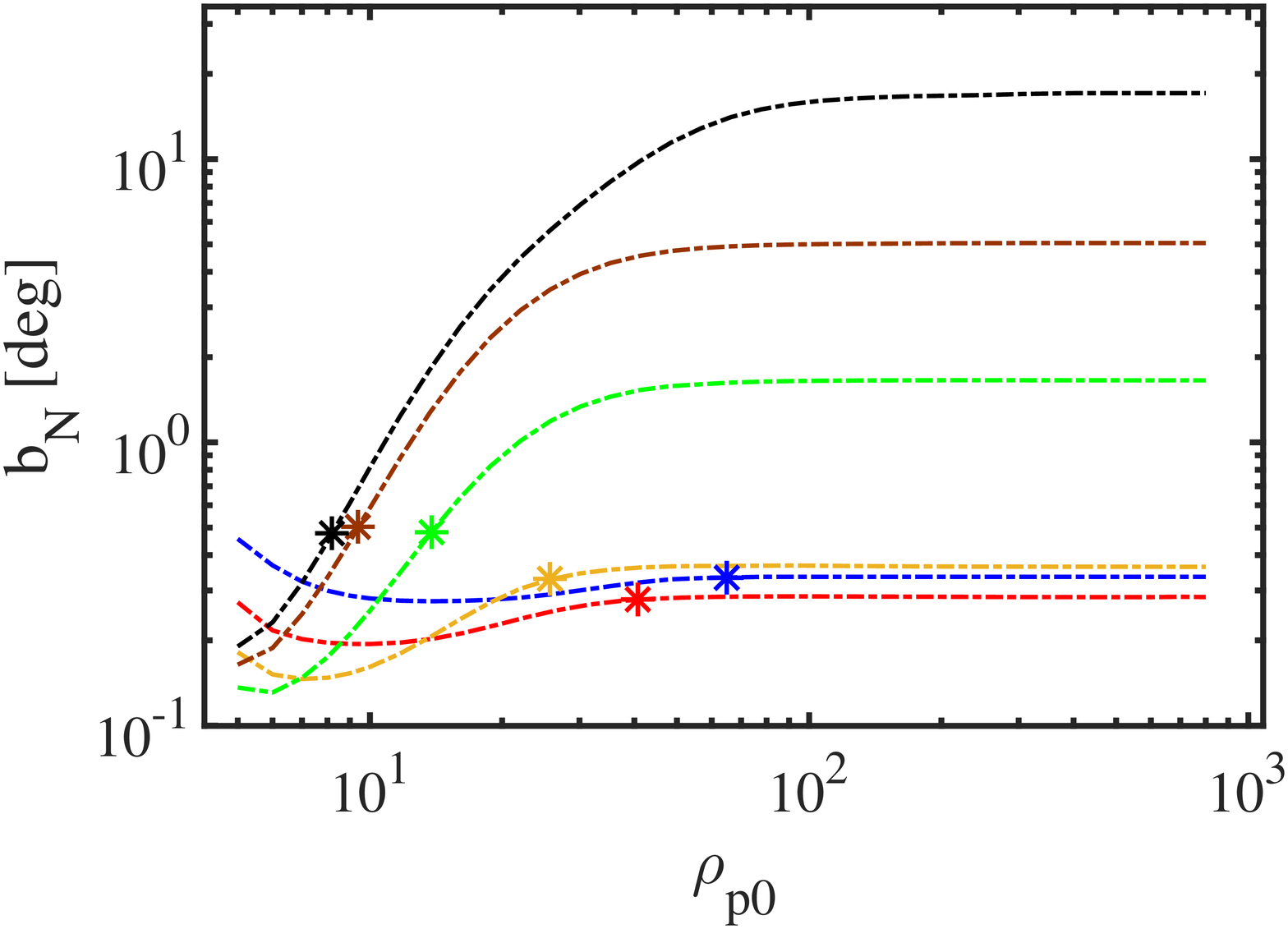}
\\    
  \includegraphics[width=80mm]{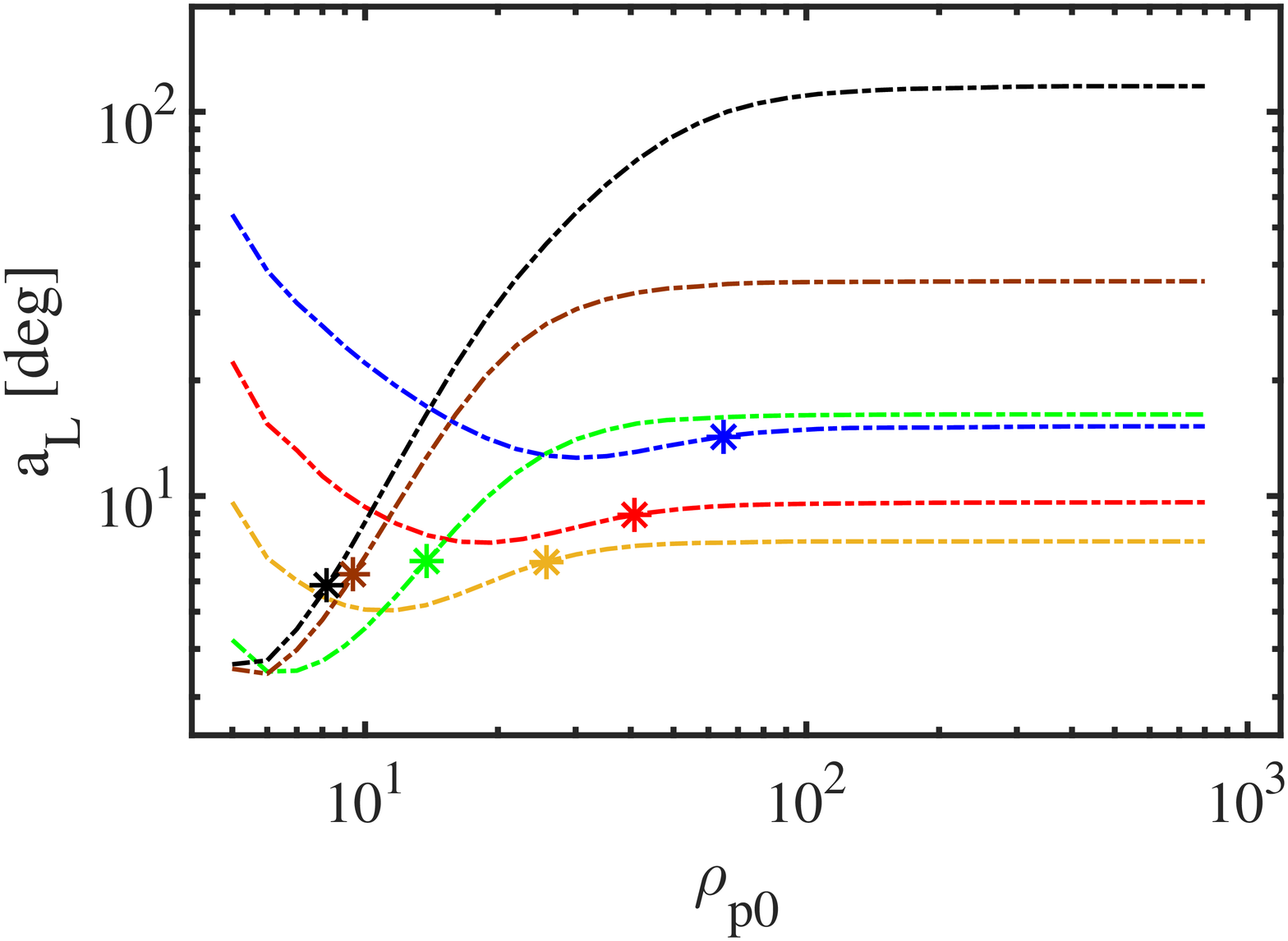}
  \includegraphics[width=80mm]{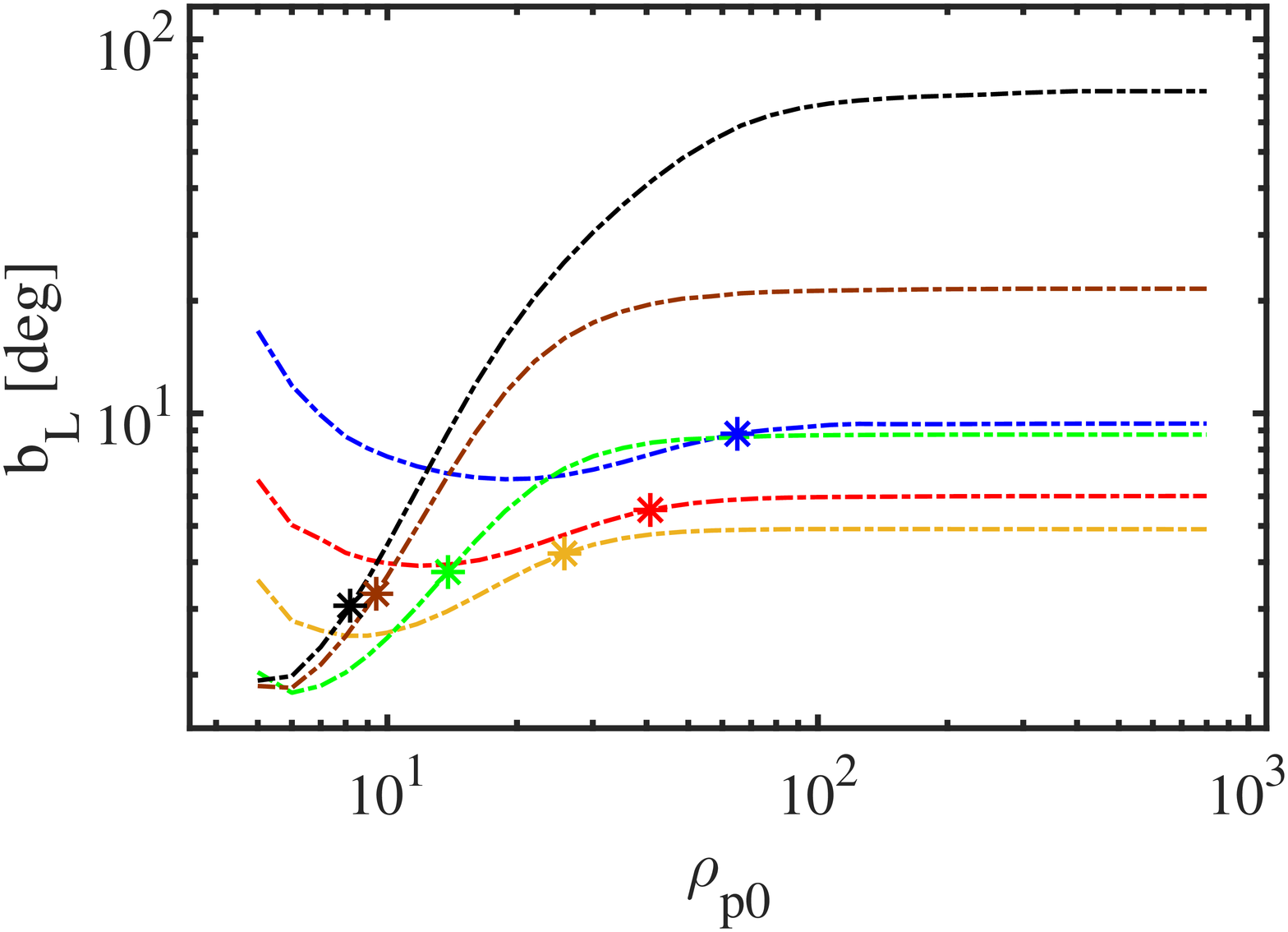}
  \caption{Same as Figure \ref{fig:ParamEst_BHNS_Circular}, but for precessing, eccentric, equal-mass \mbox{BH--BH} binaries of different masses with all binary parameters fixed as in the right panels of Figure \ref{fig:SNRtot_rho}. Binaries form in the detector's frequency band to the left of the star symbols.  }   \label{fig:ParamEst_BHBH_Circular}
\end{figure*}

\begin{figure*}
    \centering
   \includegraphics[width=80mm]{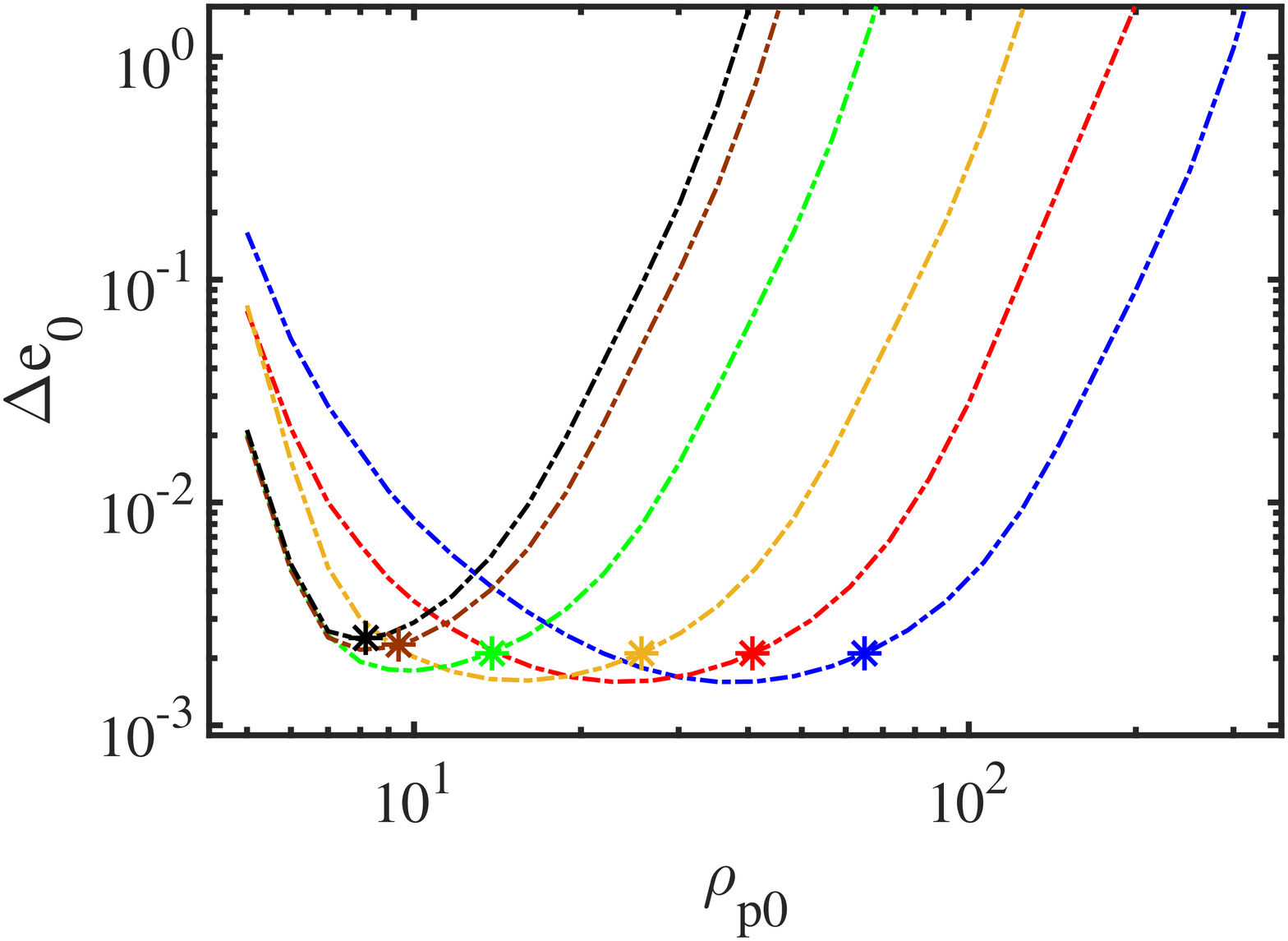}
   \includegraphics[width=80mm]{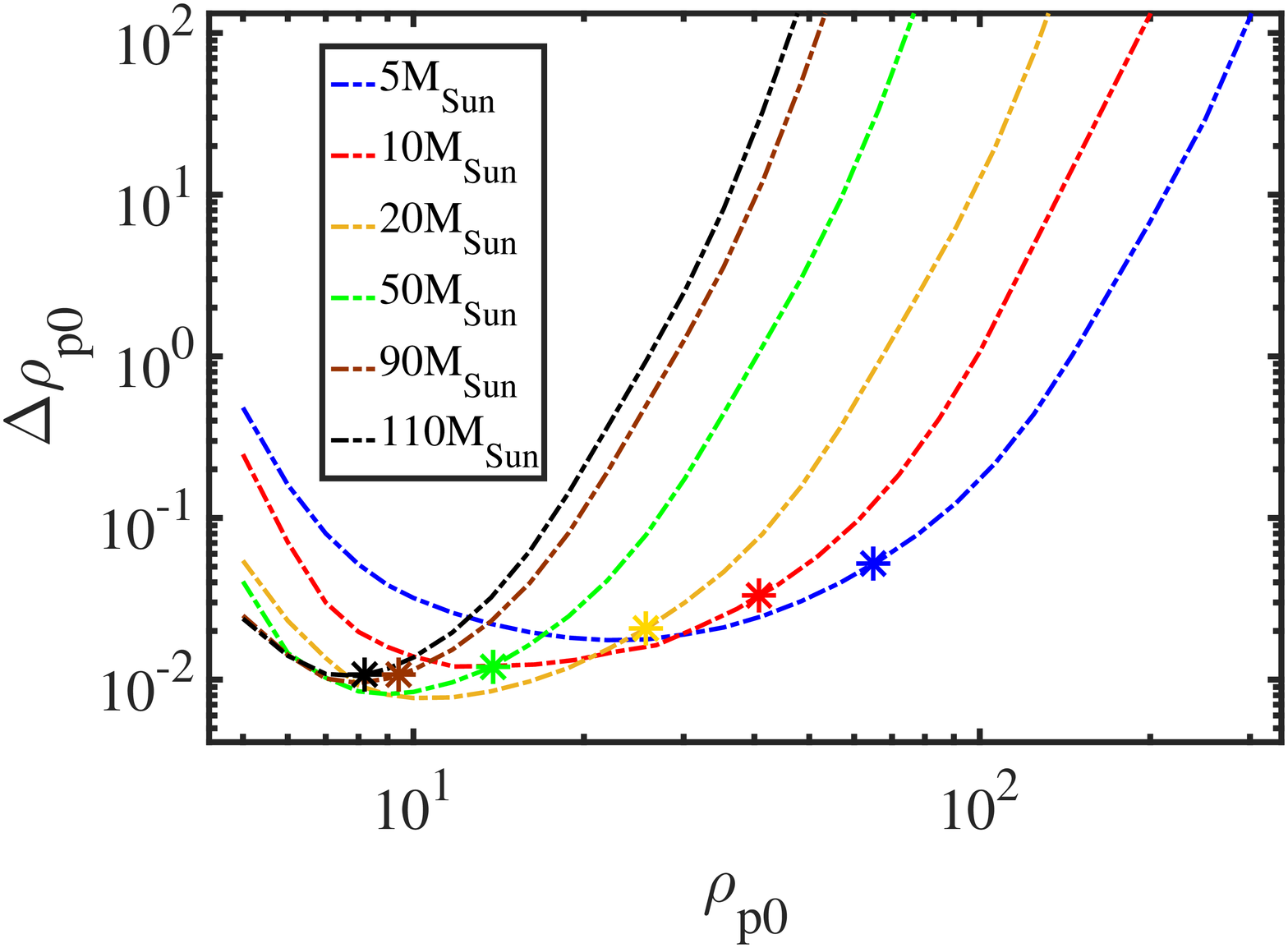}
\\    
   \includegraphics[width=80mm]{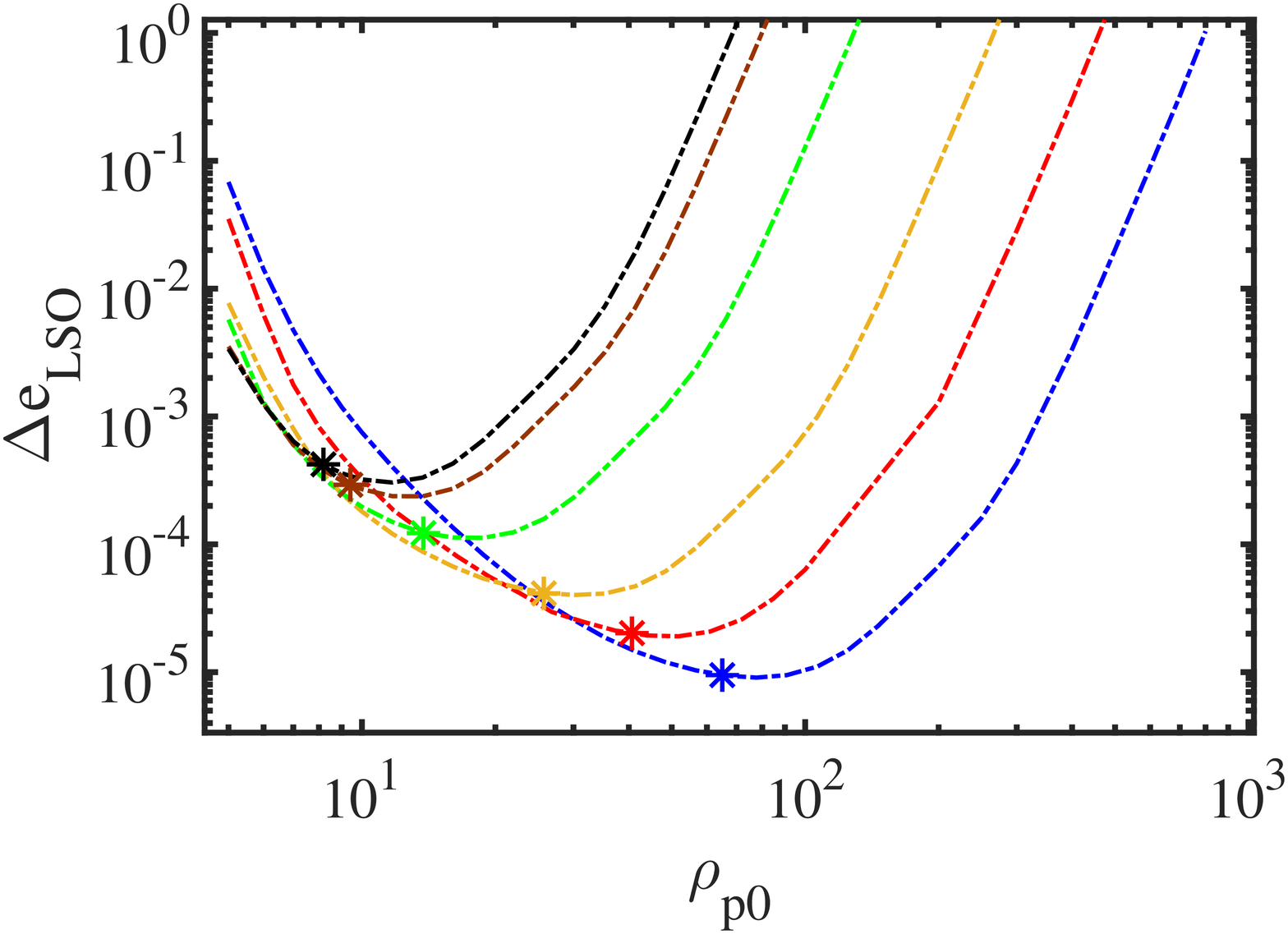}
  \caption{Same as Figure \ref{fig:ParamEst_BHNS_Eccentric}, but for precessing, eccentric, equal-mass \mbox{BH--BH} binaries with all binary parameters fixed as in the right panel of Figure \ref{fig:SNRtot_rho}. Binaries form in the detector's frequency band to the left of the stars.}   \label{fig:ParamEst_BHBH_Eccentric}
\end{figure*}

\begin{figure*}
   \centering
    \includegraphics[width=80mm]{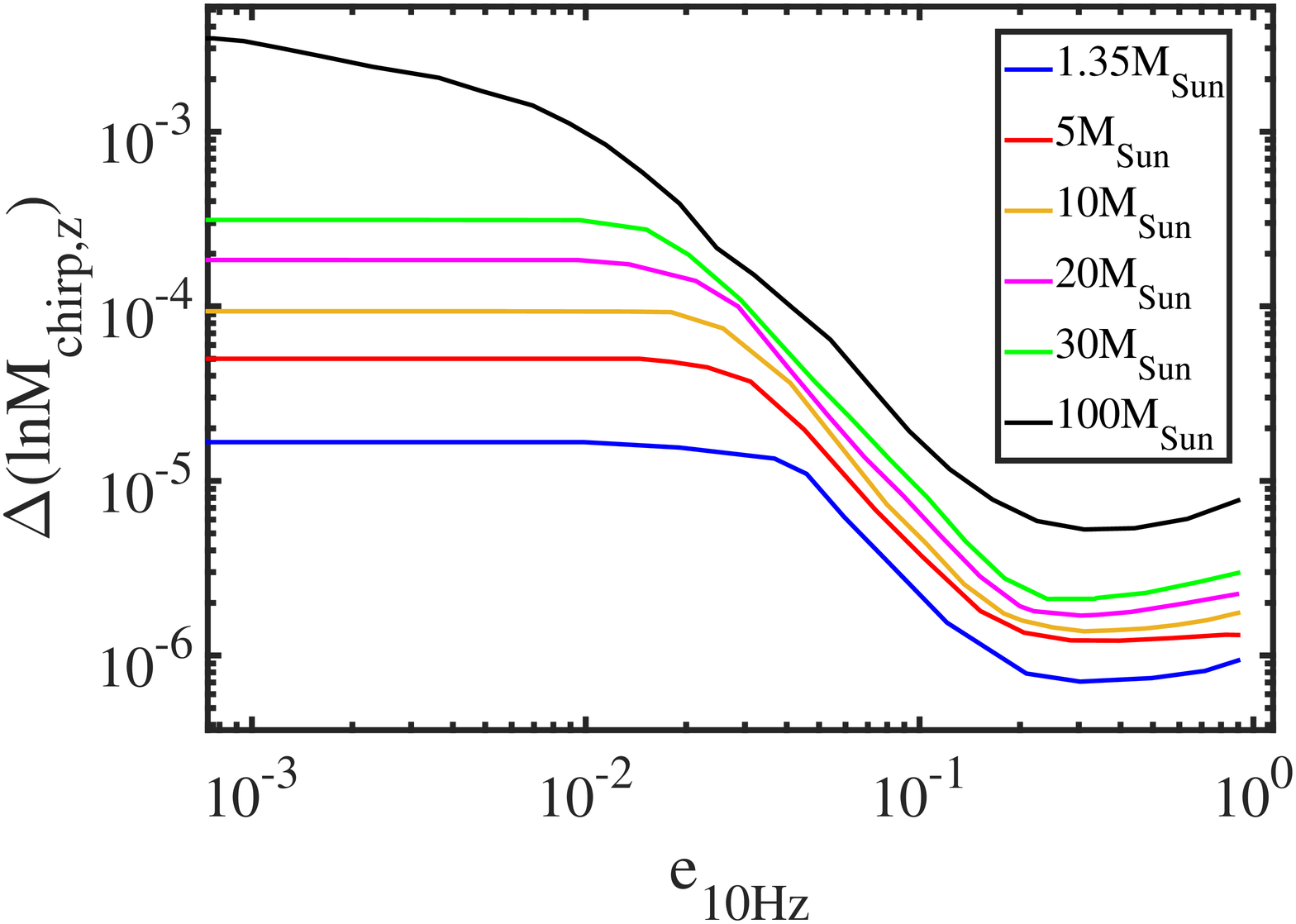}
    \includegraphics[width=80mm]{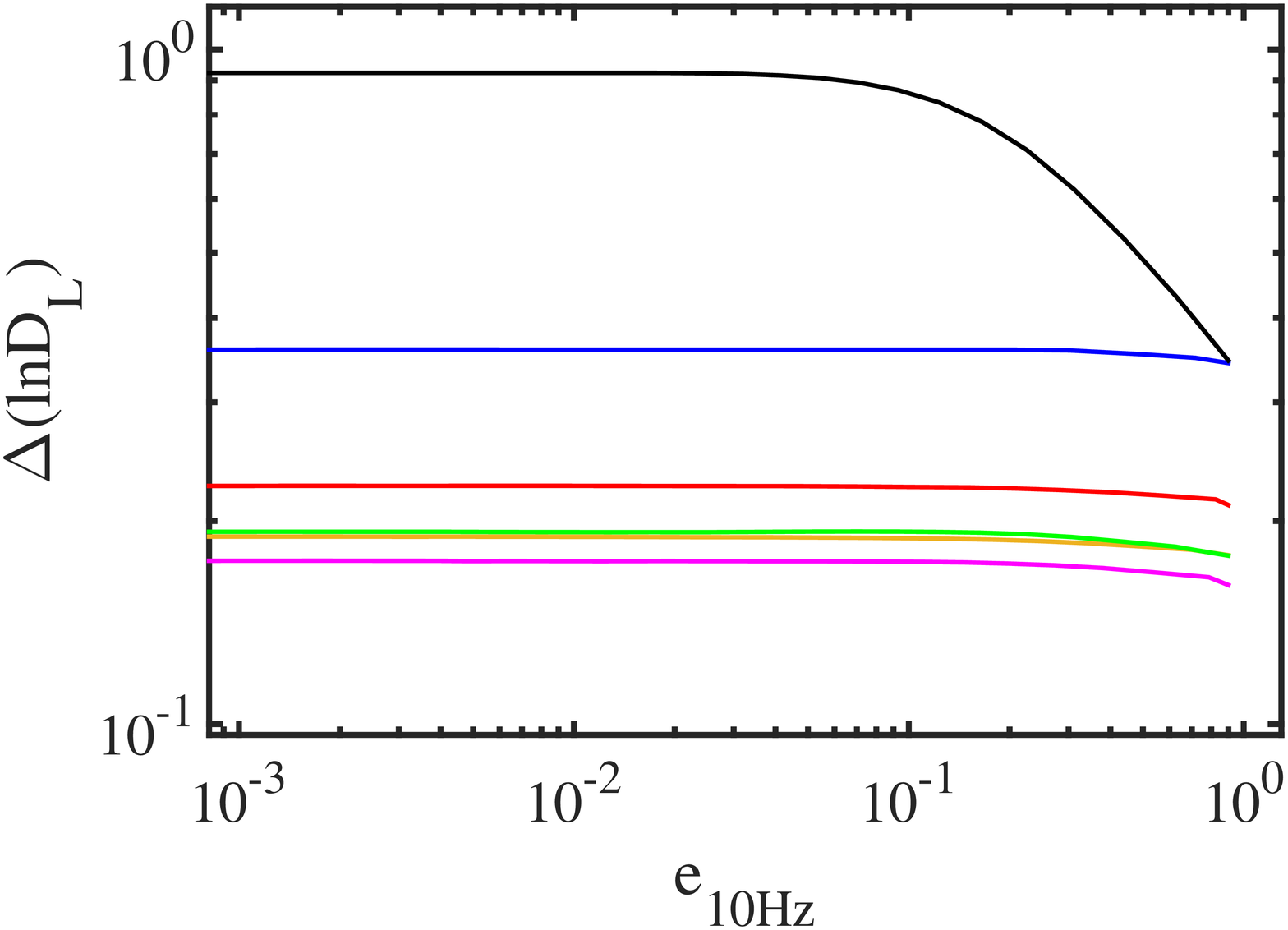}
\\
    \includegraphics[width=80mm]{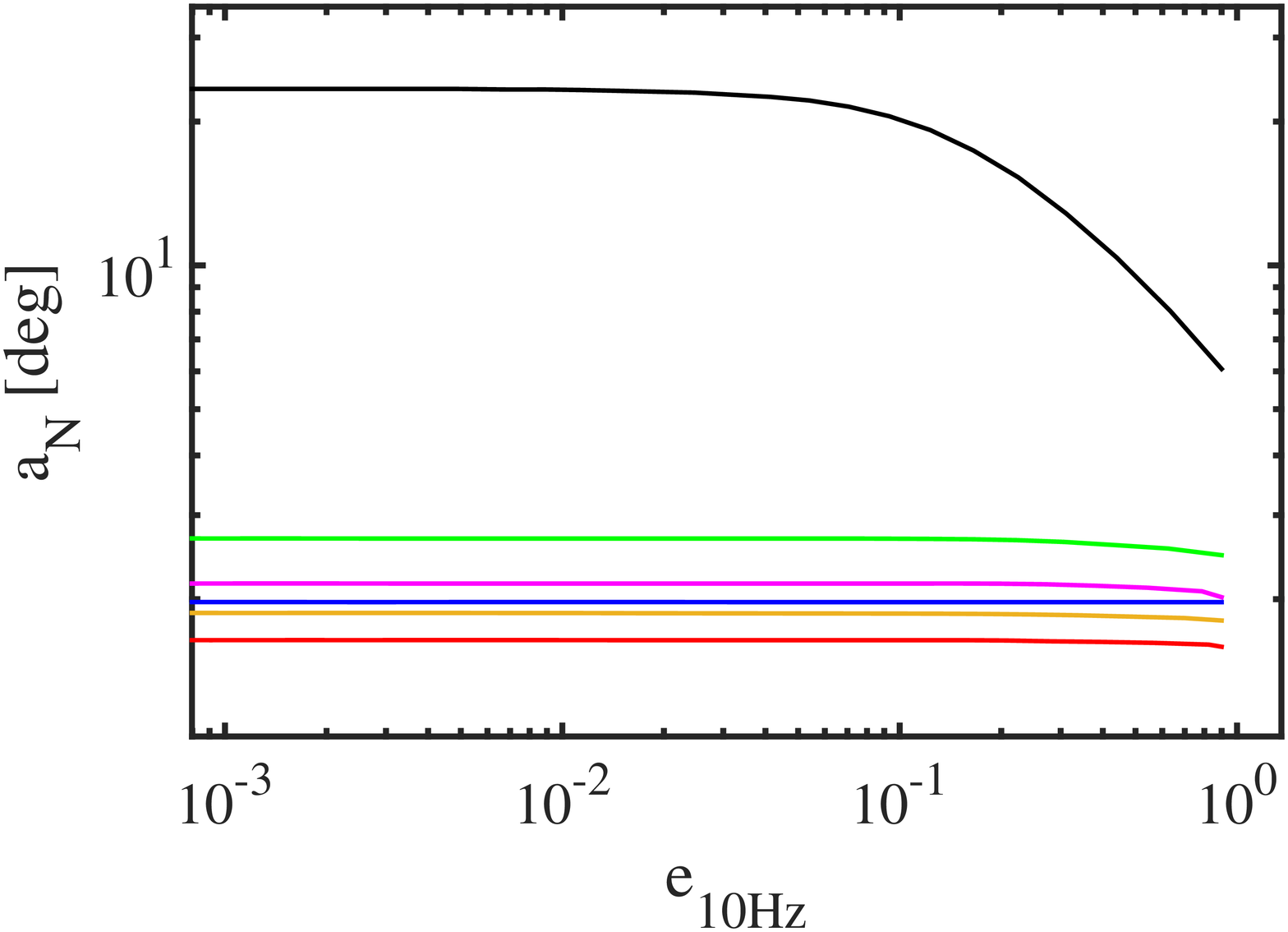}
    \includegraphics[width=80mm]{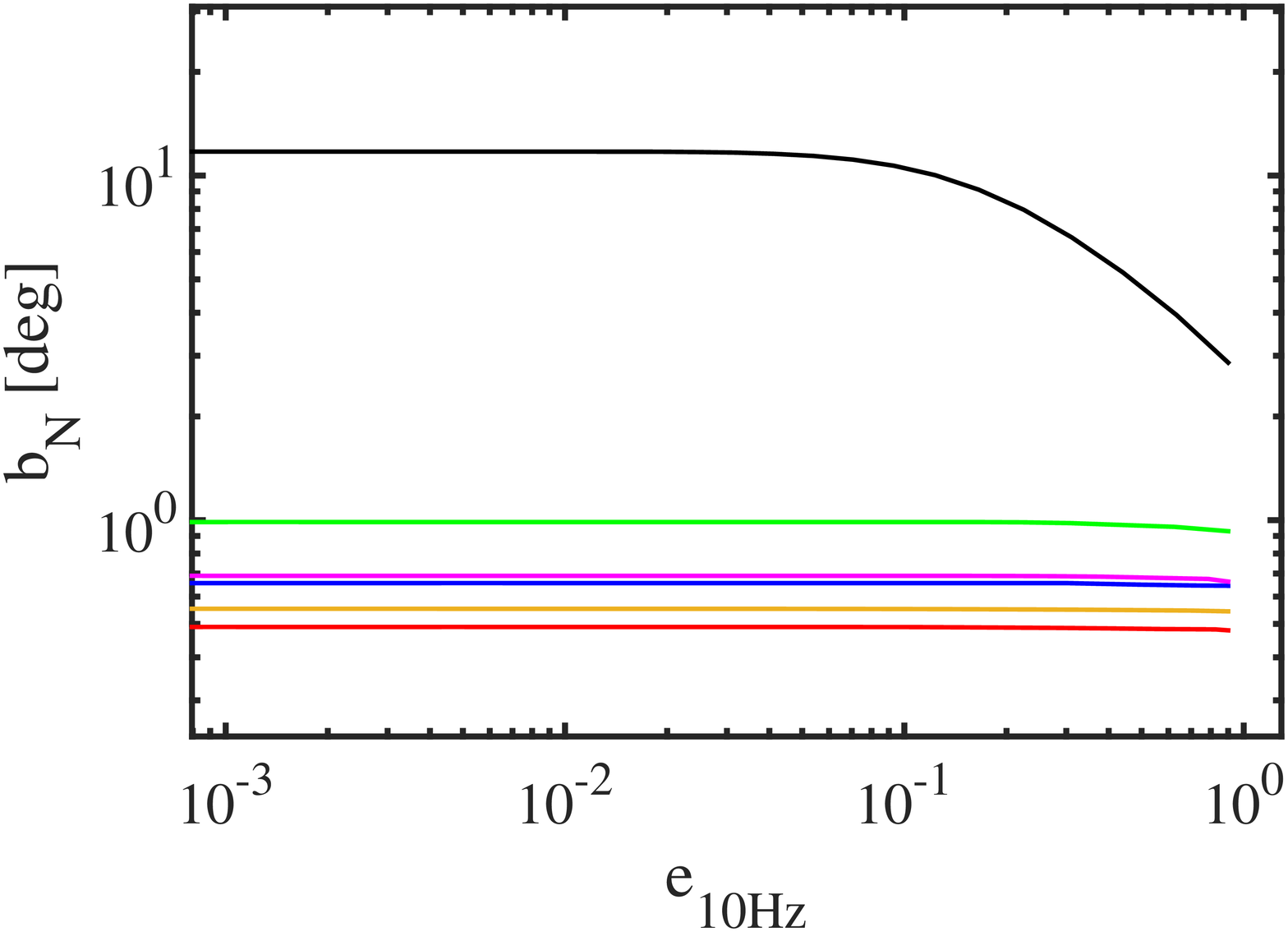}
\\    
    \includegraphics[width=80mm]{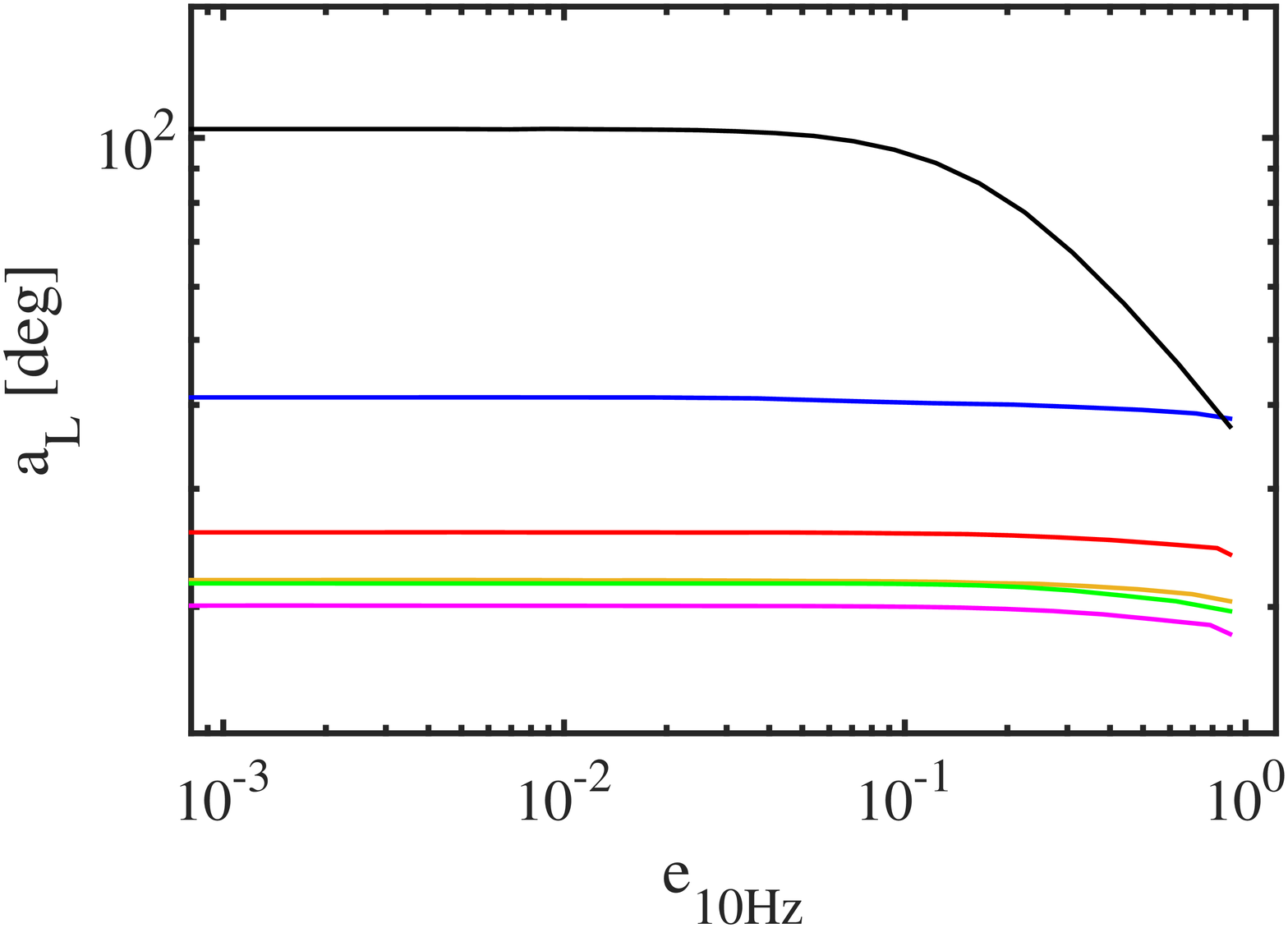}
    \includegraphics[width=80mm]{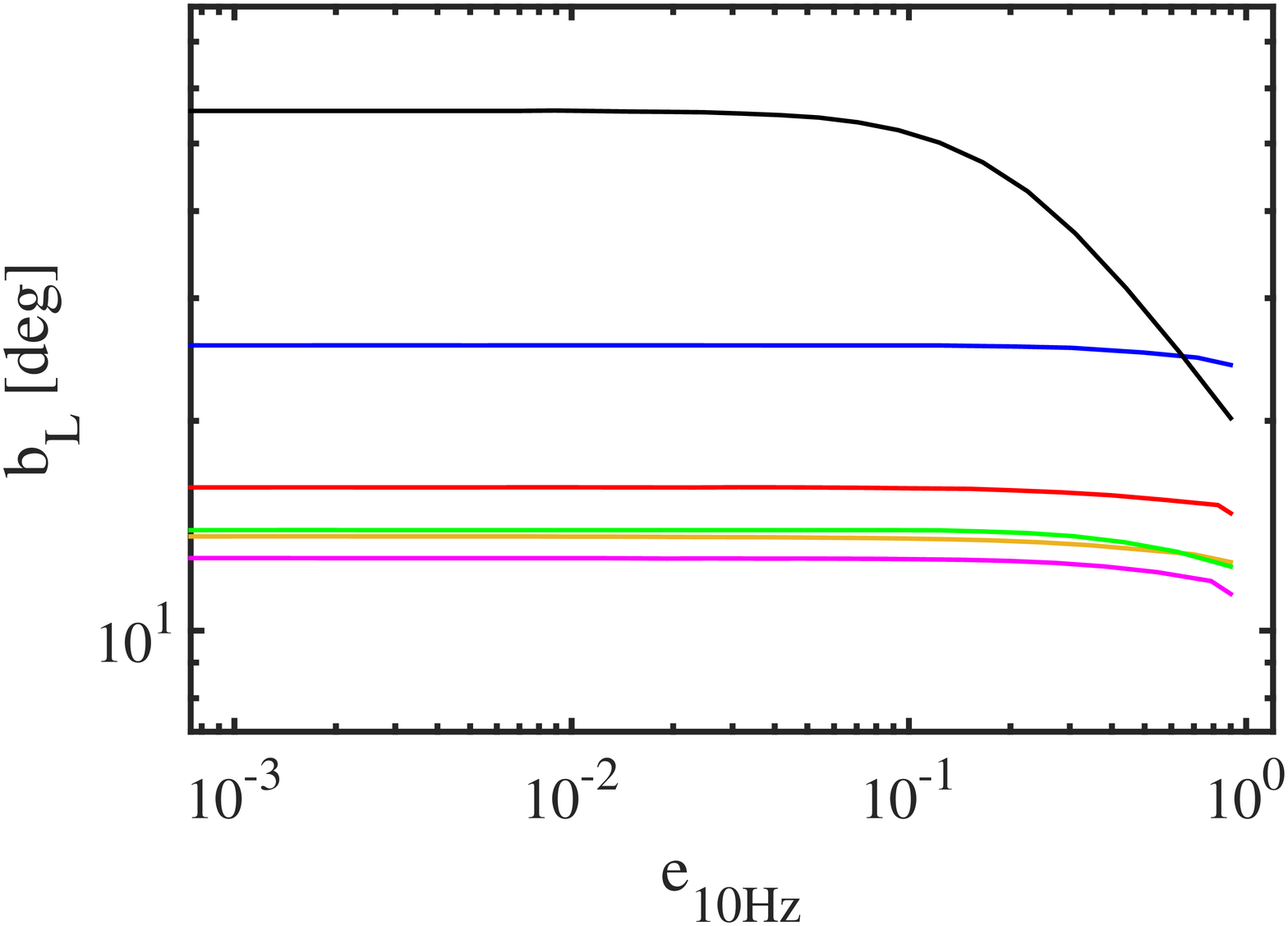}      
  \caption{Measurement errors of source parameters for precessing eccentric \mbox{NS--NS} and \mbox{NS--BH} binaries as in Figure \ref{fig:ParamEst_BHNS_Circular} but plotted as a function of $e_{\rm 10 Hz}$. Different curves show different companion masses as labeled in the legend. }  \label{fig:ParamEst_BHNS_Circular_e10Hz}
\end{figure*}

 Figure \ref{fig:SNRtot_rho} shows our findings for the $\mathrm{S/N} _\mathrm{tot}$ and Figures \ref{fig:ParamEst_BHNS_Circular}--\ref{fig:ParamEst_BHBH_Eccentric_e10Hz} show the measurement errors of binary parameters for precessing eccentric \mbox{NS--NS}, \mbox{NS--BH}, and \mbox{BH--BH} binaries using the aLIGO--AdV--KAGRA GW detector network. These figures display the dependences of measurement errors on $\rho_{\mathrm{p}0}$, $e_{\rm 10 Hz}$, and binary component masses. Figure \ref{fig:SNRtot_rho} shows the $\mathrm{S/N} _\mathrm{tot}$ as a function of $\rho_{{\rm p}0}$ and $e_{\rm 10 Hz}$ for all types of binaries of different masses. Figure \ref{fig:ParamEst_BHNS_Circular} shows the $\rho_{{\rm p}0}$ dependence of the parameter measurement errors for $(\Delta \ln \mathcal{M}_z, \Delta \ln D_L, a_N, b_N, a_L, b_L)$ for precessing eccentric \mbox{NS--NS} and \mbox{NS--BH} binaries of different BH masses, while Figure  \ref{fig:ParamEst_BHNS_Eccentric} shows that for parameters $(\Delta e_0, \Delta \rho_{{\rm p}0}, \Delta e_{\rm LSO})$. In these figures the asymptotic limit of high $\rho_{{\rm p}0}$ represents the case of circular binaries. Decreasing $\rho_{{\rm p}0}$ from infinity corresponds to binaries that form outside the GW detectors' frequency band at $f_{\rm GW} = 10 \, {\rm Hz}$ with $e_0 = 0.9$ and enter the frequency bands with some lower eccentricity $e_{10\,\rm Hz} < e_0$ (the value of $e_{10 \, \rm Hz}$ is not shown in this figure, it is higher for lower $\rho_{{\rm p}0}$), while even smaller $\rho_{{\rm p}0}$ below the star symbols in the figure represents cases where the GW-driven evolution with $e_0 = 0.9$ starts within the detector band. Similarly, Figures \ref{fig:ParamEst_BHBH_Circular} and \ref{fig:ParamEst_BHBH_Eccentric} show the $\rho_{{\rm p}0}$ dependence of parameter measurement errors for precessing eccentric \mbox{BH--BH} binaries of different masses. Figures \ref{fig:ParamEst_BHNS_Circular_e10Hz}--\ref{fig:ParamEst_BHBH_Eccentric_e10Hz} are analogous to Figures \ref{fig:ParamEst_BHNS_Circular}--\ref{fig:ParamEst_BHBH_Eccentric}, but show the parameter measurement errors as a function of $e_{10\,\rm Hz}$ instead of $\rho_{{\rm p}0}$ and include an additional panel for $\Delta e_{\rm 10 Hz}$. In particular, Figures \ref{fig:ParamEst_BHNS_Circular_e10Hz} and \ref{fig:ParamEst_BHNS_Eccentric_e10Hz} show parameter measurement errors as a function of $e_{10\,\rm Hz}$ for precessing eccentric \mbox{NS--NS} and \mbox{NS--BH} binaries, and Figures \ref{fig:ParamEst_BHBH_Circular_e10Hz} and \ref{fig:ParamEst_BHBH_Eccentric_e10Hz} show those for precessing eccentric \mbox{BH--BH} binaries. Figure \ref{fig:ParamEst_RelVelDist} displays results for the measurement accuracy of the characteristic relative velocity of the sources before GWs start driving the evolution\footnote{This is expected to be comparable to the characteristic velocity dispersion of the host environment.} (see Equation \ref{eq:rhop0-v} below) as a function of $\rho_{{\rm p}0}$ and $e_{\rm 10 Hz}$ for precessing eccentric \mbox{NS--NS}, \mbox{NS--BH}, and \mbox{BH--BH} binaries. Results for the $\mathrm{S/N} _\mathrm{tot}$ and the measurement errors of binary parameters $\Delta \lambda$ are displayed for $D_L = 100\,\rm Mpc$, but can be scaled to other $D_L$ values as $\mathrm{S/N} _\mathrm{tot}\times (100 \, \Mpc / D_L)$ and $\Delta \lambda \times (D_L / 100 \, \Mpc)$ with an inaccuracy level of $7-17 \%$ and $5-21 \%$ at $1 \, \Gpc$, respectively; see Appendix \ref{sec:DL_Dep} for details. Note that the positions of the stars correspond to $D_L = 100 \,\rm Mpc$, and they shift in the figure for higher redshift as $(1+z)^{-2/3}$ (see Equation \ref{eq:rhoDet_Gen}). 
 
 Figures \ref{fig:ParamEst_BHNS_SNRtot20} and \ref{fig:ParamEst_BHBH_SNRtot20} show the measurement errors for the parameters $(e_0, \rho_{{\rm p}0}, e_{\rm 10 Hz}, e_{\rm LSO}, \ln \mathcal{M}_z, \ln D_L, \ln w)$ for a reference $\mathrm{S/N} _\mathrm{tot}$ of $20$ and display the measurement errors as a function of $e_{\rm 10 Hz}$ for precessing eccentric NS and BH binaries. Note, further, that these results can be approximately scaled to other $\mathrm{S/N} _\mathrm{tot}$ values by multiplying the results by a factor of $20 \times \mathrm{S/N} _\mathrm{tot}^{-1}$.

\subsection{Measurement Errors of Eccentricity and Initial Dimensionless Pericenter Distance} 
\label{subsec:MeasErrEccRhoP0} 
 
\begin{figure*}
   \centering
   \includegraphics[width=80mm]{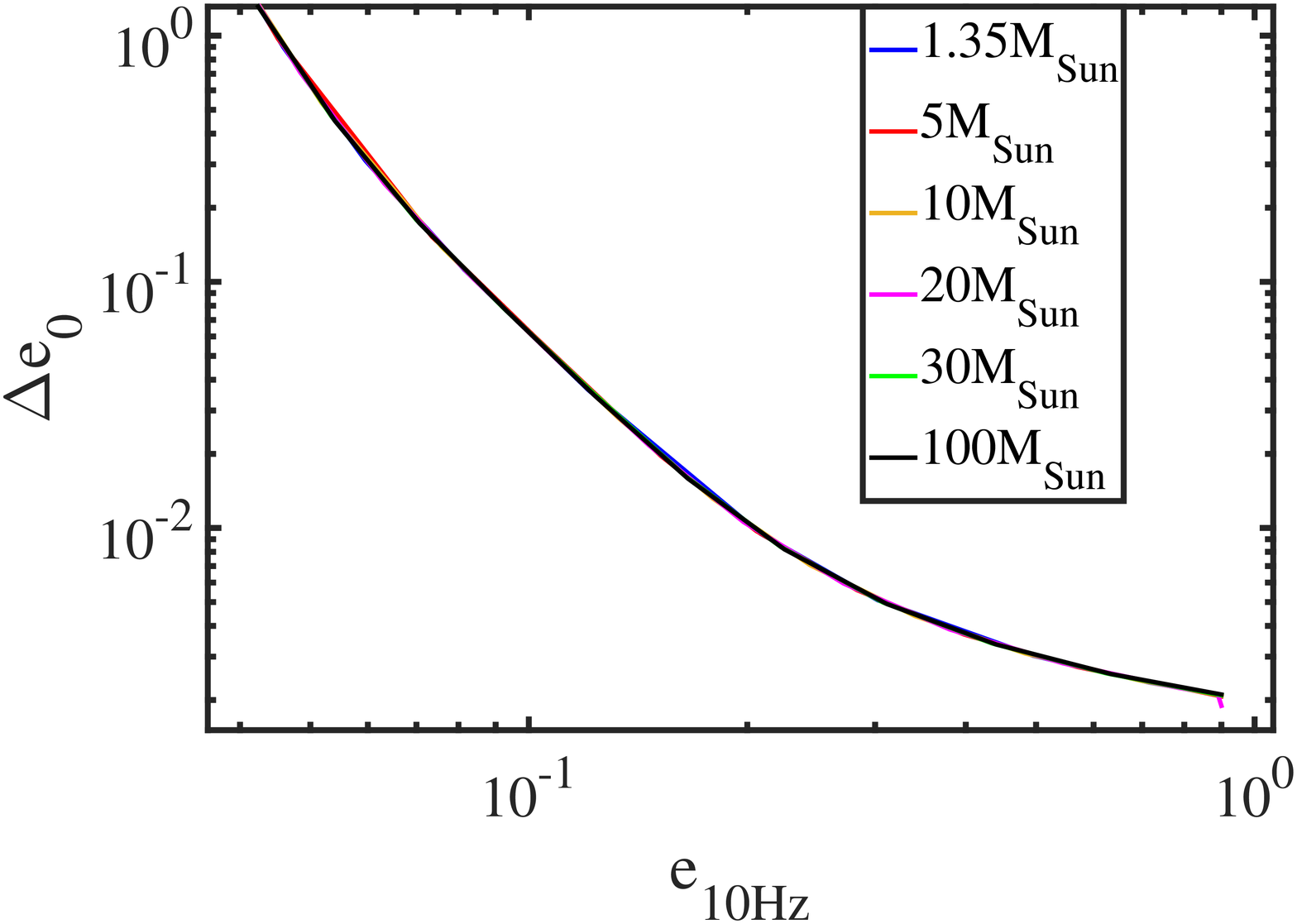}
   \includegraphics[width=80mm]{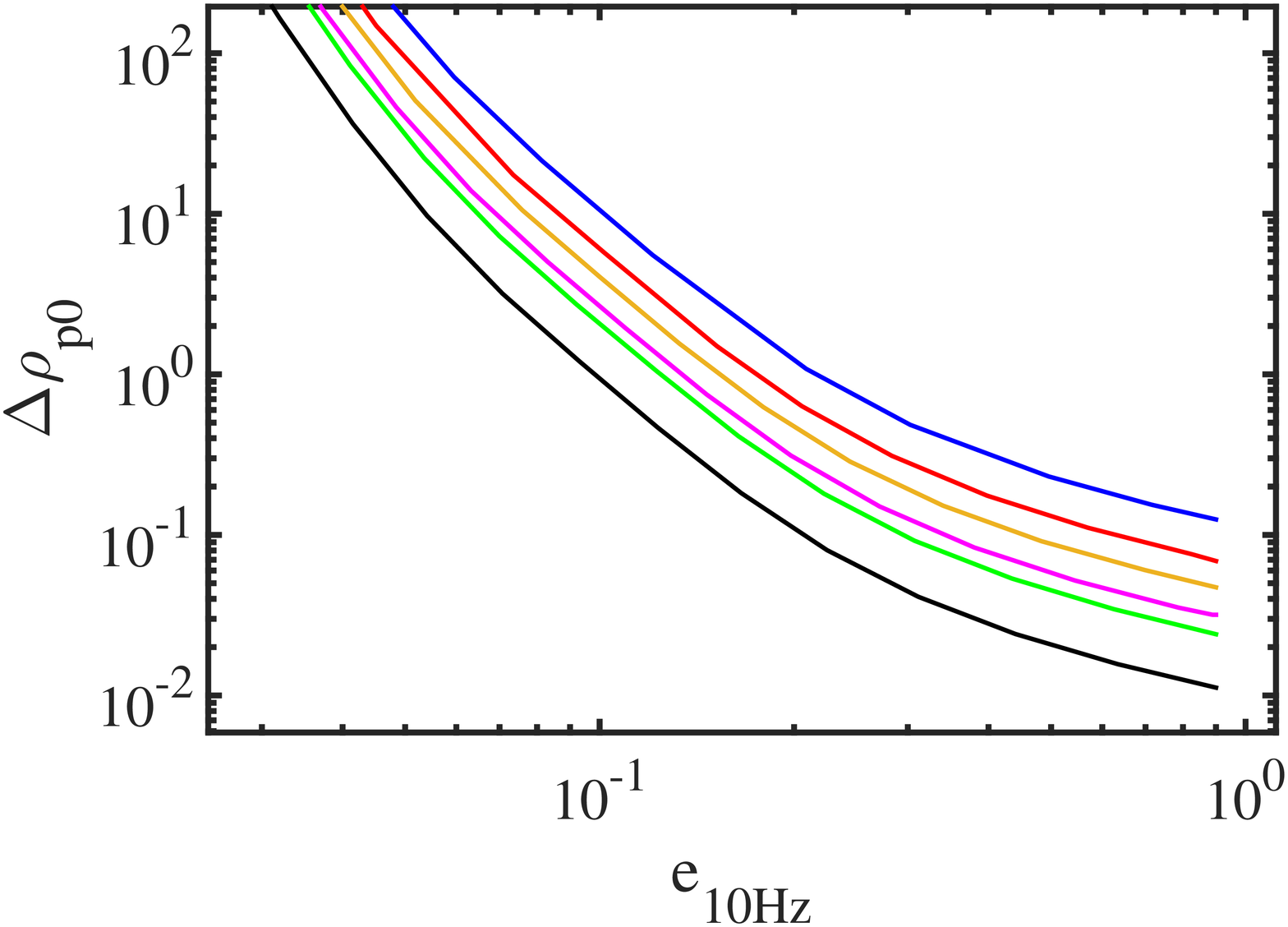}
\\
   \includegraphics[width=80mm]{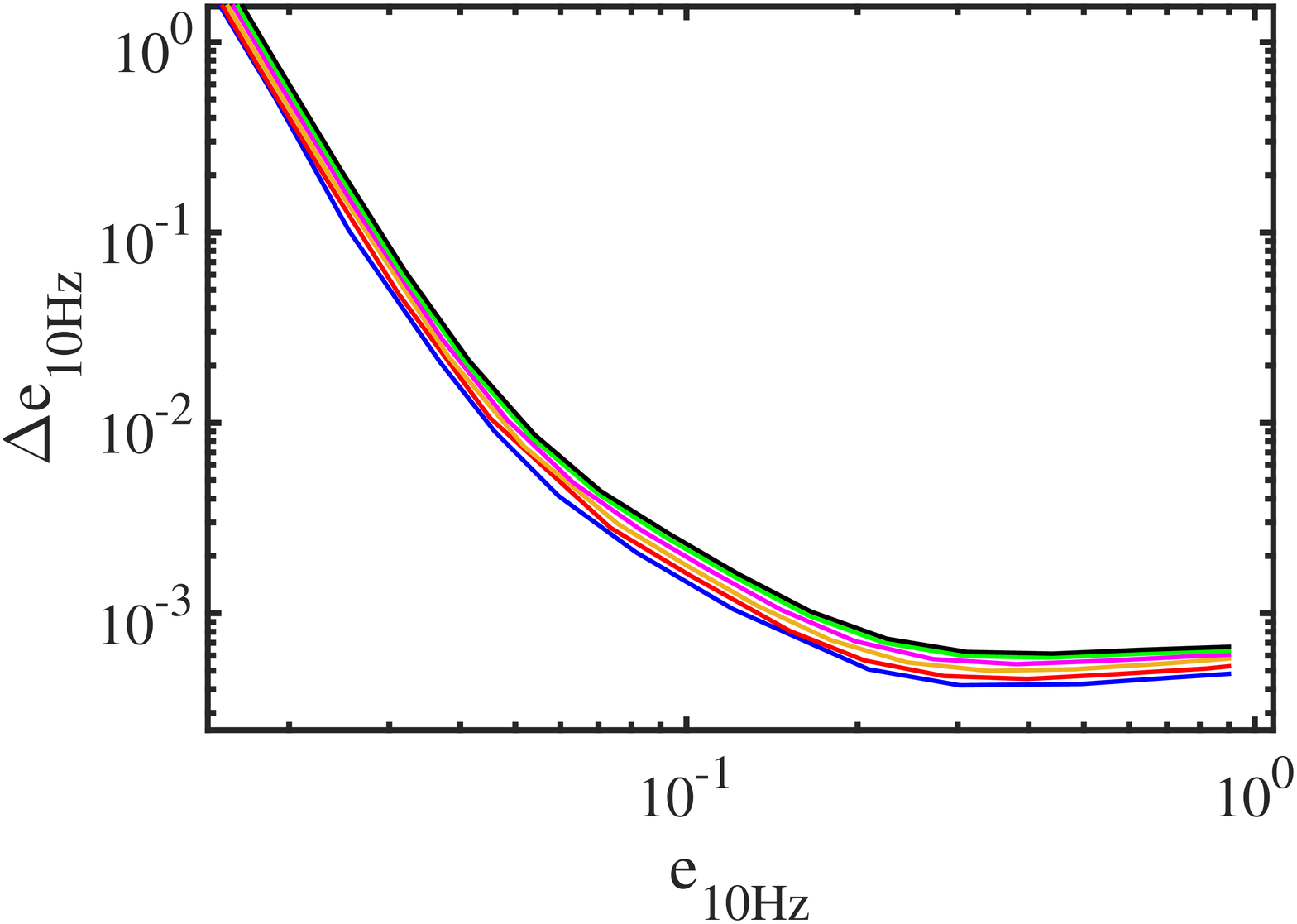}
   \includegraphics[width=80mm]{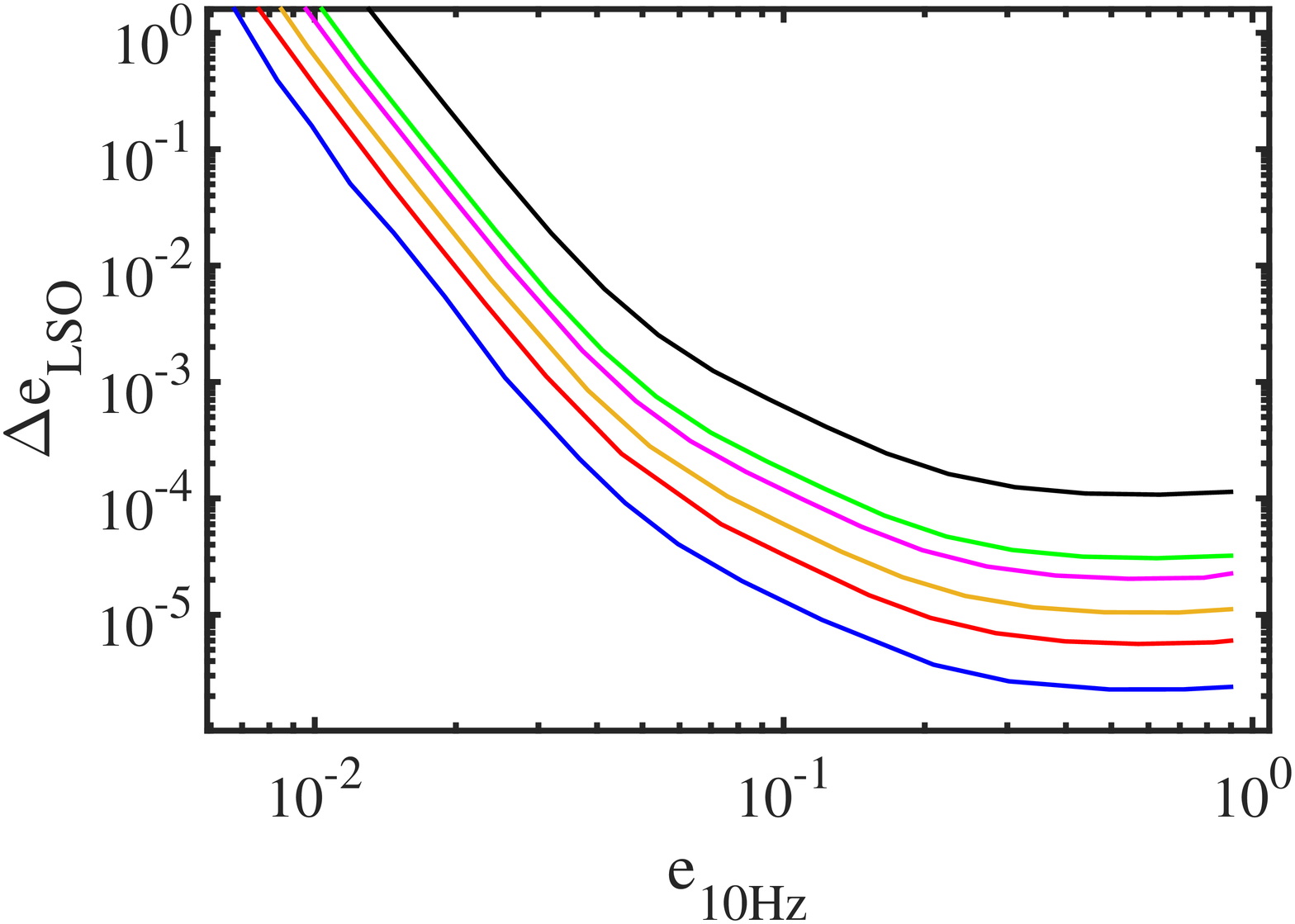}
  \caption{Measurement errors of source parameters specific for precessing eccentric \mbox{NS--NS}  and \mbox{NS--BH} binaries as in Figure \ref{fig:ParamEst_BHNS_Eccentric}, plotted as a function of  $e_{\rm 10 Hz}$ for binaries that form outside of the detectors' frequency band. Different curves  show different companion masses as labeled in the legend. }    \label{fig:ParamEst_BHNS_Eccentric_e10Hz}
\end{figure*}

\begin{figure*}
   \centering
   \includegraphics[width=80mm]{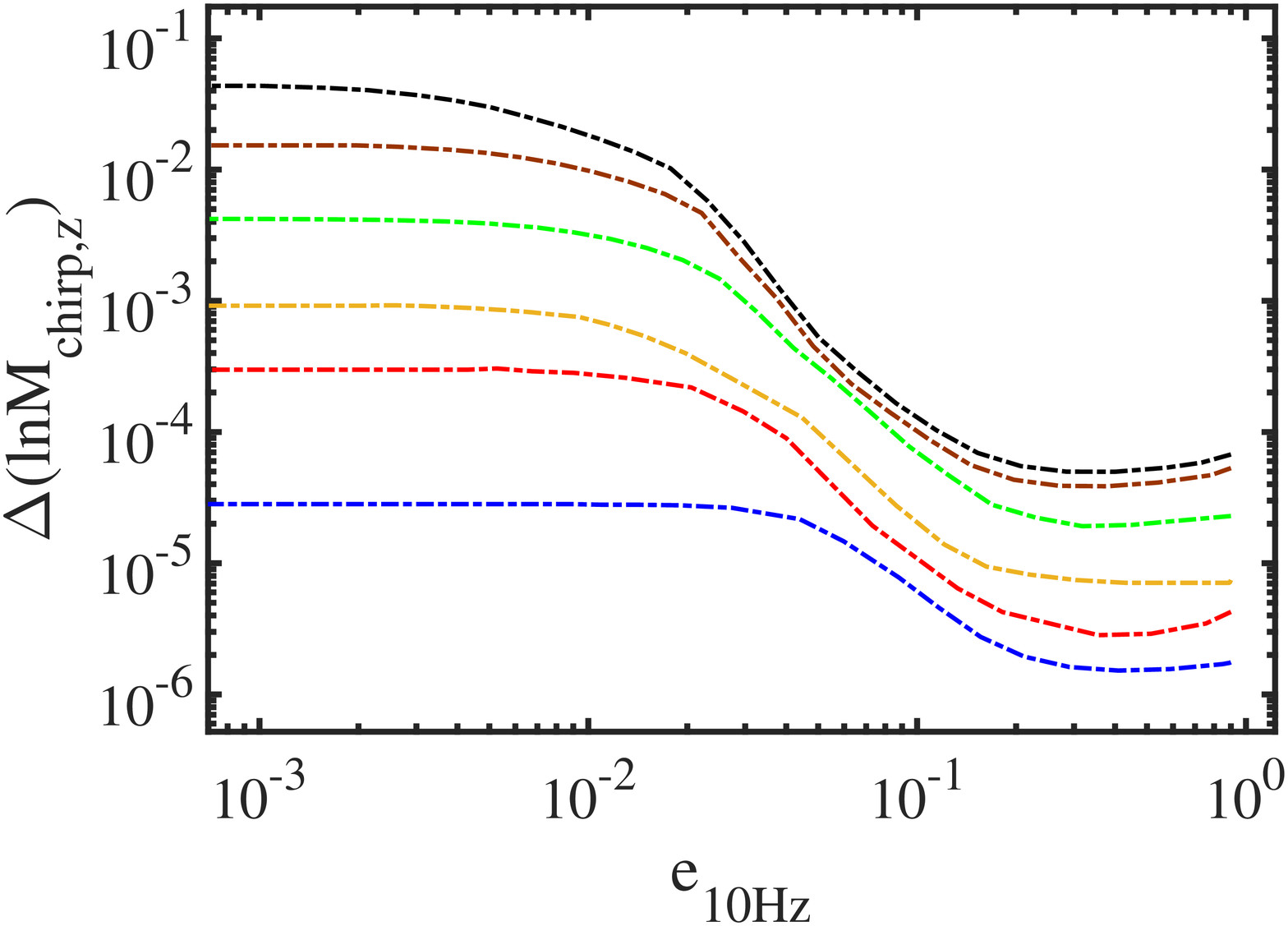}
   \includegraphics[width=80mm]{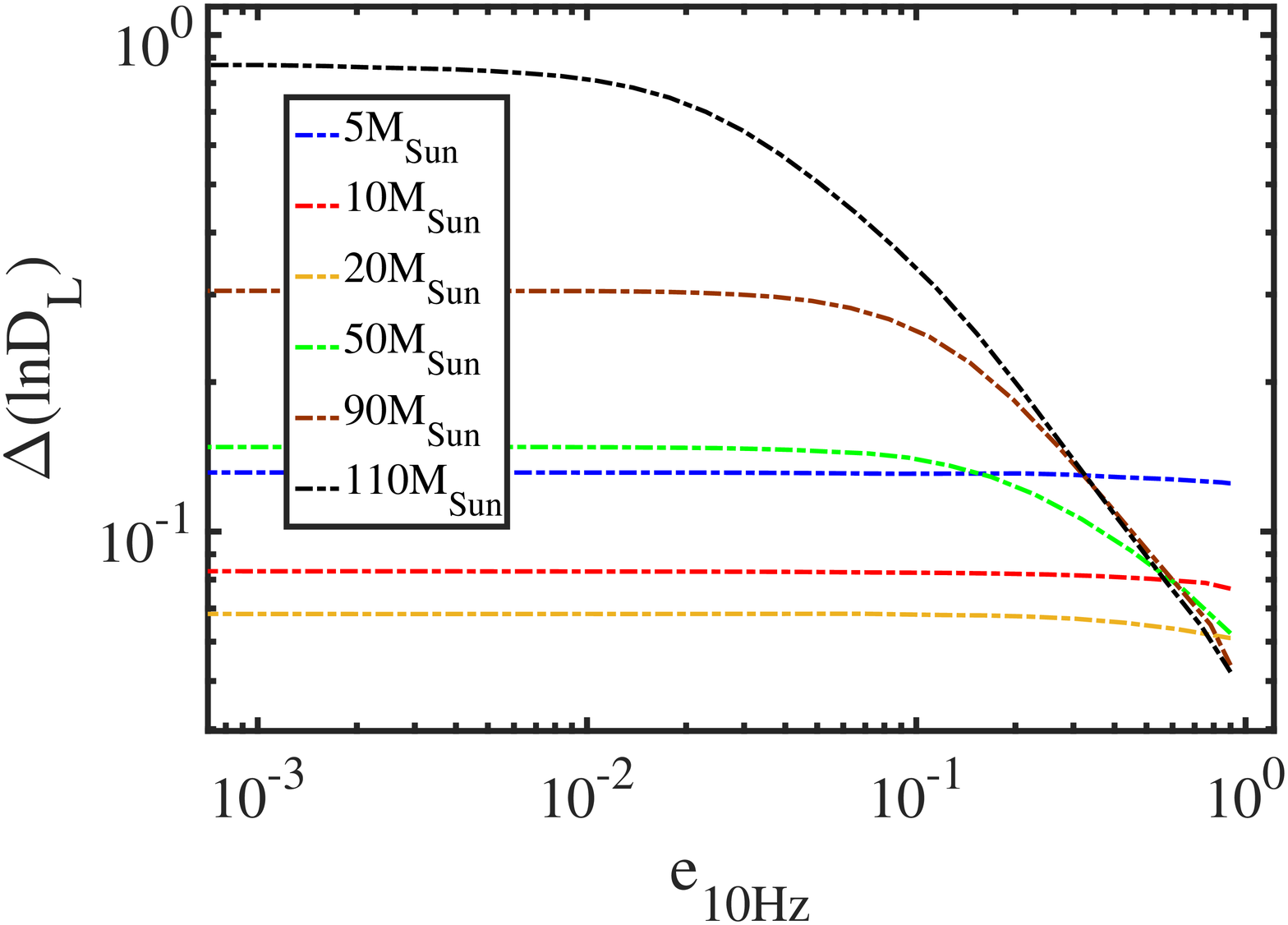}
\\
   \includegraphics[width=80mm]{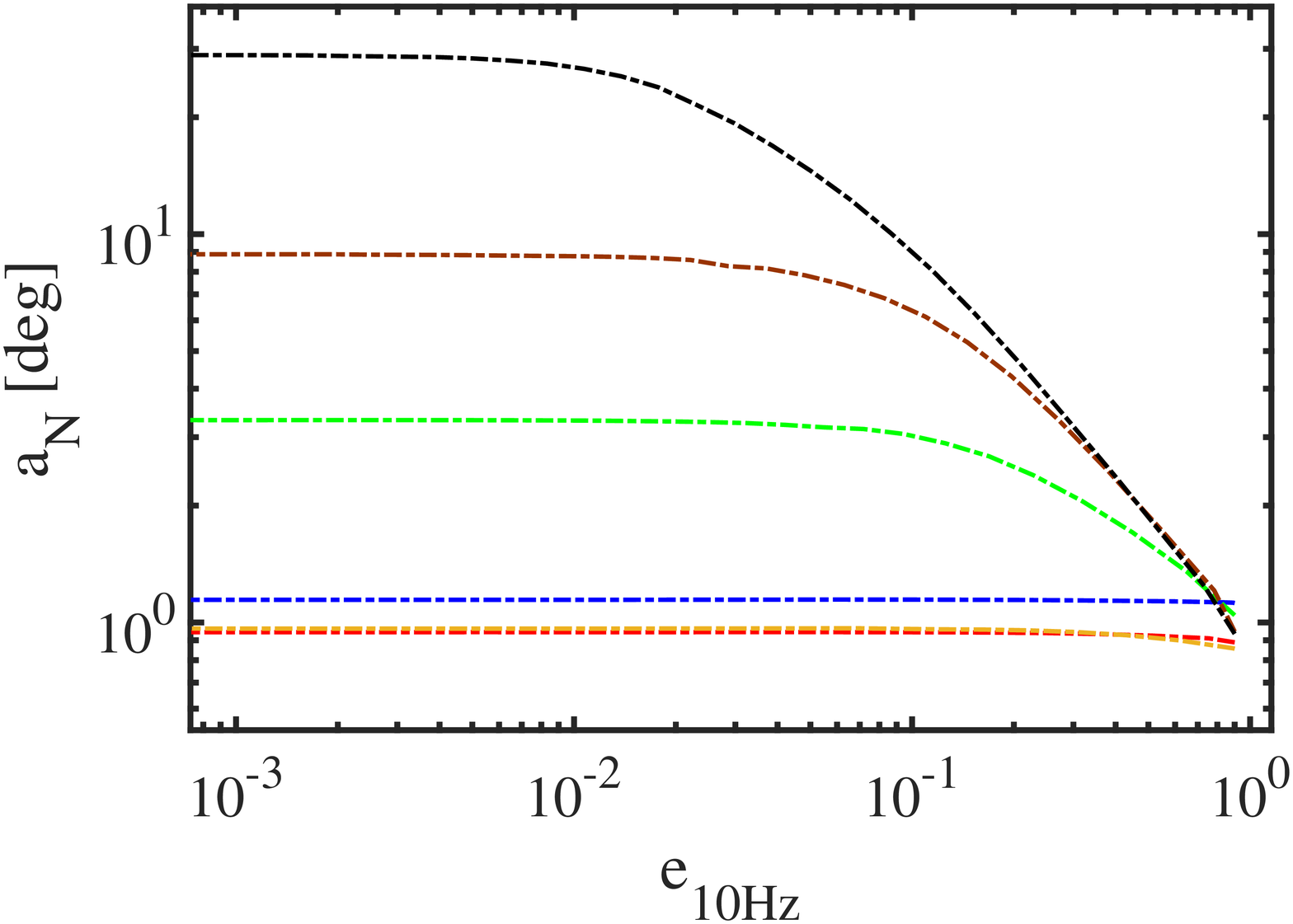}
   \includegraphics[width=80mm]{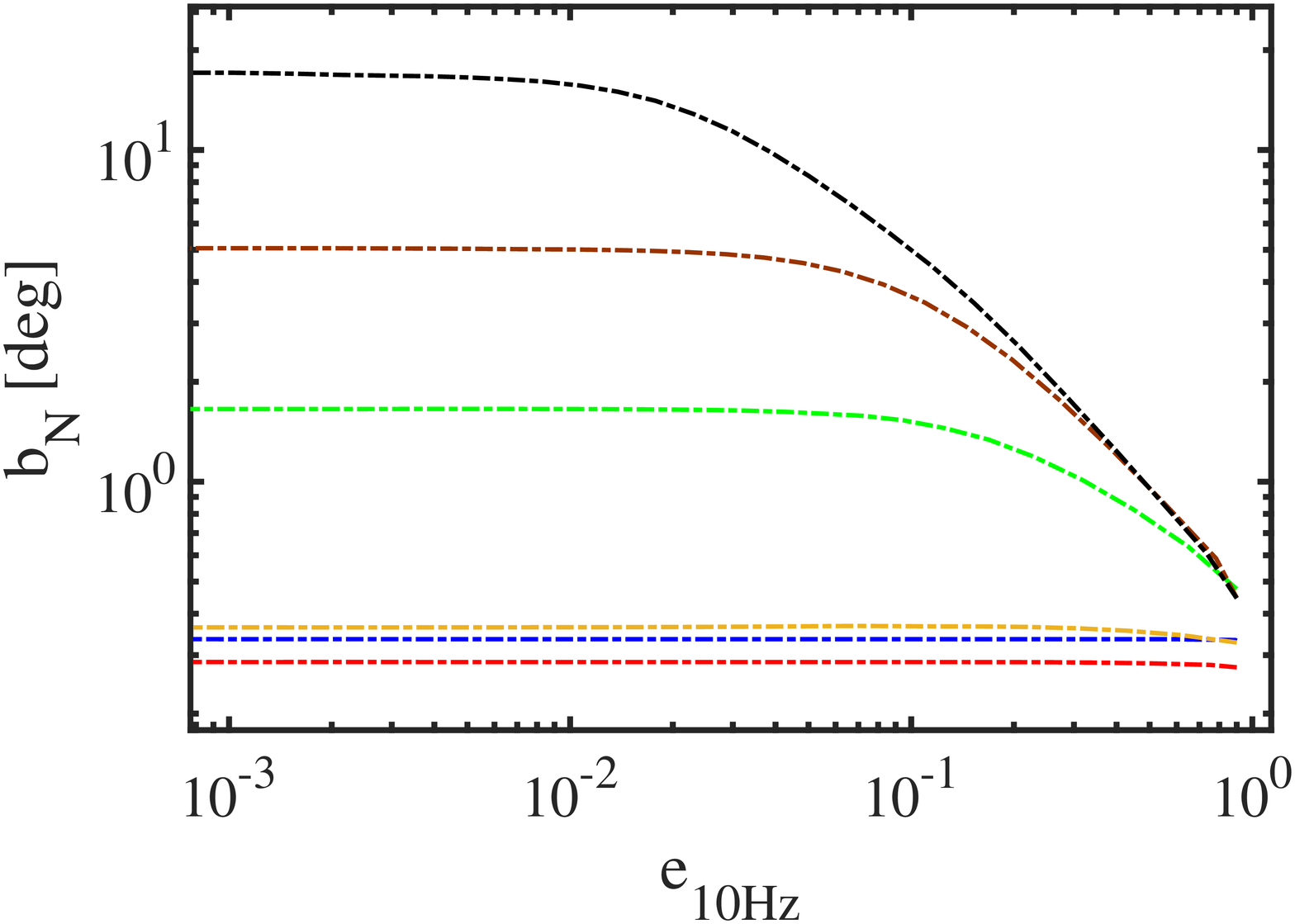}
\\    
   \includegraphics[width=80mm]{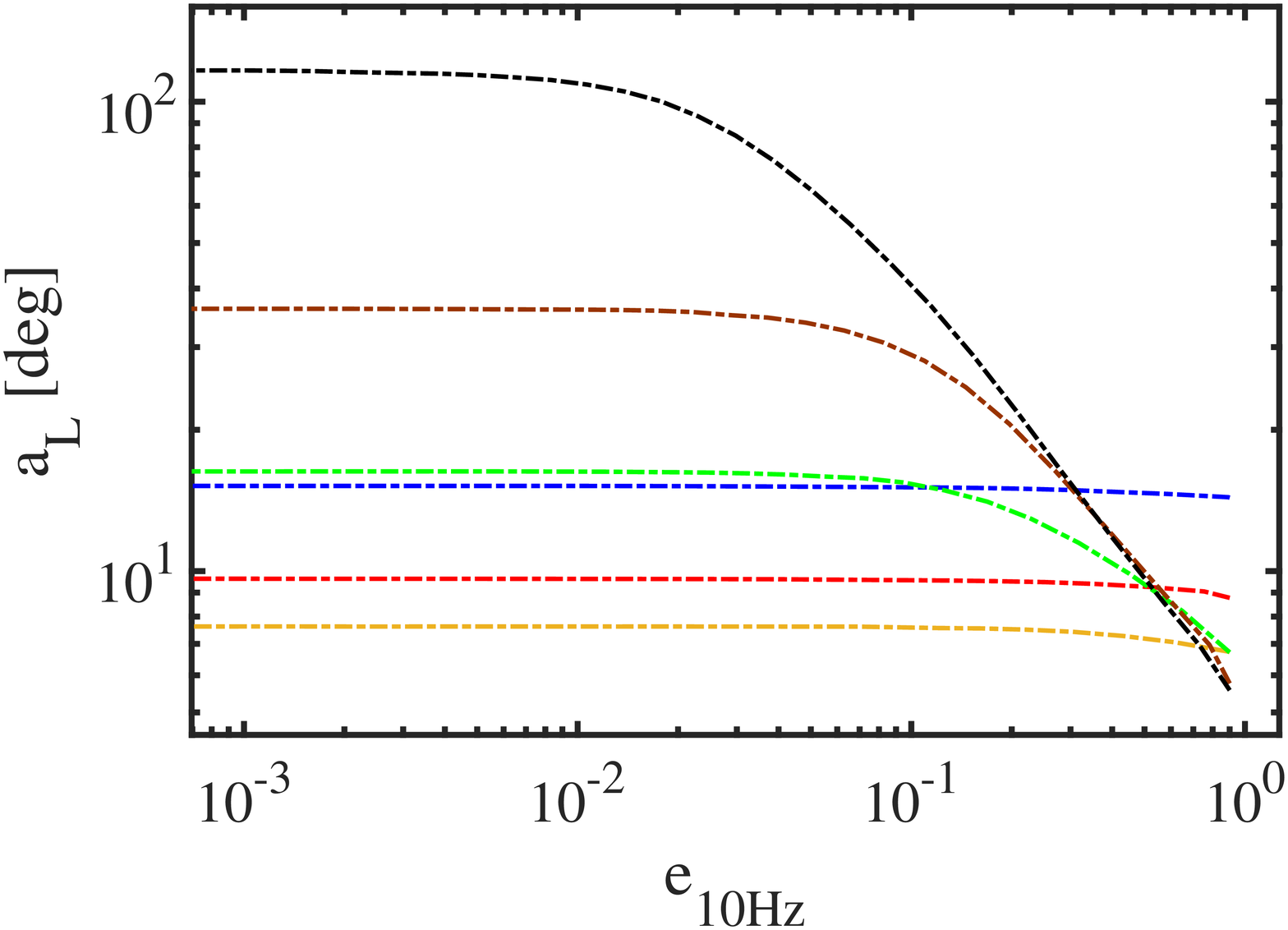}
   \includegraphics[width=80mm]{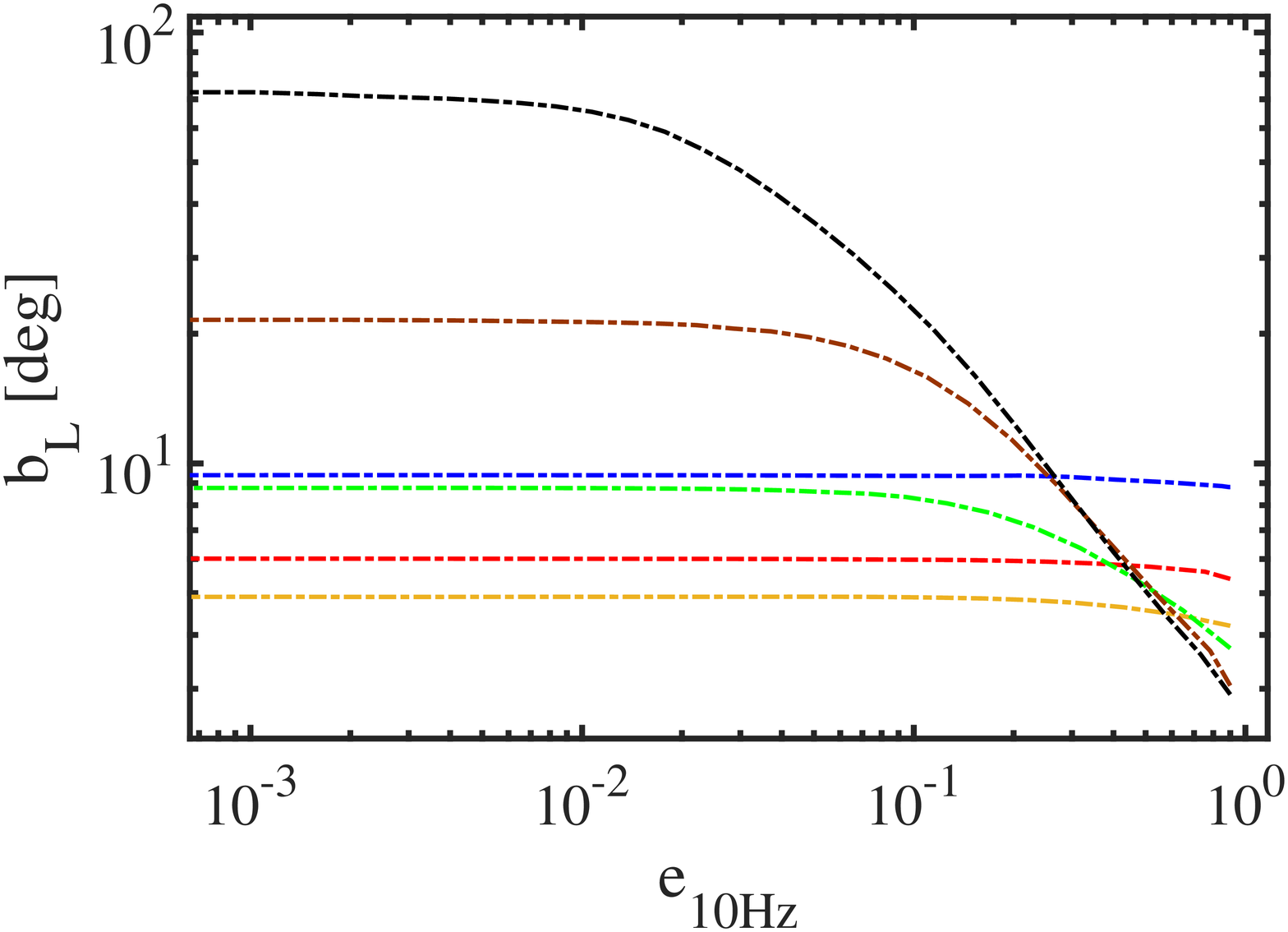}
  \caption{Same as Figure \ref{fig:ParamEst_BHNS_Circular_e10Hz}, but for precessing, eccentric, equal-mass \mbox{BH--BH} binaries of different masses, showing binaries that form outside of the detectors' frequency band. }   \label{fig:ParamEst_BHBH_Circular_e10Hz}
\end{figure*}

 Our conclusions on the eccentricity and initial dimensionless pericenter distance errors are as follows: 
\begin{itemize}

 \item The measurement error of the eccentricity at \mbox{$f_{\rm GW} = 10\,\rm Hz$} is at the level of $\Delta e_{10\,\rm Hz} \sim (10^{-4}-10^{-3}) \times (D_L / 100\,\rm Mpc)$ for precessing eccentric \mbox{NS--NS}, \mbox{NS--BH}, and \mbox{BH--BH} binaries if $e_{10\,\rm Hz} \gtrsim 0.1$ (Figures \ref{fig:ParamEst_BHNS_Eccentric_e10Hz} and \ref{fig:ParamEst_BHBH_Eccentric_e10Hz}). These errors improve systematically for lower binary masses.\footnote{This is qualitatively consistent with the findings of \citet{Sunetal2015}. Note, however, that only a qualitative comparison is possible because \citet{Sunetal2015} considered a different detector network, applied a different waveform model, and used different definitions for both $e_0$ and $\rho_{{\rm p}0}$.}
 
 \item  The orbital eccentricity of a binary may be detected to be significantly different from zero if \mbox{$e_{10 \,\rm Hz}  \gtrsim \Delta e_{10\,\rm Hz}$}. We find that the minimum eccentricity where this is satisfied can be estimated roughly as
 \begin{align}  \label{eq:e10HzLimForDist}
    e_{10 \,\rm Hz,\min} \simeq & \, 0.04 \times \left( \frac{ D_L }{ 100\,{\rm Mpc} } \right)^{0.2}
    \nonumber\\ 
   &  \times \left( \frac{ M_{\rm tot,z} }{ 100 \, \Msun } \right)^{0.1} \times \left( \frac{\eta}{0.25}\right)^{0.03}\,,
 \end{align} 
 where $\eta$ is the symmetric mass ratio defined as \mbox{$\eta = m_A m_B/ (m_A + m_B) ^2$}.\footnote{This expression is valid at least up to 1 Gpc, which is the distance limit below which we validated our codes (Section \ref{subsec:CodeValidation}) and determined numerically how the $\mathrm{S/N} _\mathrm{tot}$ and the parameter measurement errors scale with $D_L$ (Appendix \ref{sec:DL_Dep}).} This estimate is based on masses and distances of simulated binaries using the median of $\Delta e_{10 \,\rm Hz}$, obtained for the $100$ randomly generated sky position and angular momentum unit vectors (Section \ref{subsec:NumSims_2}). For eccentric sources with masses and distances that matches with the currently reported $10$ GW sources (GW150914, GW151226, GW170104, GW170608, GW170729, GW170809, GW170814, GW170817, GW170818, and GW170823; \citealt{Abbottetal2016c,Abbottetal2016b,Abbottetal2017,LIGOColl2017,Abbottetal2017_2,Abbottetal2017_3,Abbottetal2018_b}), we find that $e_{10 \,\rm Hz}$ may be measured to be nonzero using the \mbox{aLIGO--AdV--KAGRA} detector network at design sensitivity if $e_{10 \,\rm Hz} \gtrsim \{0.051, \, 0.046, \, 0.058, \, 0.043, \, 0.081, \, 0.063, \, 0.053, \, 0.023,$ $\, 0.062, \, 0.072\}$, respectively.

 We may compare these results with recent studies, which employed an overlap method \citep{Huertaetal2018,Loweretal2018} and model selection \citep{Loweretal2018}, to examine the capability to detect deviations from quasi-circularity. Our results for GW150914 match that of \citet{Loweretal2018} to within $\sim 2 \%$ and our results for GW150914, GW151226, GW170104, GW170814, and GW170608 are a factor of $\sim 3-4$ lower than those presented by \citet{Huertaetal2018}, who used different assumptions on the detector network, and a different analysis method and waveform model. In particular, \citet{Huertaetal2018} considered one aLIGO detector and spinning, quasi-circular templates in their analysis, while we considered the aLIGO-AdV-KAGRA detector network and a non-spinning waveform model, which may be responsible for a substantial part of the discrepancy.

 \item Figures \ref{fig:ParamEst_BHNS_Eccentric_e10Hz} and \ref{fig:ParamEst_BHBH_Eccentric_e10Hz} show that the eccentricity at the LSO can be measured to an exquisite precision with the aLIGO--AdV--KAGRA detector network at design sensitivity at the level of \mbox{$\Delta e_{\rm LSO} \sim (10^{-5}-10^{-4}) \times (D_L / 100\,\rm Mpc)$} and \mbox{$\Delta e_{\rm LSO} \sim 10^{-3} \times (D_L /100 \,\rm Mpc)$} for precessing eccentric \mbox{BH--BH} binaries with component masses below and above $50 \, \Msun$, respectively, if $e_{10\,\rm Hz} \gtrsim 0.1$. The measurement errors for $e_{\rm LSO}$ are even better for precessing eccentric \mbox{NS--NS} binaries and \mbox{NS--BH} binaries with BH masses below $\sim 10 \, \Msun$ and reach $\Delta e_{\rm LSO} \sim (10^{-6}-10^{-5}) \times (D_L / 100\,\rm Mpc)$. The eccentricity errors improve systematically for lower binary masses and higher $e_{10\,\rm Hz}$ values for all types of binaries. The magnitude of $\Delta e_{\rm LSO}$ is somewhat worse for binaries that form well inside the advanced GW detectors' frequency band than those that form at the lower edge of the band (i.e., \mbox{$f_{\rm GW} \sim 10 \, \Hz$}) as shown in Figures \ref{fig:ParamEst_BHNS_Eccentric} and \ref{fig:ParamEst_BHBH_Eccentric} (i.e., left of stars therein). 
 
 \item $\Delta e_{\rm LSO}$ and $\Delta e_{10\,\rm Hz}$ have a weak $e_{10\,{\rm Hz}}$ dependence for all types of binaries if $e_{10\,{\rm Hz}} \gtrsim 0.2$ (Figures \ref{fig:ParamEst_BHNS_Eccentric_e10Hz} and \ref{fig:ParamEst_BHBH_Eccentric_e10Hz}). The eccentricity measurement errors increase if decreasing $e_{10\,{\rm Hz}}$ to lower values. Asymptotically for $e_{10\,\rm Hz} < 0.04$, Figures \ref{fig:ParamEst_BHNS_Eccentric_e10Hz} and \ref{fig:ParamEst_BHBH_Eccentric_e10Hz} show that both $\Delta e_{\rm LSO}$ and $\Delta e_{10\,\rm Hz}$ can be fitted roughly with a power-law profile as \mbox{$\Delta e_{\rm LSO} \propto e_{10\,\rm Hz}^{-\alpha_{\rm LSO}}$} and \mbox{$\Delta e_{10\,\rm Hz} \propto e_{10\,\rm Hz}^{-\alpha_{\rm 10 Hz}}$}, respectively, where the exponents are within the range \mbox{$\alpha_{\rm LSO} \approx 4.7-5.3$} and \mbox{$\alpha_{\rm 10 Hz} \approx 3.3-3.6$}, depending on the component masses and the type of the binary. 
 
 \item Similar to $\Delta e_{10\,\rm Hz}$, the measurement error of the eccentricity at formation $\Delta e_0$ is lowest for binaries that form with $f_{\rm GW} \sim 10 \, \Hz$, and somewhat worse for binaries that form well inside the band or enter the advanced GW detectors' frequency band with $e_{10\,\rm Hz} \leqslant e_0$ (Figures \ref{fig:ParamEst_BHNS_Eccentric}, \ref{fig:ParamEst_BHBH_Eccentric}, \ref{fig:ParamEst_BHNS_Eccentric_e10Hz}, and \ref{fig:ParamEst_BHBH_Eccentric_e10Hz}). For either precessing eccentric BH or NS binaries, the measurement error when the binary forms with \mbox{$f_{\rm GW} \sim 10 \, \Hz$} with $e_0 = e_{10\,\rm Hz} = 0.9$ is at the level of $\Delta e_0 \sim 10^{-3} \times (D_L / 100\,\rm Mpc)$, which decreases to $\Delta e_0 \sim 10^{-1} \times (D_L / 100 \,\rm Mpc)$ when \mbox{$e_{10\,\rm Hz} \sim 0.1$} or when binaries form with \mbox{$\rho_{{\rm p}0} \lesssim 10$}. Further, we find that $\Delta e_0$ has a weak mass dependence for binaries that enter the advanced GW detectors' frequency bands with similar $e_{10\,\rm Hz}$ values. 
 
 \item An important parameter that characterizes the velocity dispersion of the host environment or the orbital velocity of a binary before a hardening encounter is $\rho_{{\rm p}0}$ \citep{OLearyetal2009,Gondanetal2017}. This parameter can be measured at the level of \mbox{$\Delta \rho_{{\rm p}0} \sim (10^{-2}-10) \times (D_L / 100\,\rm Mpc)$} for $e_{10\,\rm Hz} \gtrsim 0.1$, where the errors decrease with increasing mass and $e_{10\,\rm Hz}$ (Figures \ref{fig:ParamEst_BHNS_Eccentric_e10Hz} and \ref{fig:ParamEst_BHBH_Eccentric_e10Hz}). This parameter may be measured typically more accurately for binaries that form in the advanced GW detectors' frequency band than those that form outside of the bands (i.e., to the left of the stars in Figures \ref{fig:ParamEst_BHNS_Eccentric} and \ref{fig:ParamEst_BHBH_Eccentric}) at the level of $\Delta \rho_{{\rm p}0} \sim (0.005-0.5) \times (D_L / 100\,\rm Mpc)$. 
 
\end{itemize}

\subsection{Measurement Errors of Source Distance, Direction, Binary Orientation, and the Redshifted Chirp Mass} 
\label{subsec:MeasErrSlowParamChirpMass} 
 
 The measurement errors of source distance, sky location, and binary orientation (i.e., slow parameters) are lowest for binaries that form in the advanced GW detectors' frequency band (Figures \ref{fig:ParamEst_BHNS_Circular} and \ref{fig:ParamEst_BHBH_Circular}), worse for binaries that enter the band with \mbox{$e_{10\,\rm Hz} \lesssim e_0$} (Figures \ref{fig:ParamEst_BHNS_Circular_e10Hz} and \ref{fig:ParamEst_BHBH_Circular_e10Hz}), and highest for binaries in the circular limit (i.e., \mbox{$\rho_{\mathrm{p}0} \rightarrow \infty$} or equivalently \mbox{$e_{10\,\rm Hz} \ll 1$}). The measurement errors of these parameters slightly improve for precessing eccentric \mbox{NS--NS}, \mbox{NS--BH}, and \mbox{BH--BH} binaries forming in the band or entering the band with \mbox{$e_{10\,\rm Hz} \lesssim e_0$} compared to similar binaries in the circular limit unless the BH masses exceed $\sim 50 \, \Msun$. This is expected because the measurement errors of slow parameters depend on the GW amplitude, which is set by the $\mathrm{S/N} _\mathrm{tot}$ \citep{Kocsisetal2007}. Similar trends characterize the mass dependence of the $\mathrm{S/N} _\mathrm{tot}$ (Figure \ref{fig:SNRtot_rho}). For BH masses in the range $50 \, \Msun - 110 \, \Msun$, the measurement errors of slow parameters can improve by up to a factor of $2-4$ and $2-20$ for precessing eccentric \mbox{NS--BH} and \mbox{BH--BH} binaries, respectively, for $e_{10\,\rm Hz} \gtrsim 0.1$ compared to similar binaries in the circular limit. The errors improve systematically for higher $e_{10\,\rm Hz}$ values and for higher BH masses in the range $50 \, \Msun - 110 \, \Msun$. For binaries that form in the advanced GW detectors' frequency band, the measurement precision further improves to factors of $2-6$ and $2-50$ for precessing eccentric \mbox{NS--BH} and \mbox{BH--BH} binaries, respectively. 
 
 The measurement error of redshifted chirp mass $\Delta (\ln \mathcal{M}_z)$ is lowest for binaries that form in the advanced GW detectors' frequency band near the detectors' low-frequency cutoff, implying that they enter the band at $e_{10\,\rm Hz} \sim e_0$ (Figures \ref{fig:ParamEst_BHNS_Circular} and \ref{fig:ParamEst_BHBH_Circular}). The measurement precision can improve by up to a factor of $\sim 20$ and $\sim 50-100$ for precessing eccentric \mbox{NS--NS} and \mbox{NS--BH} binaries with BH companion masses $\lesssim 40 \, \Msun$, respectively, compared to similar binaries in the circular limit. Even higher precision may be achieved for BH companion masses above $\sim 40 \, \Msun$, in particular, a factor of $\sim 300$ may be achieved for precessing eccentric \mbox{NS--BH} binaries with BH masses of $\sim 100 \, \Msun$ because for these sources a quasi-circular inspiral terminates outside of the frequency band. In comparison to circular binaries, the measurement accuracy can improve by up to a factor of $\sim 10-10^2$ and $\sim 10^2-10^3$ for precessing eccentric \mbox{BH--BH} binaries with BH masses below $\sim 40 \, \Msun$ and BH masses in the range $40 - 110 \, \Msun$, respectively. In all cases, higher precision corresponds to higher $e_{10\,\rm Hz}$ values and BH masses.
 
 We find in agreement with \citet{OLearyetal2009} that the $\mathrm{S/N} _\mathrm{tot}$ can be significantly greater for high-mass binaries with relatively low $\rho_{{\rm p}0}$ than for similar binaries in the circular limit, which implies that eccentric binaries with higher masses may be detected to larger distances, and generally greater BH mass binaries may be detected if eccentric than similar binaries in the circular mergers; see Figure \ref{fig:SNRtot_rho} for examples. Similarly, we find that measurement errors of slow binary parameters are comparable for low-mass and high-mass precessing eccentric \mbox{BH--BH} binaries and for precessing eccentric \mbox{NS--BH} binaries with BH companion masses $\lesssim 30 \, \Msun $ when binaries form in the advanced GW detectors' frequency band (Figures \ref{fig:ParamEst_BHNS_Circular} and \ref{fig:ParamEst_BHBH_Circular}) or enter the $10 \, \Hz$ frequency band with $0.1 \lesssim e_{10\,\rm Hz}$ and $e_{10\,\rm Hz} \leqslant e_0$, respectively (Figures \ref{fig:ParamEst_BHNS_Circular_e10Hz} and \ref{fig:ParamEst_BHBH_Circular_e10Hz}).

\subsection{Measurement Accuracy of the Characteristic Relative Velocity} 
\label{subsec:MeasErrCharVelDispOfHost} 
 
 For GW capture events with a fixed initial relative velocity between two compact objects, the initial pericenter distance $\rho_{{\rm p}0}$  is uniformly distributed up to a critical value where the system forms a marginally bound binary \citep{Kocsisetal2006,OLearyetal2009,Gondanetal2017}. Similarly, for resonant binary--single encounters \citep{Rodriguezetal2018,Rodriguezetal2018_2,Samsingetal2018_2}, the GW emission may form a merging binary whose $\rho_{{\rm p}0}$ is expected to be approximately uniformly distributed up to a critical value set by the orbital velocity before a hardening encounter $v_{\rm bin}$ plus the initial velocity of the scattering object. A crude relation between the initial pericenter distance and the orbital velocity of the binary may be derived from Equation (18) in \citet{OLearyetal2009},
 \begin{equation}\label{eq:rhop0-v}
    \rho_{\mathrm{p}0} \sim \left(\frac{85\pi\eta}{6 \sqrt{2}}\right)^{2/7} v_{\rm bin}^{-4/7} = 175 \, (4\eta)^{2/7} \left(\frac{v_{\rm bin}}{100\,\rm km/s}\right)^{-4/7} \, 
 \end{equation}
 where the symmetric mass ratio $\eta$ is defined above in Equation (\ref{eq:e10HzLimForDist}).
 Here $v_{\rm bin}$ can be expressed by the measurable quantities $\eta$ and $\rho_{\mathrm{p}0}$ as
 \begin{equation} \label{eq:v-rhop0eta}
   v_{\rm bin} \sim 266 \, \rm km/s \times (4\eta)^{1/2} \left(\frac{\rho_{\mathrm{p}0}}{100}\right)^{-7/4}  \, .
 \end{equation} 
 The initial pericenter distance $\rho_{\mathrm{p}0}$ sets the eccentricity at the LSO (see Equations 30 and (32) in \citealt{Gondanetal2017}).
   
 Similar arguments apply for single--single GW capture sources in GNs and GCs, where $v_{\rm bin}$ is replaced by $2 \sqrt{2} \sigma$,\footnote{For single--single GW capture sources, $\rho_{\mathrm{p}0}$ depends on the maximum relative velocity between the objects in Equation (\ref{eq:rhop0-v}), which is $v_{\rm rel,max} = 2v_{\rm max}$ where $v_{\rm max}$ is the maximum velocity of the individual objects in the system. Given the velocity dispersion, this may be expressed as  $\sigma \sim v_{\rm max} / \sqrt{2}$.}, where $\sigma$ refers to the characteristic velocity dispersion of the host environment \citep{Gondanetal2017}. Thus, $\sigma$\footnote{Note that Equation (63) in \citet{Gondanetal2017} relates the eccentricity at the LSO to $\sigma$ as
 \begin{equation}
   \sigma \sim 258 \,\rm km/s \times (4\eta)^{1/2} \left(\frac{e_{\rm LSO}}{0.01}\right)^{35/32} \, ,
 \end{equation}
 which can be given by substituting $\rho_{\mathrm{p}0}$ with $e_{\rm LSO}$ in Equation (\ref{eq:sigma_etarhop0}) using Equation (32) in \citet{Gondanetal2017}.} can be expressed by $\eta$ and $\rho_{\mathrm{p}0}$ as 
 \begin{equation}  \label{eq:sigma_etarhop0}
   \sigma \sim 94 \, \rm km/s \times (4\eta)^{1/2} \left(\frac{\rho_{\mathrm{p}0}}{100}\right)^{-7/4} \, .
 \end{equation} 
 
 From Equation (\ref{eq:v-rhop0eta}) and using Equations (\ref{eq:deltarhop0}) and (\ref{eq:deltae10Hz}), the measurement error of \mbox{$v_{\rm bin} = v_{\rm bin}(\eta, \rho_{\mathrm{p}0} )$}, can be expressed as
\begin{align}  \label{eq:errdeltav}
  ( \Delta v_{\rm bin} )^2  = & \left( \frac{\partial v_{\rm bin} (\eta, \rho_{\mathrm{p}0} )}{\partial \eta} \right)^2  (\Delta \eta )^2
   \nonumber\\ 
   & + \left( \frac{\partial v_{\rm bin} (\eta, \rho_{\mathrm{p}0} )}{\partial \rho_{\mathrm{p}0} } \right)^2 (\Delta \rho_{\mathrm{p}0} )^2 
  \nonumber\\
  & + 2 \frac{\partial v_{\rm bin} (\eta, \rho_{\mathrm{p}0} )}{\partial \eta} \frac{\partial v_{\rm bin} (\eta, 
  \rho_{\mathrm{p}0} )}{\partial \rho_{\mathrm{p}0} } \left\langle \Delta \eta \Delta \rho_{\mathrm{p}0} \right\rangle \, ,
\end{align} 
 which simplifies as
\begin{equation}   \label{eq:relveldist}
 \frac{ \Delta v_{\rm bin} }{ v_{\rm bin} } = \sqrt{ \left( \frac{1}{2} \right)^2 \frac{ (\Delta \eta)^2 }{\eta ^2} + \left( \frac{7}{4} \right)^2 \frac{ (\Delta \rho_{\mathrm{p}0} )^2 }{ \rho_{\mathrm{p}0}^2 }}\, ,
\end{equation} 
 and an identical equation holds for $\Delta \sigma / \sigma$. Here we used the fact that the cross-correlation coefficient $\left\langle \Delta \eta \Delta \rho_{\mathrm{p}0} \right\rangle$ is negligible compared to $(\Delta \eta )^2$ and $(\Delta \rho_{\mathrm{p}0} )^2$, as found from numerical investigations. $\Delta \eta / \eta$ can be given using Equation (54) in \citet{Gondanetal2018}, and we find from numerical investigation that crudely $\Delta \eta/\eta \sim 10\,\Delta \mathcal{M}_z / \mathcal{M}_z$; see Table 4 in \citet{Gondanetal2018} for examples. Thus, since \mbox{$\Delta v_{\rm bin} / v_{\rm bin} = \Delta \sigma / \sigma$}, we refer to both $v_{\rm bin}$ and $\sigma$ as the ``characteristic relative velocity'' $w$ and denote $\Delta v_{\rm bin} / v_{\rm bin}$ and $\Delta \sigma / \sigma$ by $\Delta w / w$ in this paper.

 Results for $\Delta w /w$ are displayed in Figure \ref{fig:ParamEst_RelVelDist}. Generally, $\Delta w /w$ is at the level of $\Delta w /w \sim (10^{-3} - 10^{-1}) \times (D_L / 100\,\rm Mpc)$ for $0.04 \lesssim e_{10\,\rm Hz}$, where lower errors correspond to higher $e_{10\,\rm Hz}$ values. In particular, $\Delta w /w$ is lowest for binaries that form in the advanced GW detectors' frequency band near the low-frequency cutoff of detectors or enter the band at \mbox{$e_{10\,\rm Hz} \sim e_0$}, and it is higher for binaries that form well inside the band or form with \mbox{$e_{10\,\rm Hz} < e_0$}. Further, we find that $\Delta w /w$ weakly depends on the binary mass for different $e_{10\,\rm Hz}$ values for all types of binaries. We conclude that the characteristic relative velocity can be estimated with high accuracy with GW observations of eccentric mergers.

\begin{figure*}
  \centering
  \includegraphics[width=80mm]{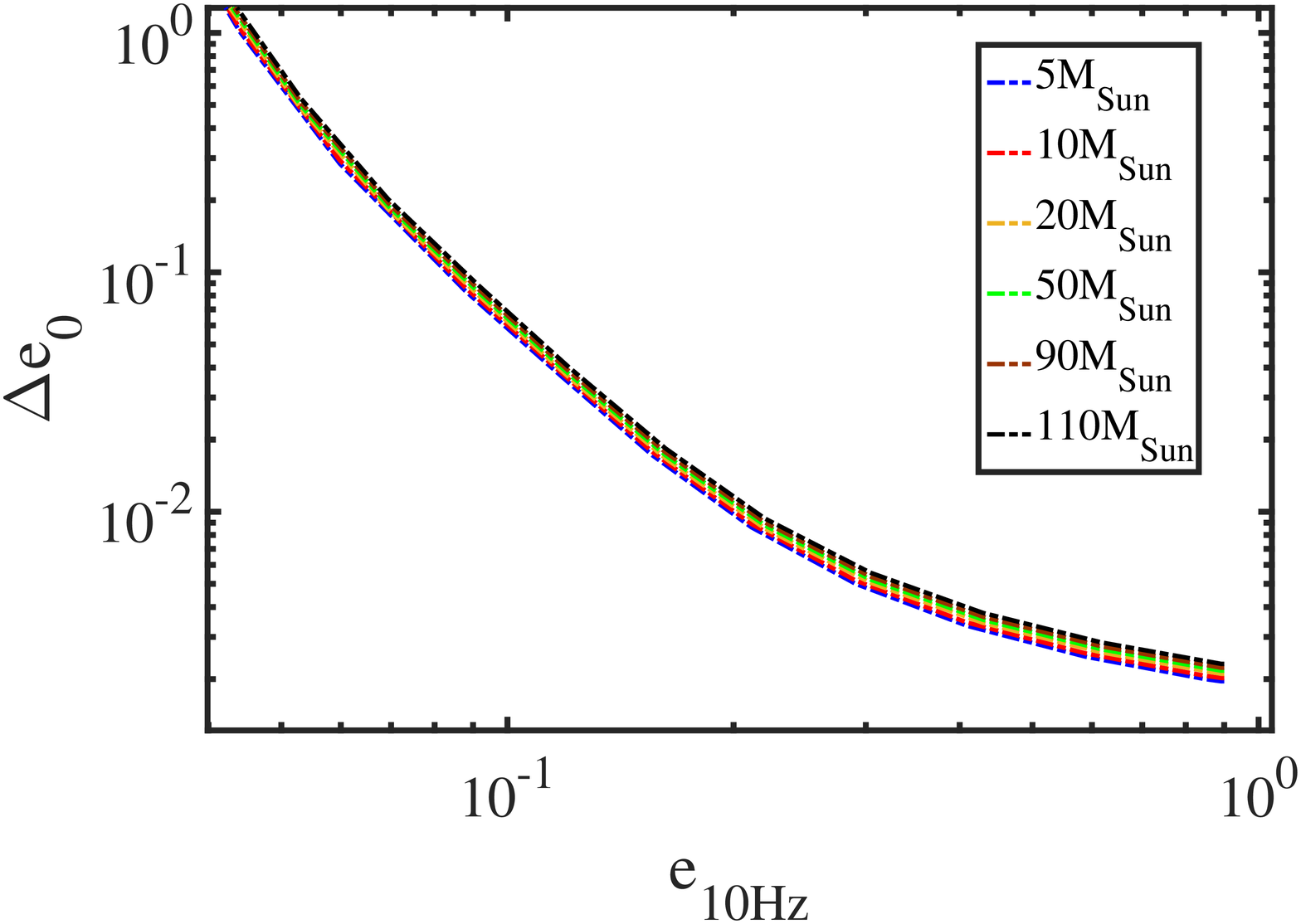}
  \includegraphics[width=80mm]{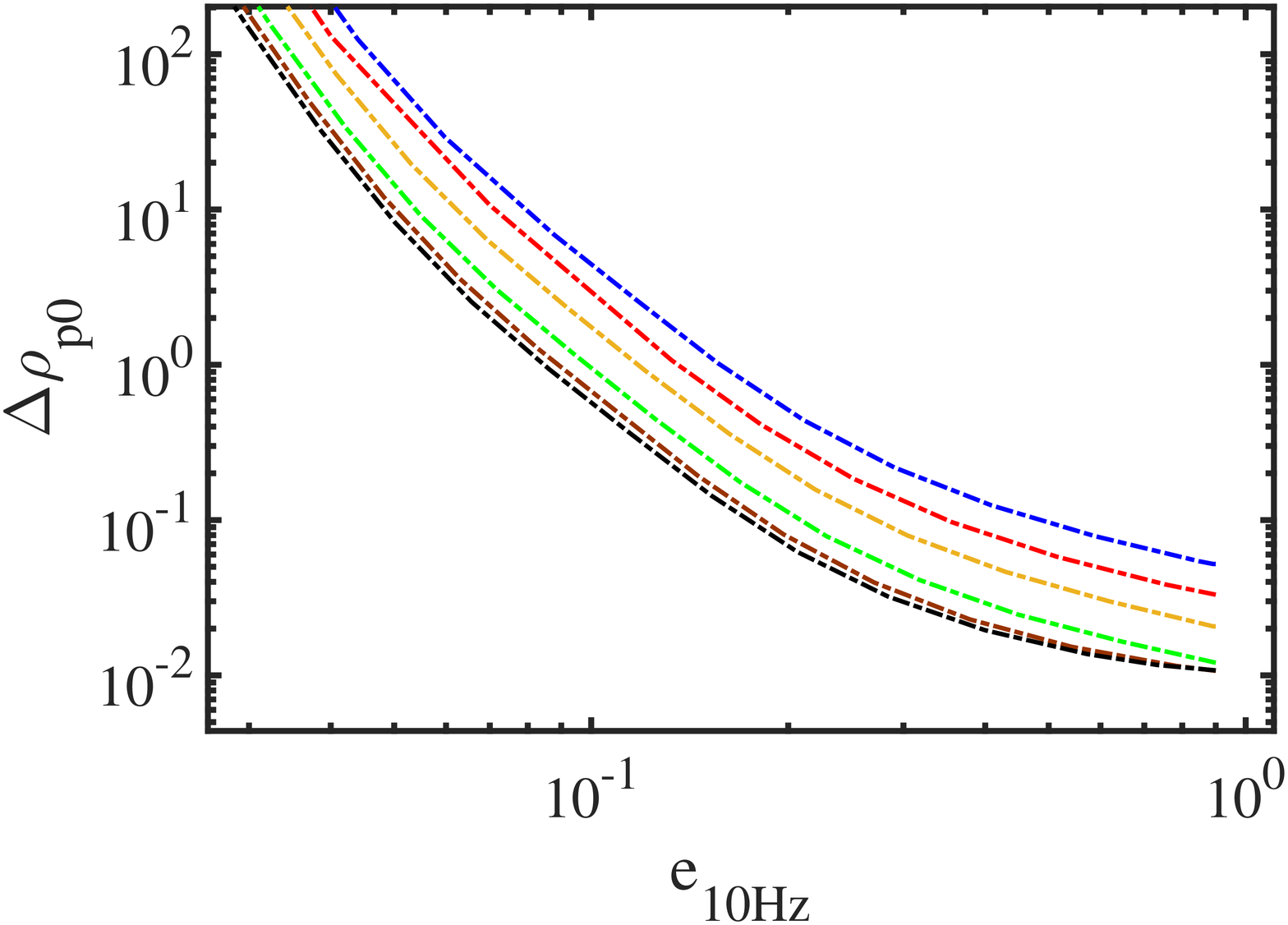}
\\    
  \includegraphics[width=80mm]{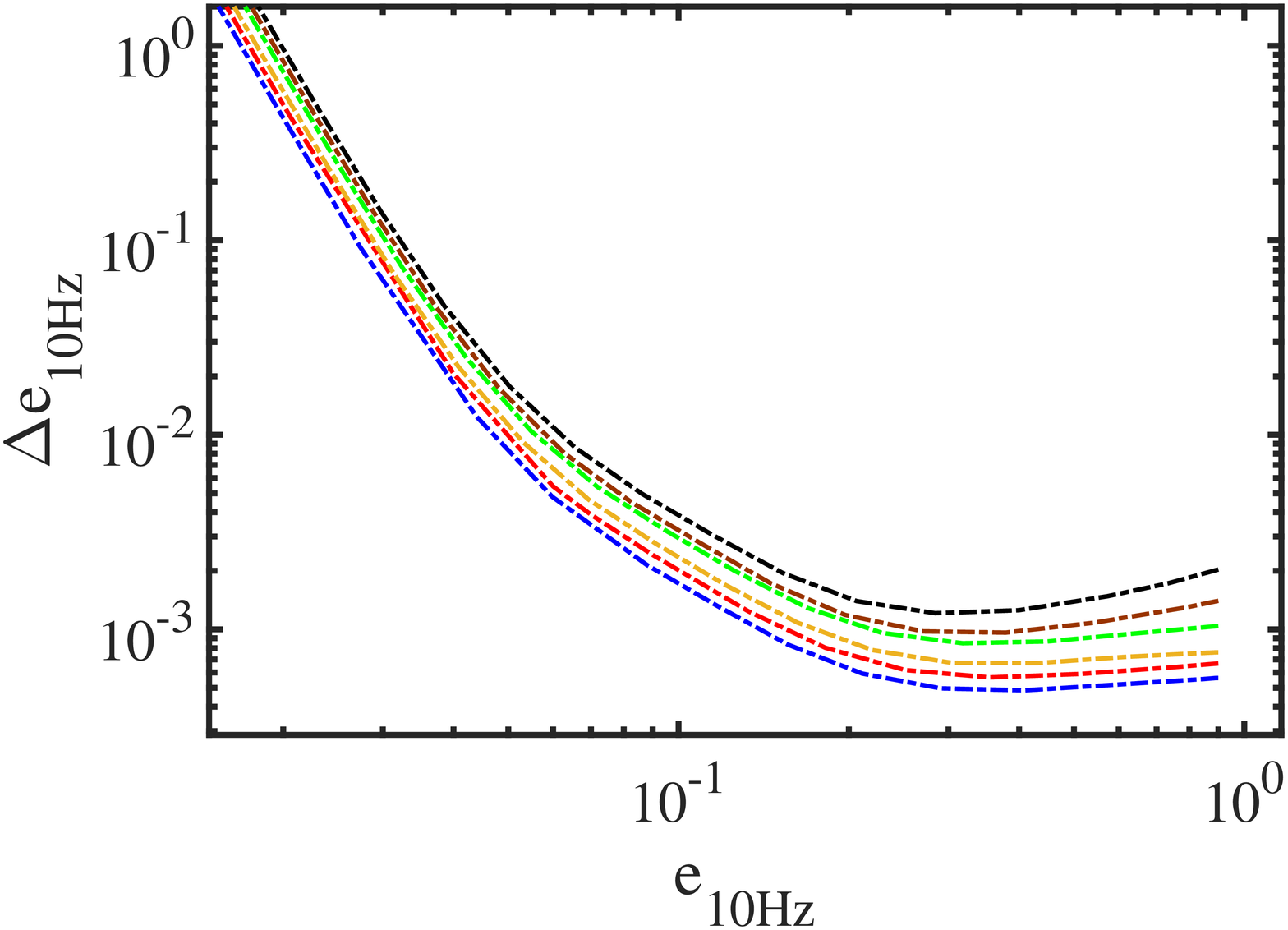}
  \includegraphics[width=80mm]{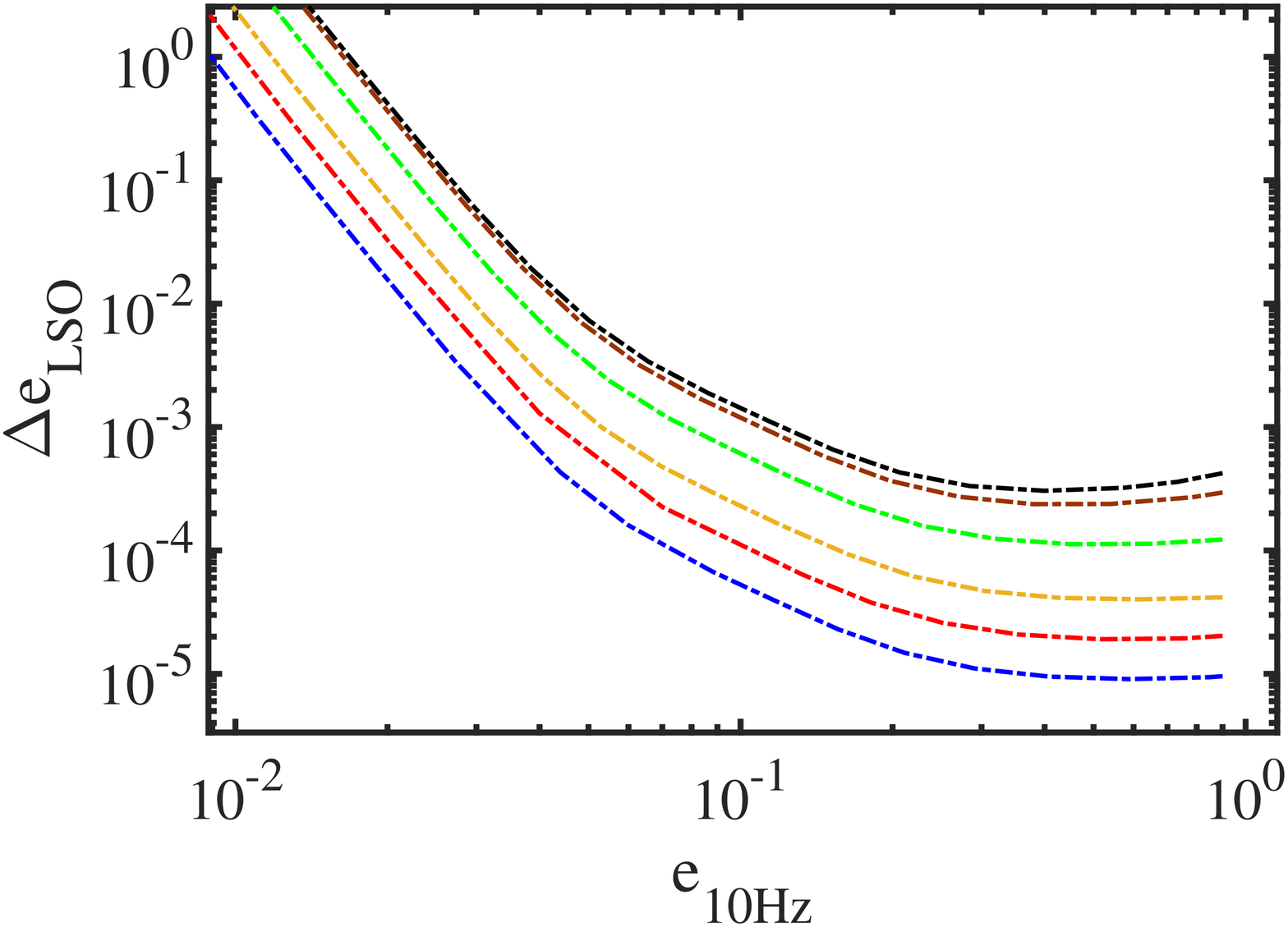}
  \caption{Same as Figure \ref{fig:ParamEst_BHNS_Eccentric_e10Hz}, but for precessing, eccentric, equal-mass \mbox{BH--BH} binaries. }   \label{fig:ParamEst_BHBH_Eccentric_e10Hz}
\end{figure*}

\begin{figure*}
   \centering
   \includegraphics[width=80mm]{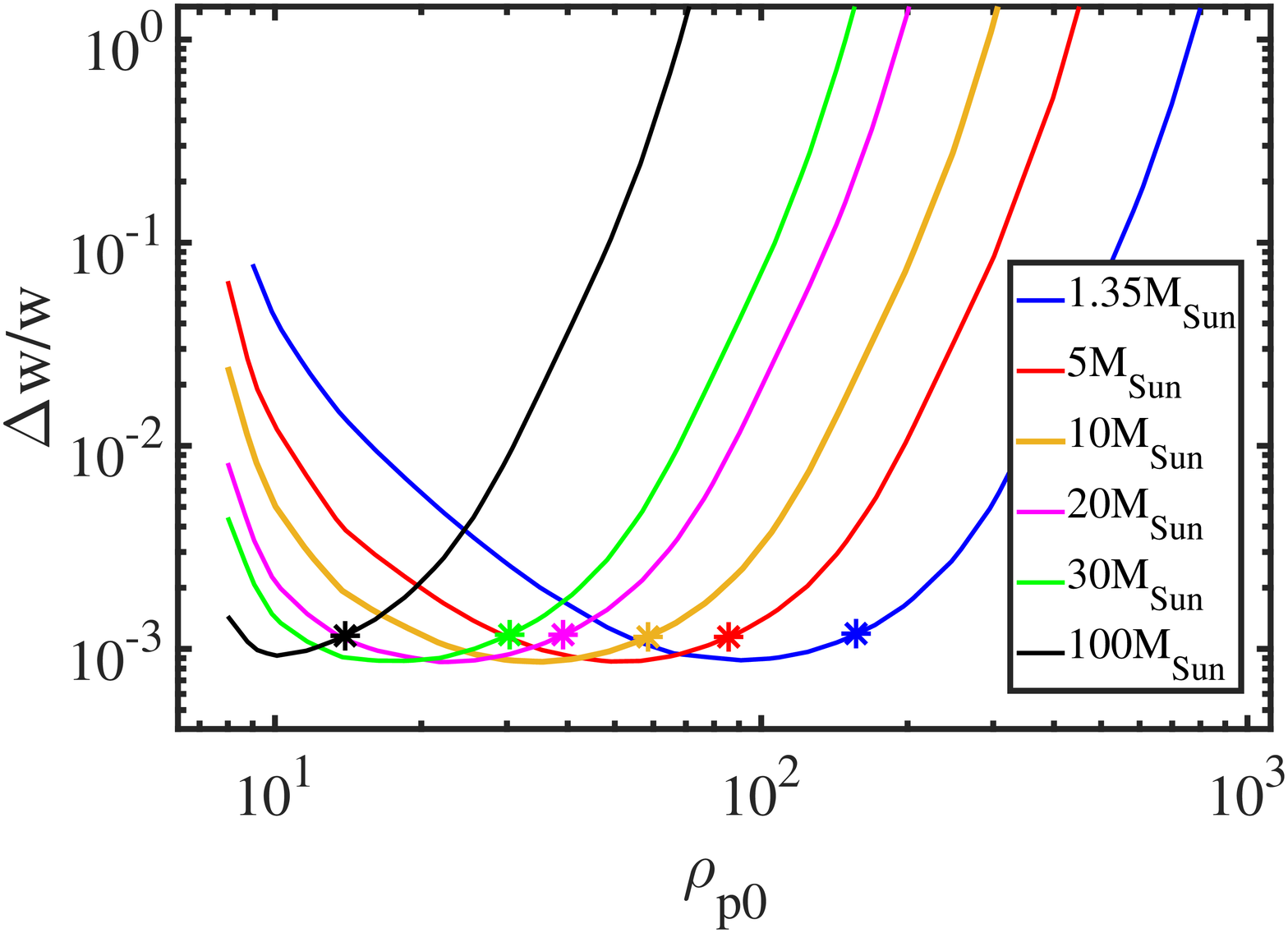}
   \includegraphics[width=80mm]{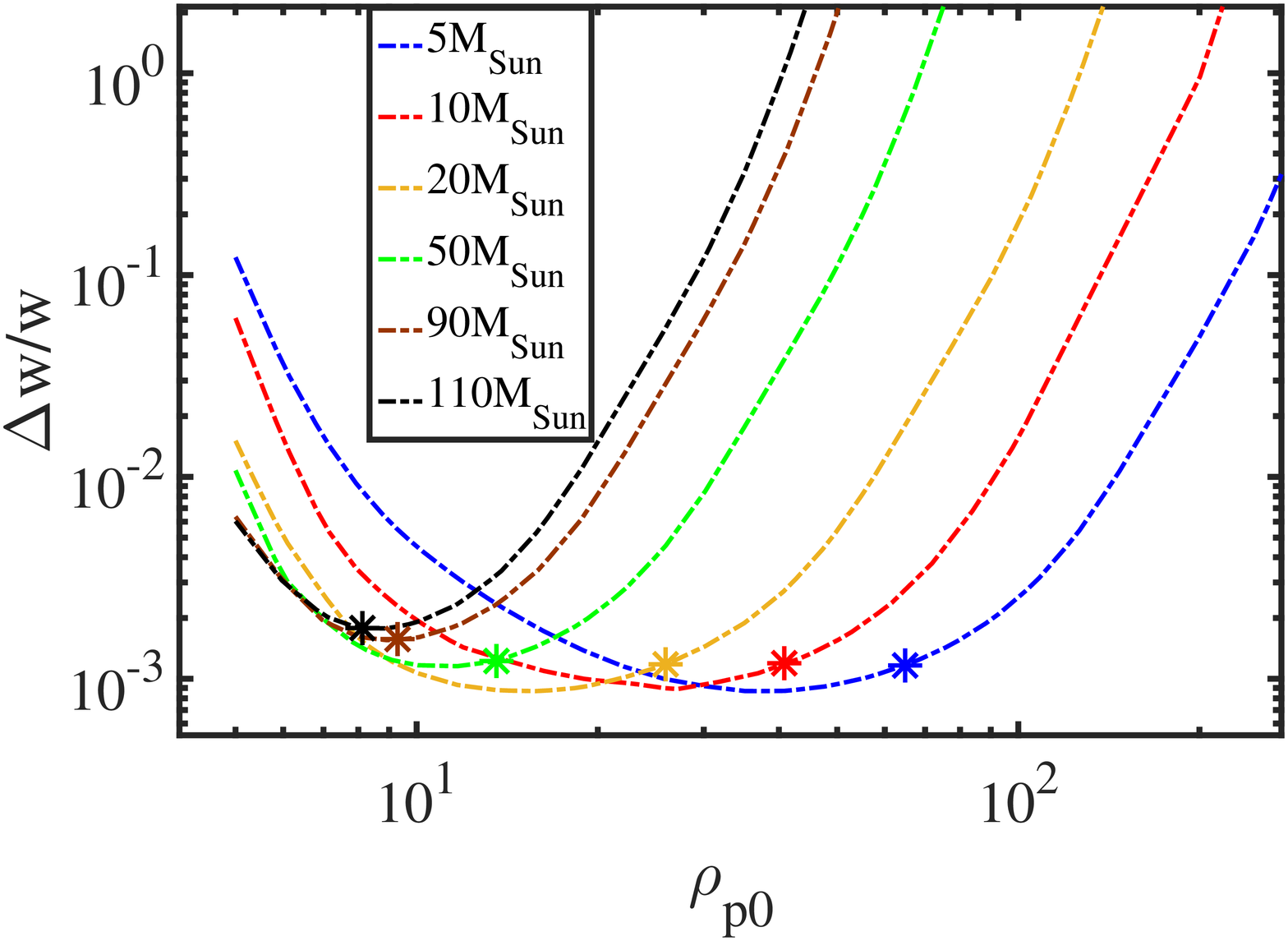}
\\
   \includegraphics[width=80mm]{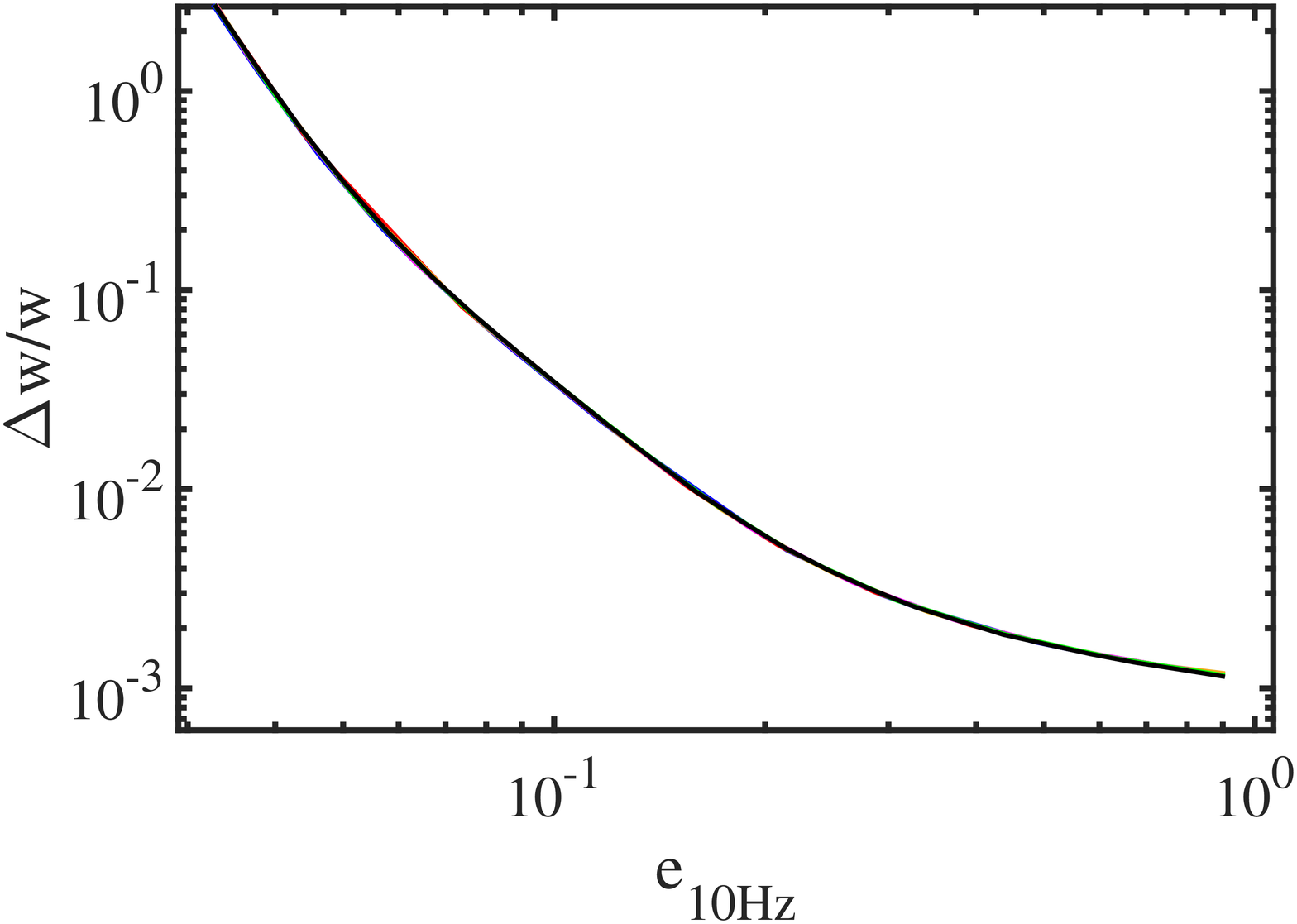}
   \includegraphics[width=80mm]{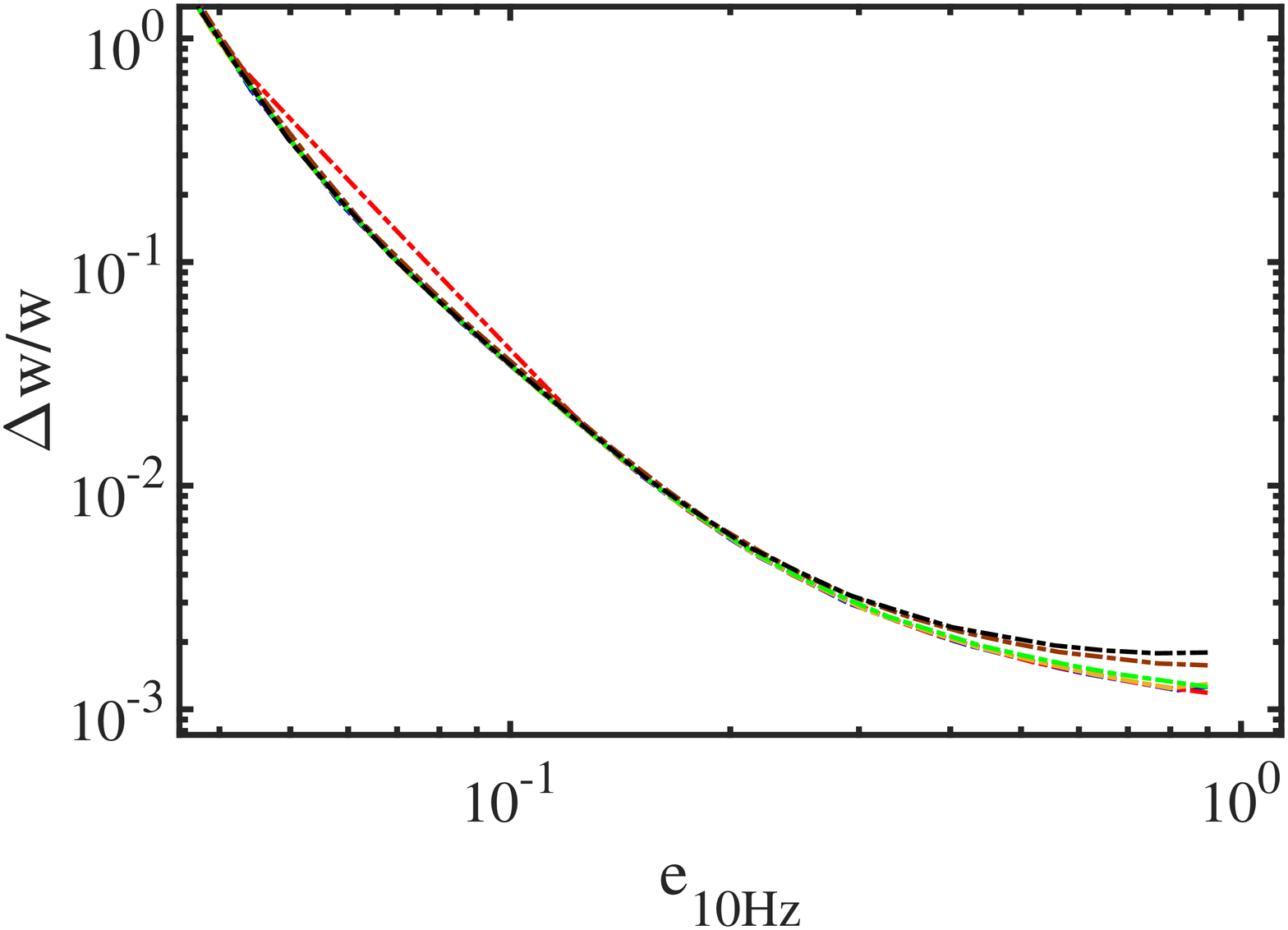}
  \caption{Measurement accuracy of the characteristic relative velocity of the sources before GWs start driving the evolution for precessing eccentric \mbox{NS--NS} and \mbox{NS--BH} binaries (left) and for precessing eccentric \mbox{BH--BH} binaries (right) as a function of $\rho_{\mathrm{p}0}$ (top) and $e_{\rm 10 Hz}$ (bottom). This is expected to be comparable to the velocity dispersion of the host environment. Measurement errors are calculated by combining the chirp mass errors with the $e_{\rm LSO}$ errors as in Equation (\ref{eq:relveldist}). Different curves show different companion masses as labeled in the legends. Parameters of the detector network and all other binary parameters are fixed as in Figure \ref{fig:SNRtot_rho}. Similar to Figures \ref{fig:ParamEst_BHNS_Circular} and \ref{fig:ParamEst_BHNS_Eccentric}, measurement errors are displayed for $100 \, \Mpc$ but can be scaled to other $D_L$ values by multiplying the displayed results by a factor of $D_L / 100 \, \Mpc$. Stars indicate the case where the binary forms at $f_{\rm GW} = 10\,\rm Hz$ as in Figure \ref{fig:SNRtot_rho}. We find similar trends with $\rho_{\mathrm{p}0}$ and $e_{\rm 10Hz}$ for other random choices of binary direction and orientation (not shown). }    \label{fig:ParamEst_RelVelDist}
\end{figure*}

\newpage 
 
\subsection{Initial Orbital Parameter Dependence of Measurement Errors} 
\label{subsec:MeasErrorsInitialOrbParams} 
 
 We identify the following trends for the $\rho_{{\rm p}0}$ dependence of measurement errors in Figures  \ref{fig:SNRtot_rho}--\ref{fig:ParamEst_BHBH_Eccentric}.
\begin{itemize}
 
 \item The $\rho_{\mathrm{p}0}$ dependence of the slow parameter errors $(\Delta \ln D_L, a_N, b_N, a_L, b_L)$ is qualitatively similar to that of $1/(\mathrm{S/N}_\mathrm{tot})$ as expected \citep{Kocsisetal2007}. The errors have a minimum at moderate $\rho_ {\mathrm{p}0}$ between 10 and 100. The errors increase rapidly with decreasing $\rho_{\mathrm{p}0}$ for lower $\rho_{\mathrm{p}0}$, and converge asymptotically for $\rho_{\mathrm{p}0}$ to the value of the circular limit.
 
 \item  The measurement errors of fast parameters (Equation \ref{eq:fast}) are qualitatively similar to those of slow parameters (Equation \ref{eq:slow}) in terms of their $\rho_{\mathrm{p}0}$ dependence, except that they have a minimum at a higher $\rho_{\mathrm{p}0}$, which corresponds to a characteristic GW frequency of \mbox{$f_{ \rm GW} \sim 7 \, {\rm Hz}$}. Further, $\Delta e_\mathrm{LSO}$ is inaccurately estimated by our method for high $\rho_{\mathrm{p}0}$ as the Fisher matrix algorithm becomes invalid in that regime.
 
 \item The initial parameters $\Delta e_0$ and $\Delta \rho_{\mathrm{p}0}$ increase rapidly with $\rho_{\mathrm{p}0}$ for high $\rho_{\mathrm{p}0}$ and become indeterminate in the circular limit. Indeed, for high $\rho_{\mathrm{p}0}$, the binary forms outside of the advanced GW detectors' frequency band, and the measured GW signal cannot specify the initial condition. Similarly, $\Delta e_0$ and $\Delta \rho_{\mathrm{p}0}$ increase rapidly in the limit of low $\rho_{\mathrm{p}0}$ because of a low $\mathrm{S/N}_\mathrm{tot}$ and possible degeneracies among the parameters.
 
\end{itemize}
 We refer the readers for further discussion to \citet{Gondanetal2018}. 
 
 Furthermore, we find that measurement accuracies do not improve significantly for the parameter ranges shown in Figures \ref{fig:SNRtot_rho}--\ref{fig:ParamEst_BHBH_Eccentric} if $e_0$ is increased beyond $e_0 \sim 0.9$. For instance, measurement accuracies of fast parameters (Equation \ref{eq:fast}), slow parameters (Equation \ref{eq:slow}), the parameters ($e_0, \rho_{\mathrm{p}0}$), and $e_{10\,\rm Hz}$ improve by up tu a factor of $\sim 2$, $\sim 1.1$, $\sim 2$, and $\sim 2$, respectively, if $e_0$ is increased from $0.9$ to $0.99$.

\subsection{Mass Dependence of Measurement Errors} 
\label{subsec:MassDepOfErrors}
 
 The mass dependence of the measurement accuracies can be examined by comparing the different curves in \mbox{Figures \ref{fig:ParamEst_BHNS_Circular}--\ref{fig:ParamEst_BHBH_Eccentric_e10Hz}}. This shows that $(\Delta \ln D_L, a_N, b_N, a_L, b_L)$ have a minimum as a function of mass for fixed $\rho_{{\rm p}0}$. The mass dependency of $(\Delta \ln D_L, a_N, b_N, a_L, b_L)$ can be understood from the scaling $1/(S/N_{\rm tot})$, which has a maximum at fixed $\rho_{{\rm p}0}$ \citep{OLearyetal2009}. The characteristic GW spectral amplitude per logarithmic frequency bin (i.e \mbox{$h_{\rm char} = f \tilde{h} \propto \mathcal{M}^{5/6} f^{-1/6}$}) is an increasing function of the total binary mass and is roughly independent of frequency for frequencies lower than the ISCO $f_{\rm ISCO} \propto M_{{\rm tot},z}^{-1}$. The detector noise  per logarithmic frequency bin increases for lower frequencies. The combination of these effects is to increase/decrease the $\mathrm{S/N} _\mathrm{tot}$ for total masses lower/higher than roughly $50 \Msun$ for low-eccentricity or circular binaries. The total binary mass at which the $\mathrm{S/N} _\mathrm{tot}$ peaks $M_{\rm tot,z,peak}$ depends on $\rho_{\mathrm{p}0}$; see Figure $11$ in \citet{OLearyetal2009} for an example. Additionally, $\mathrm{S/N} _\mathrm{tot} \propto \eta^{1/2}$ at fixed $M_{{\rm tot},z}$, which implies that the slow parameter errors are proportional to $\eta^{-1/2}$ so that nearly equal-mass binaries are localized better than unequal-mass binaries. There are small differences with respect to the $( \mathrm{S/N} _\mathrm{tot} )^{-1}$ scaling for $\Delta \ln D_L$, $(a_N, b_N)$, and $(a_L, b_L)$, respectively, as the $\mathrm{S/N} _\mathrm{tot}$ is the scalar product of the waveform with itself, while $\ln D_L$, $(\theta_N, \phi_N)$, and $(\theta_L, \phi_L)$ are set by the inverse of the Fisher matrix, which have elements given by the scalar product of the parameter  derivatives of the waveform. The nonzero correlation between these slow parameters and the masses introduces small differences in the mass dependences of the inverse matrix elements (see Sections $4$ and $5$ in \citealt{Gondanetal2018}). 
 
 The measurement errors of fast parameters (e.g., $\ln \mathcal{M}_z$ and $e_{\rm LSO}$) and $(e_0, e_{10\,\rm Hz})$ increase with the binary mass because a higher amount of GW phase accumulates in the detector band for lower binary mass. Indeed, the frequency-dependent GW phase is proportional to $\mathcal{M}^{-5/3}$ at leading order \citep{CutlerFlanagan1994}, and the effect of pericenter precession does not modify this dependence significantly.

 We compare our results with a previous study of \citet{AjithBose2009} for the $M_{\rm tot}$ dependence of $\Delta (\ln D_L)$ and $\Delta (\ln \mathcal{M}_z)$ (i.e., for the mutual parameters) for fixed $D_L$ in the circular limit. We find that $\Delta (\ln D_L)$ has one global minimum for a certain $M_{\rm tot}$ value \footnote{We find this global minimum to be between $50 \, \Msun$ and $60 \, \Msun$ for the aLIGO--AdV--KAGRA detector network at design sensitivity.} and $\Delta (\ln \mathcal{M}_z)$ systematically increases with $M_{\rm tot}$ (Figures \ref{fig:ParamEst_BHNS_Circular} and \ref{fig:ParamEst_BHBH_Circular}). These findings are qualitatively in agreement with results presented for the considered waveform model in the inspiral phase in Figures 2--4 in \citet{AjithBose2009}.\footnote{Note that only a qualitative comparison is possible because \citet{AjithBose2009} considered different detectors with noise curves and applied different waveform models.}

\begin{figure*}
    \centering
    \includegraphics[width=80mm]{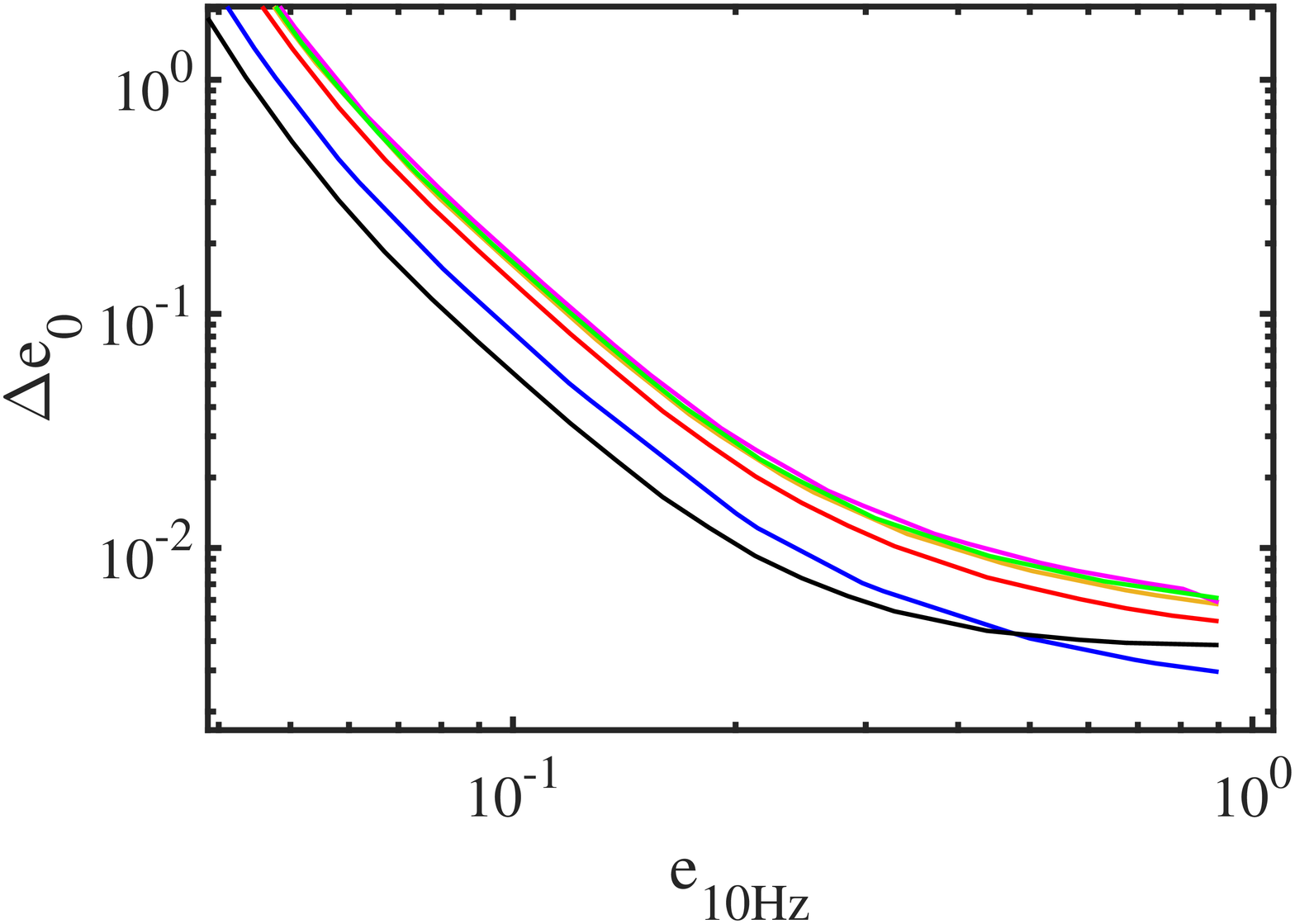}
    \includegraphics[width=80mm]{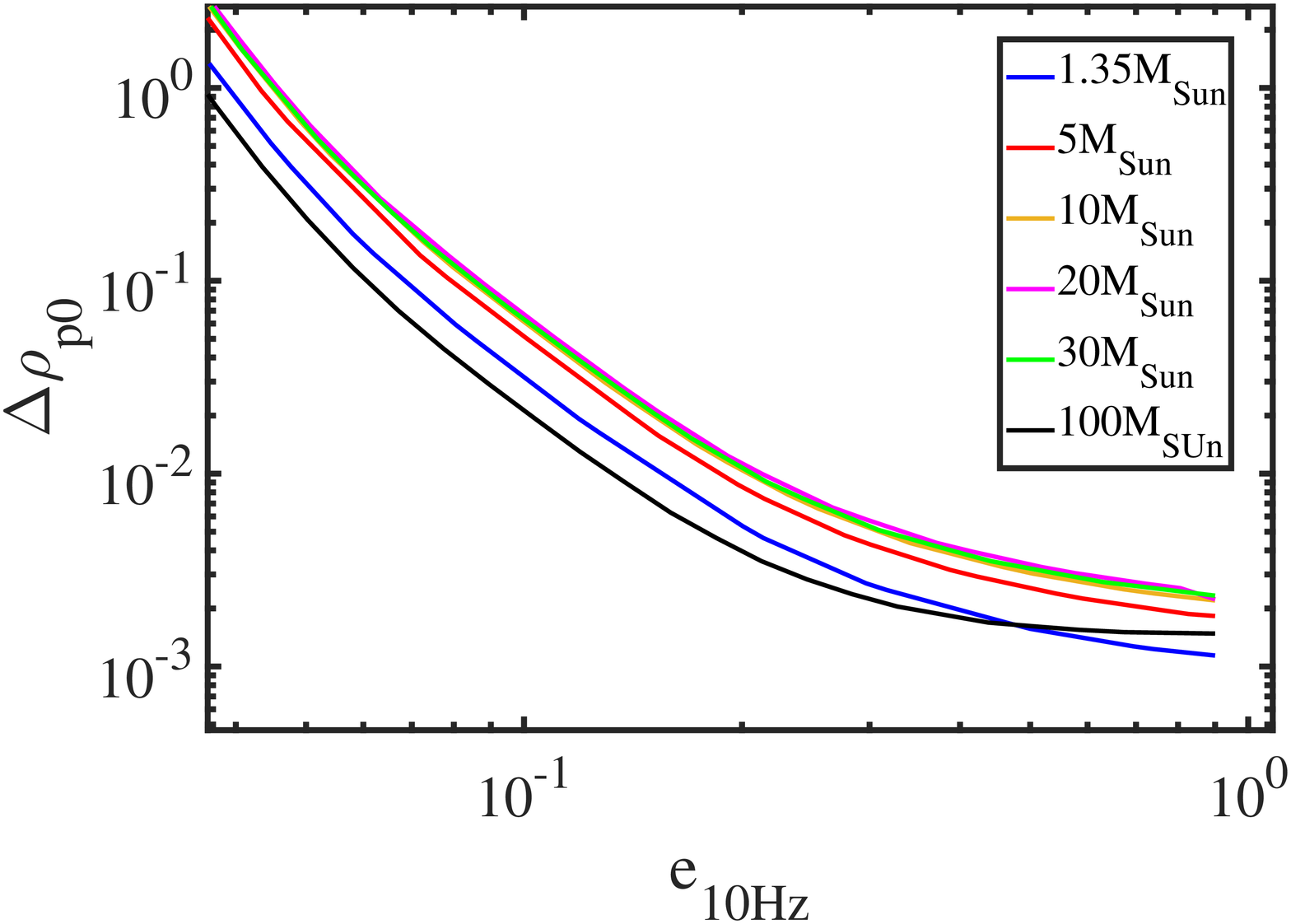}
\\
    \includegraphics[width=80mm]{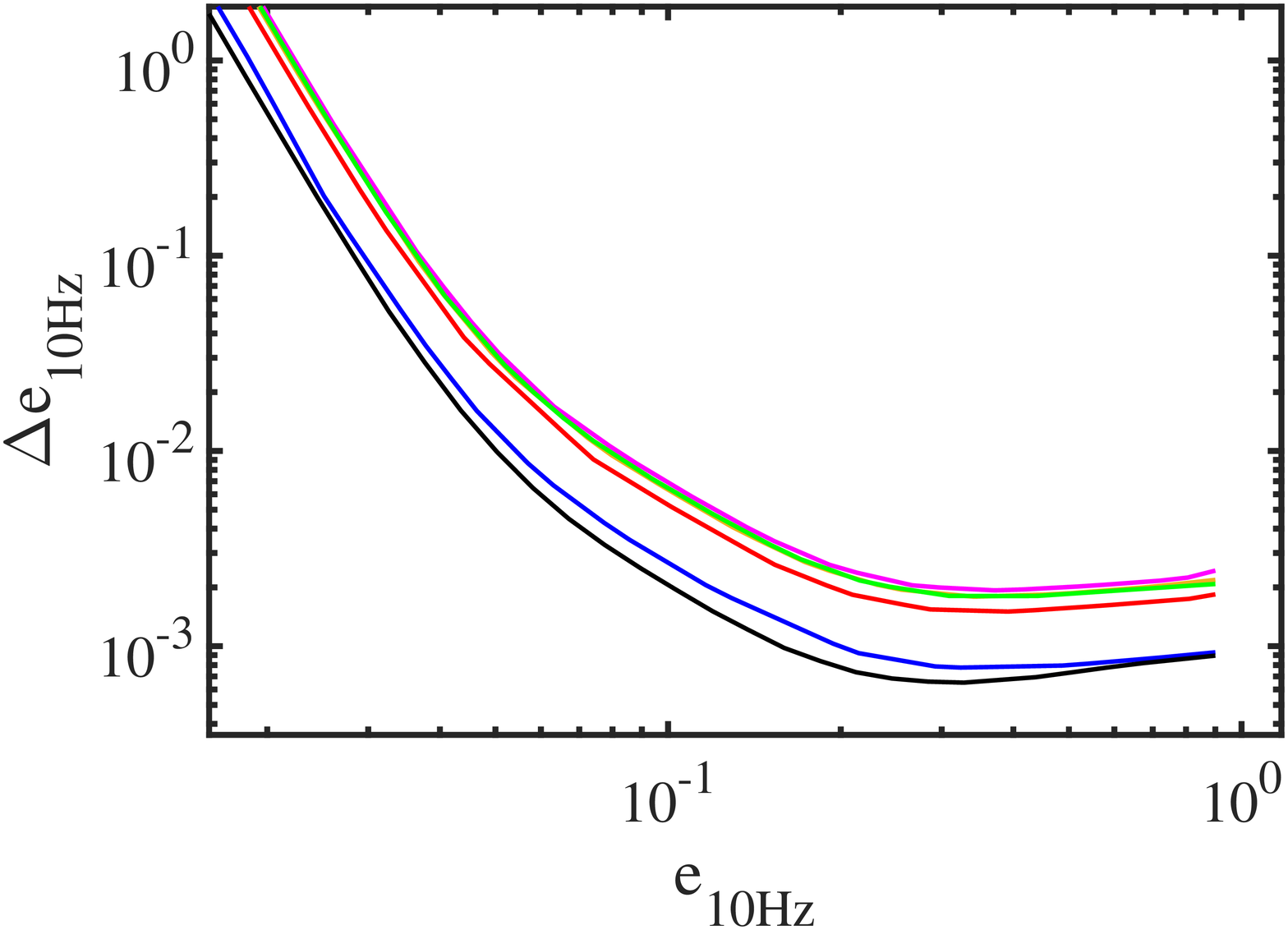}
    \includegraphics[width=80mm]{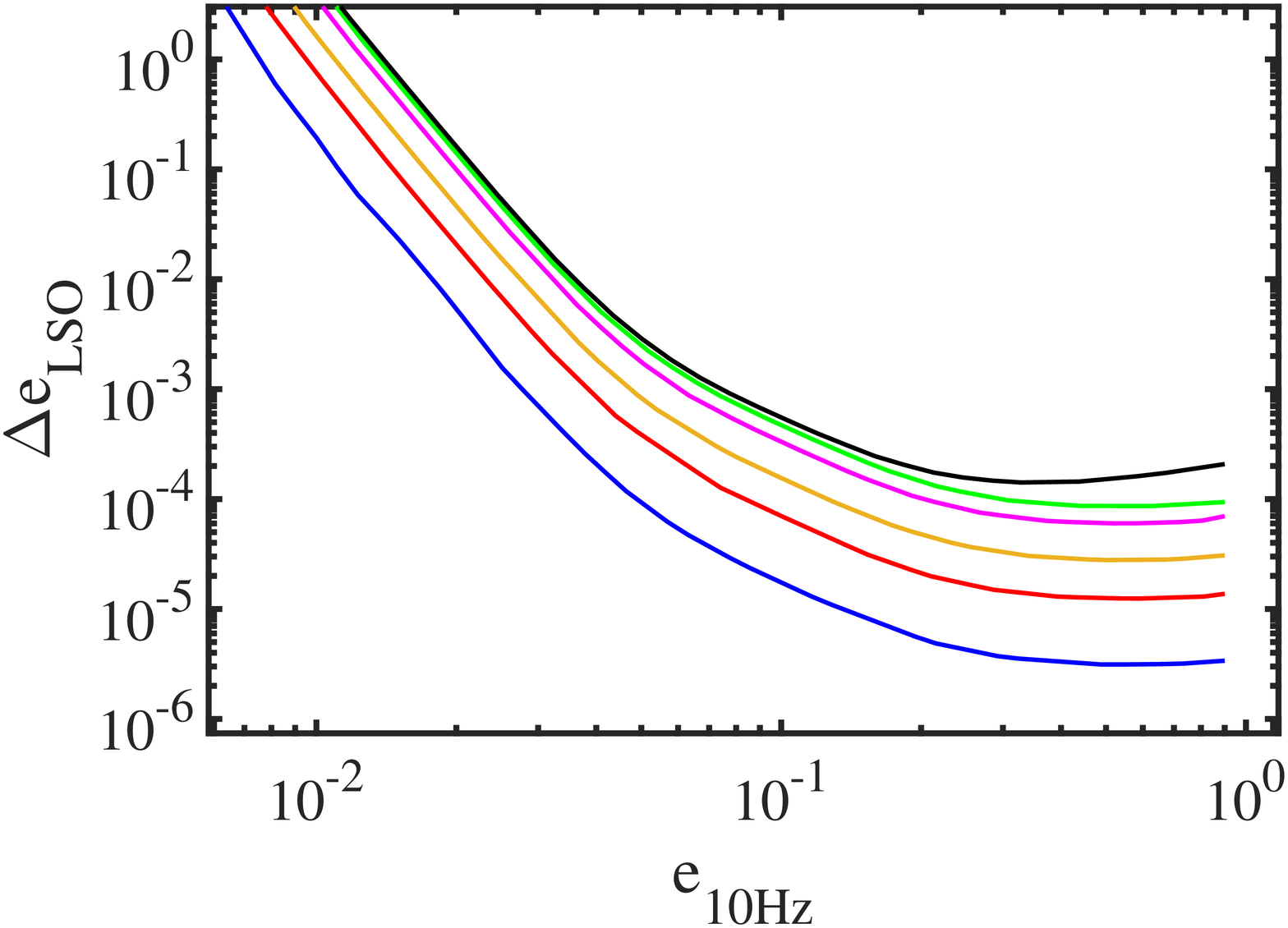}
\\    
    \includegraphics[width=80mm]{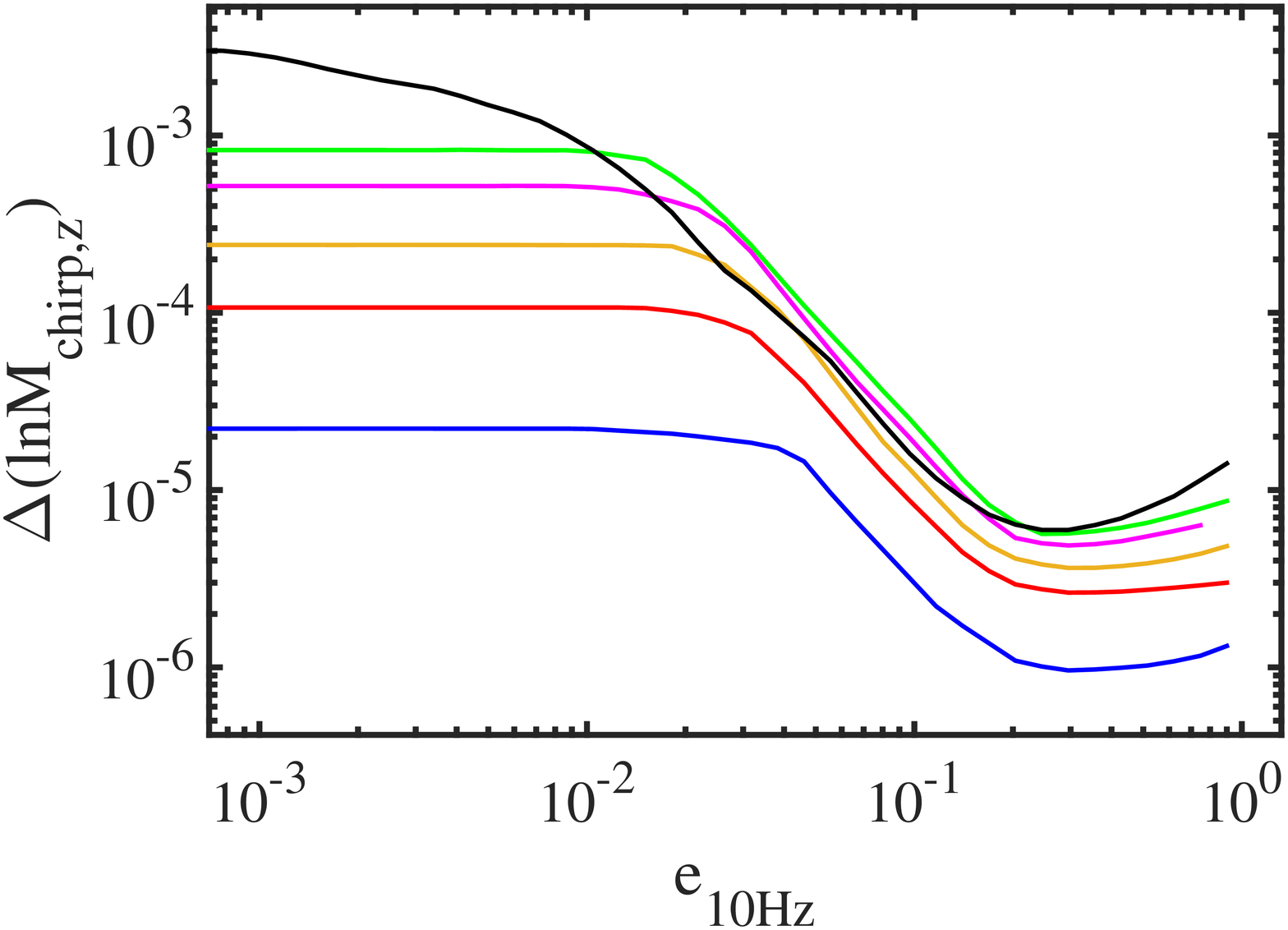}
    \includegraphics[width=80mm]{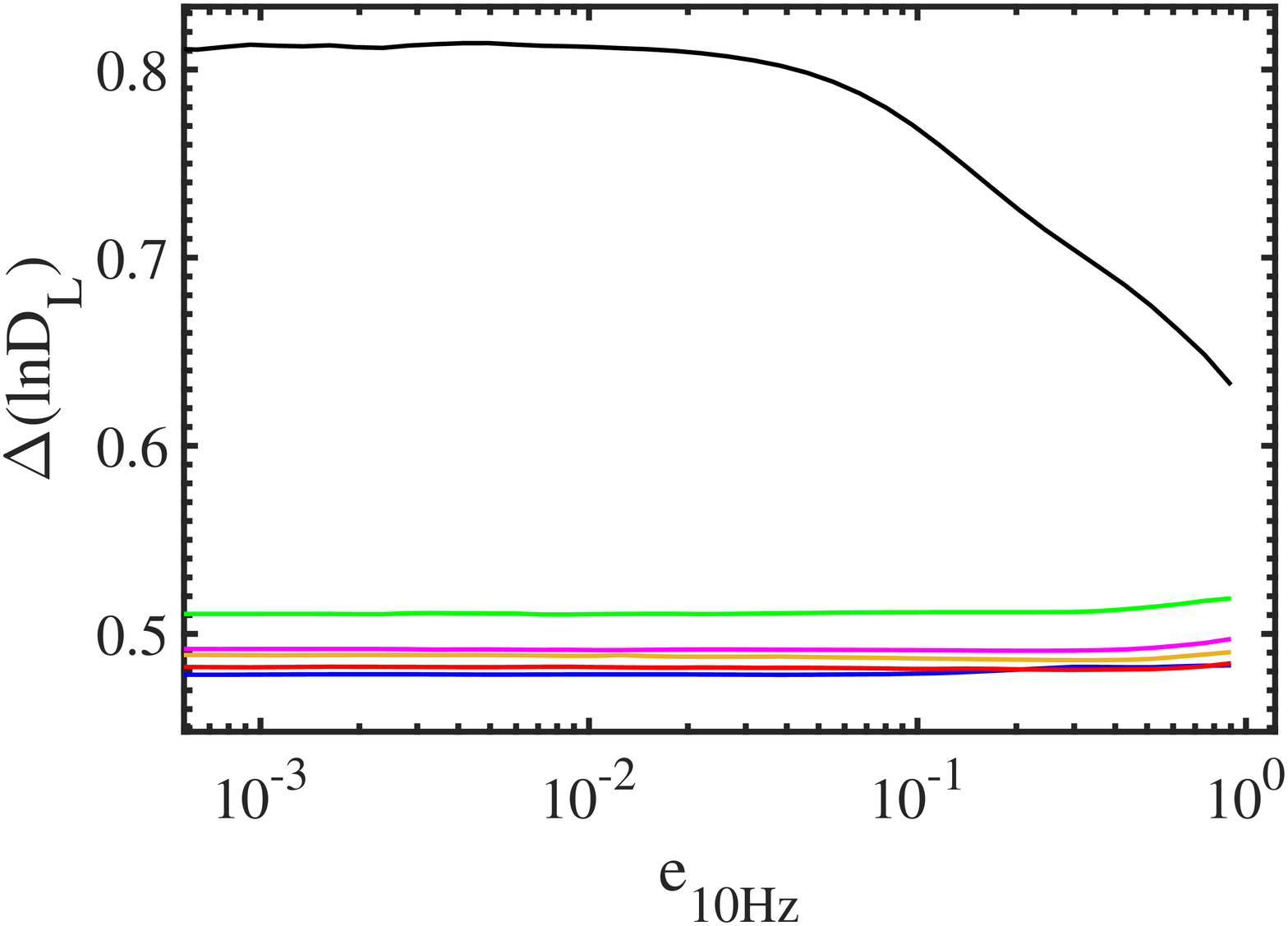}  %
\\
    \includegraphics[width=80mm]{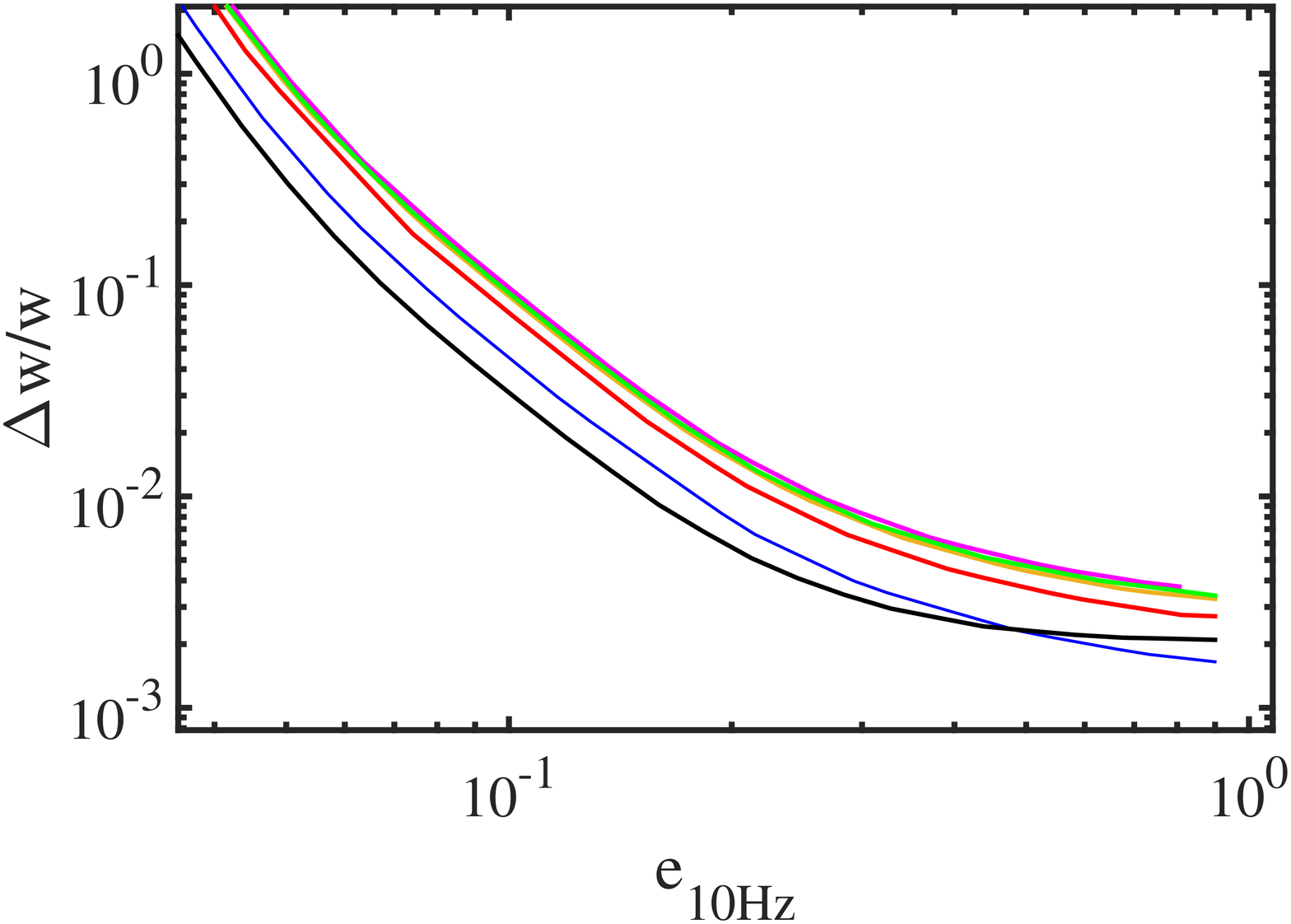} 
  \caption{Measurement errors of various source parameters as a function of $e_{\rm 10Hz}$ for precessing eccentric \mbox{NS--NS} and \mbox{NS--BH} binaries with different companion masses as labeled in the legend, showing binaries that form outside of the detectors' frequency band. Results are displayed for a reference $\mathrm{S/N}_\mathrm{tot}$ of $20$, but can be scaled to other $\mathrm{S/N}_\mathrm{tot}$ values by multiplying the displayed results by a factor of $20 \times (\mathrm{S/N} _\mathrm{tot})^{-1}$. Parameters of the detector network and all other binary parameters are fixed as in Figure \ref{fig:SNRtot_rho}. First row left:  Initial eccentricity $\Delta e_0$. First row right: Initial dimensionless pericenter distance $\Delta \rho_{\mathrm{p}0}$. Second row left: Eccentricity at a GW frequency of $10 \, {\rm Hz}$ $\Delta e_{\rm 10Hz}$. Second row right: Eccentricity at the LSO $\Delta e_{LSO}$. Third row left: Redshifted chirp mass, $\Delta \mathcal{M}_z / \mathcal{M}_z= \Delta (\mathrm{ln} \mathcal{M}_z)$. Third row right: Luminosity distance $\Delta D_L / D_L = \Delta (\mathrm{ln} D_L)$. Fourth row: Characteristic relative velocity $\Delta w / w$. We find similar trends with $e_{\rm 10Hz}$ for other random choices of binary direction and orientation (not shown).  }  \label{fig:ParamEst_BHNS_SNRtot20}
\end{figure*}

\begin{figure*}
    \centering
    \includegraphics[width=80mm]{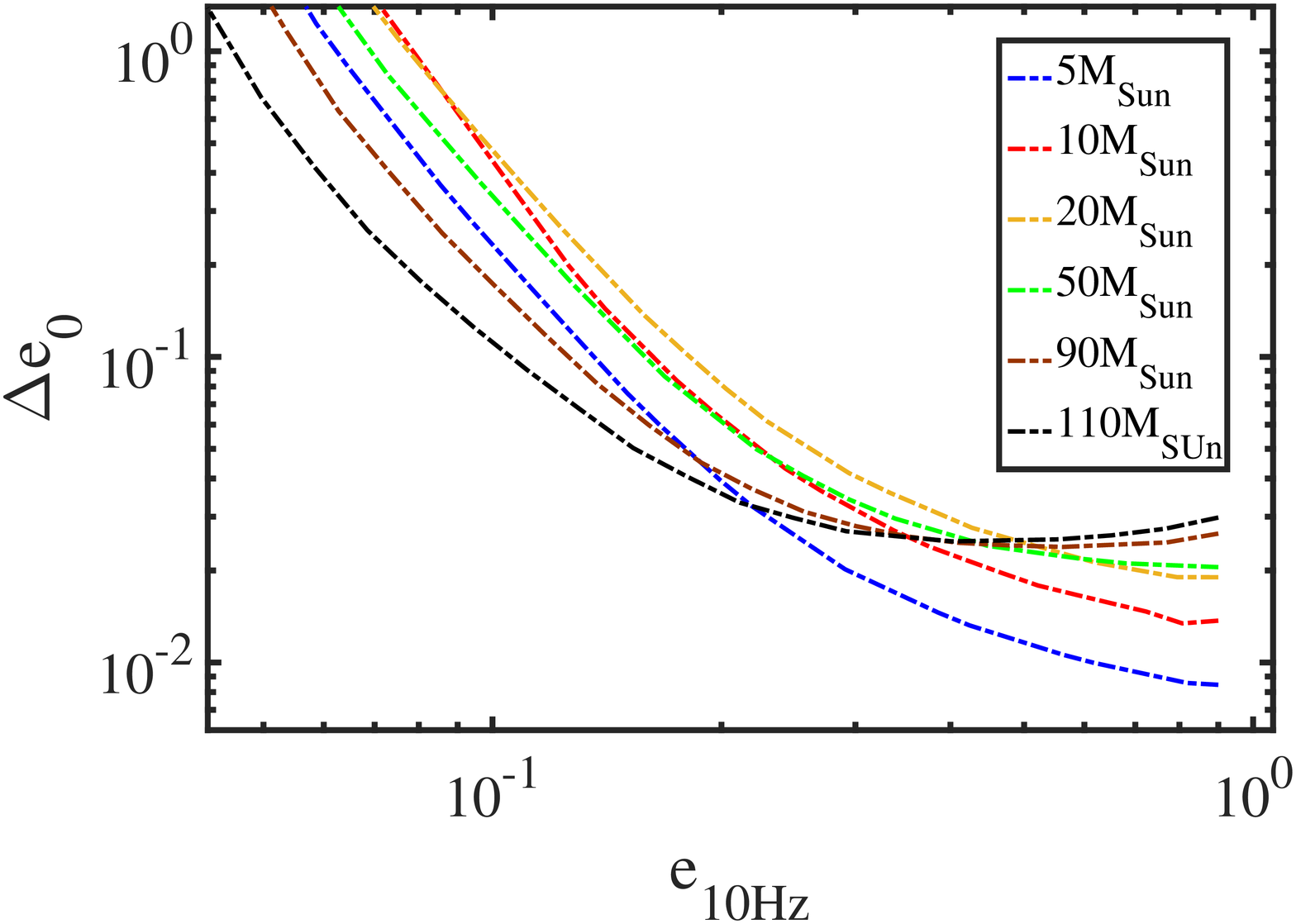}
    \includegraphics[width=80mm]{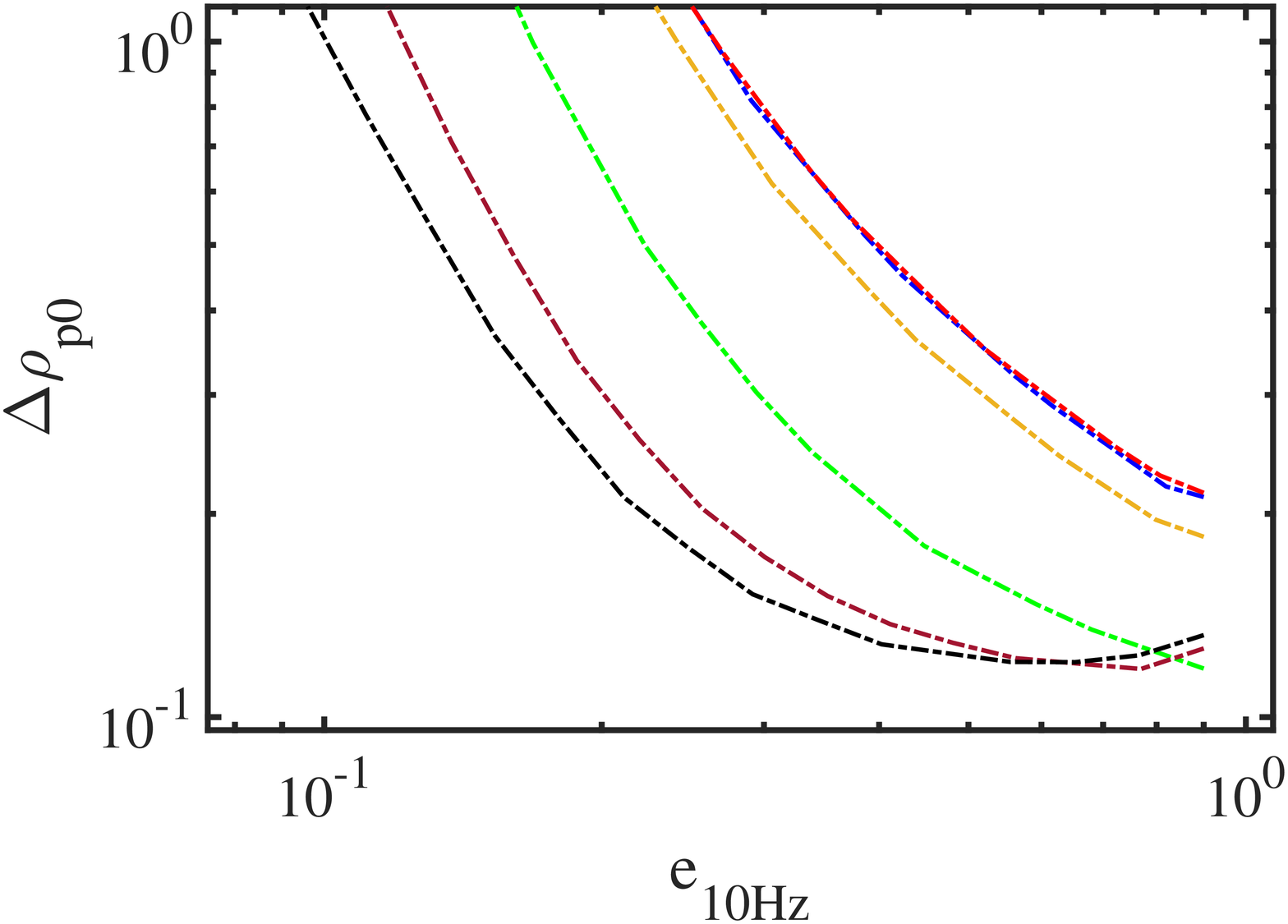}
\\
    \includegraphics[width=80mm]{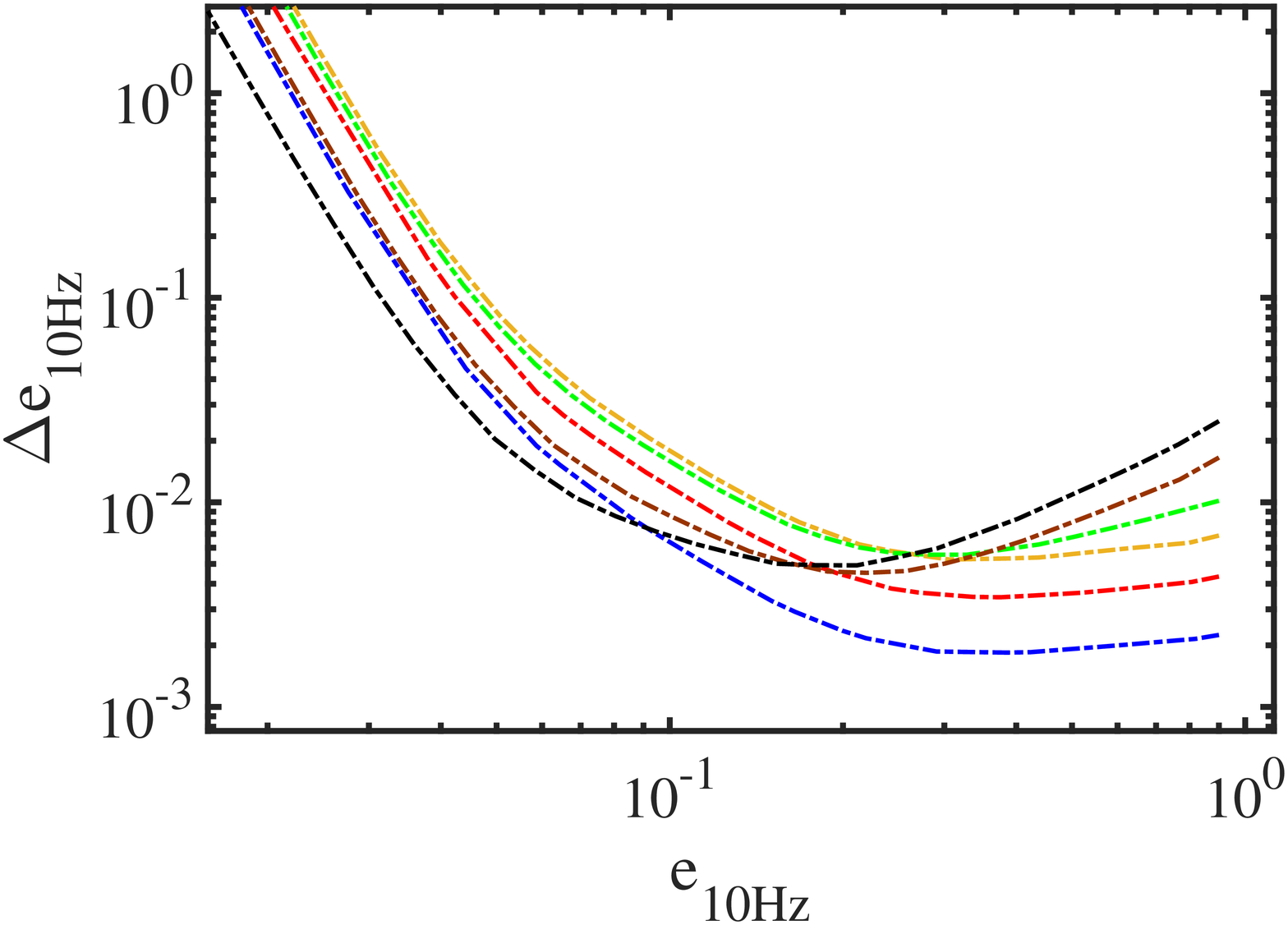}
    \includegraphics[width=80mm]{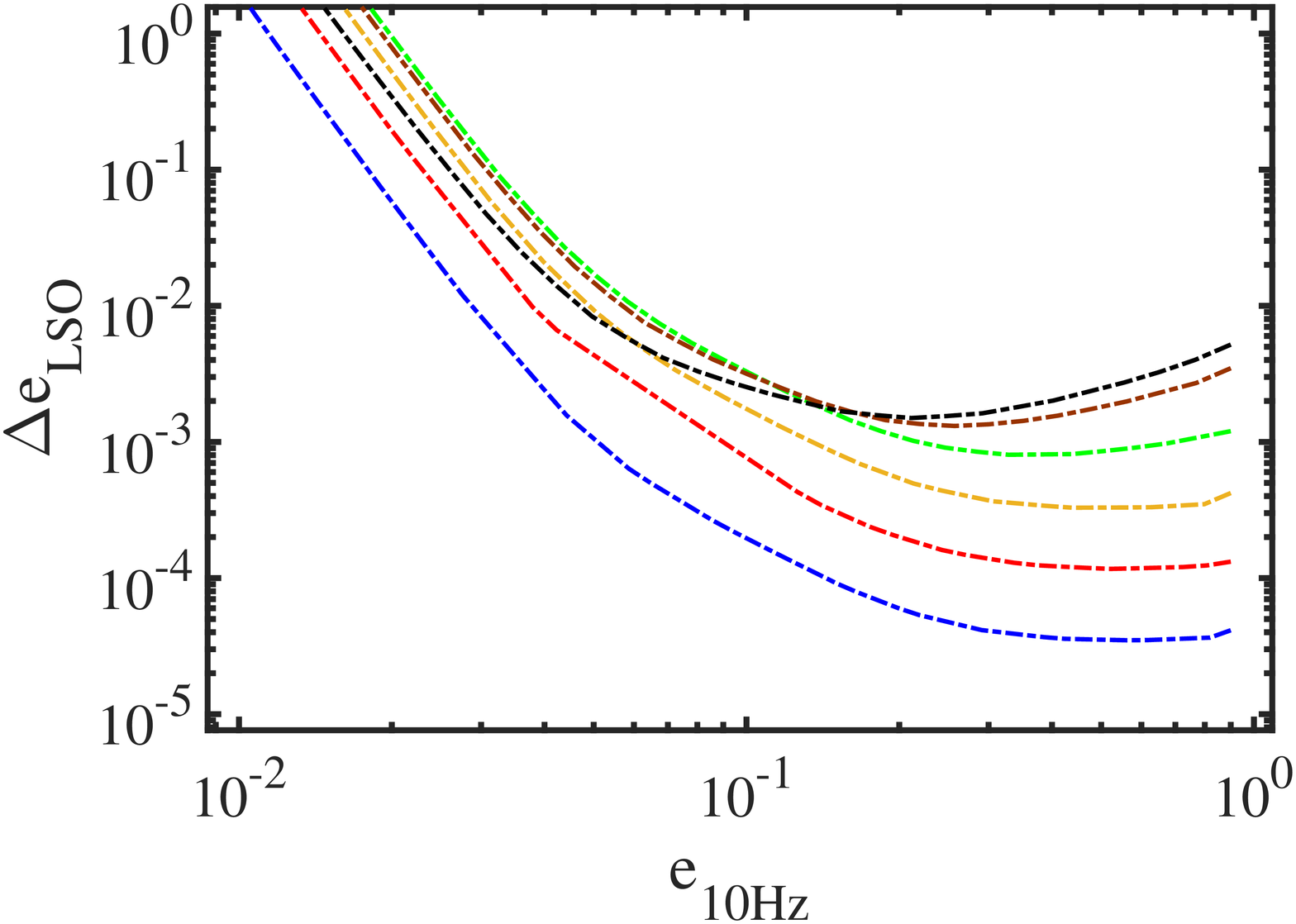}
\\    
    \includegraphics[width=80mm]{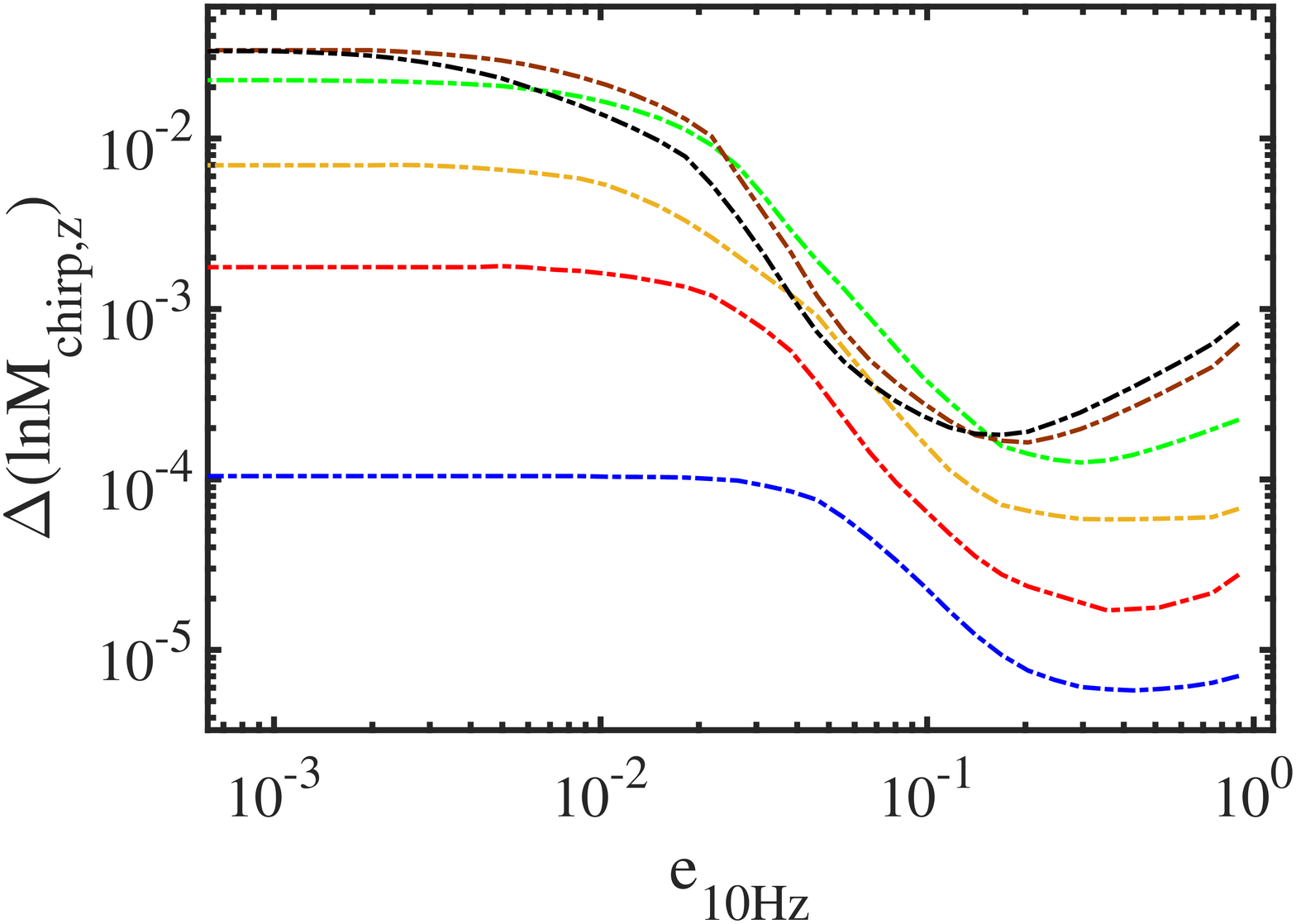}
    \includegraphics[width=80mm]{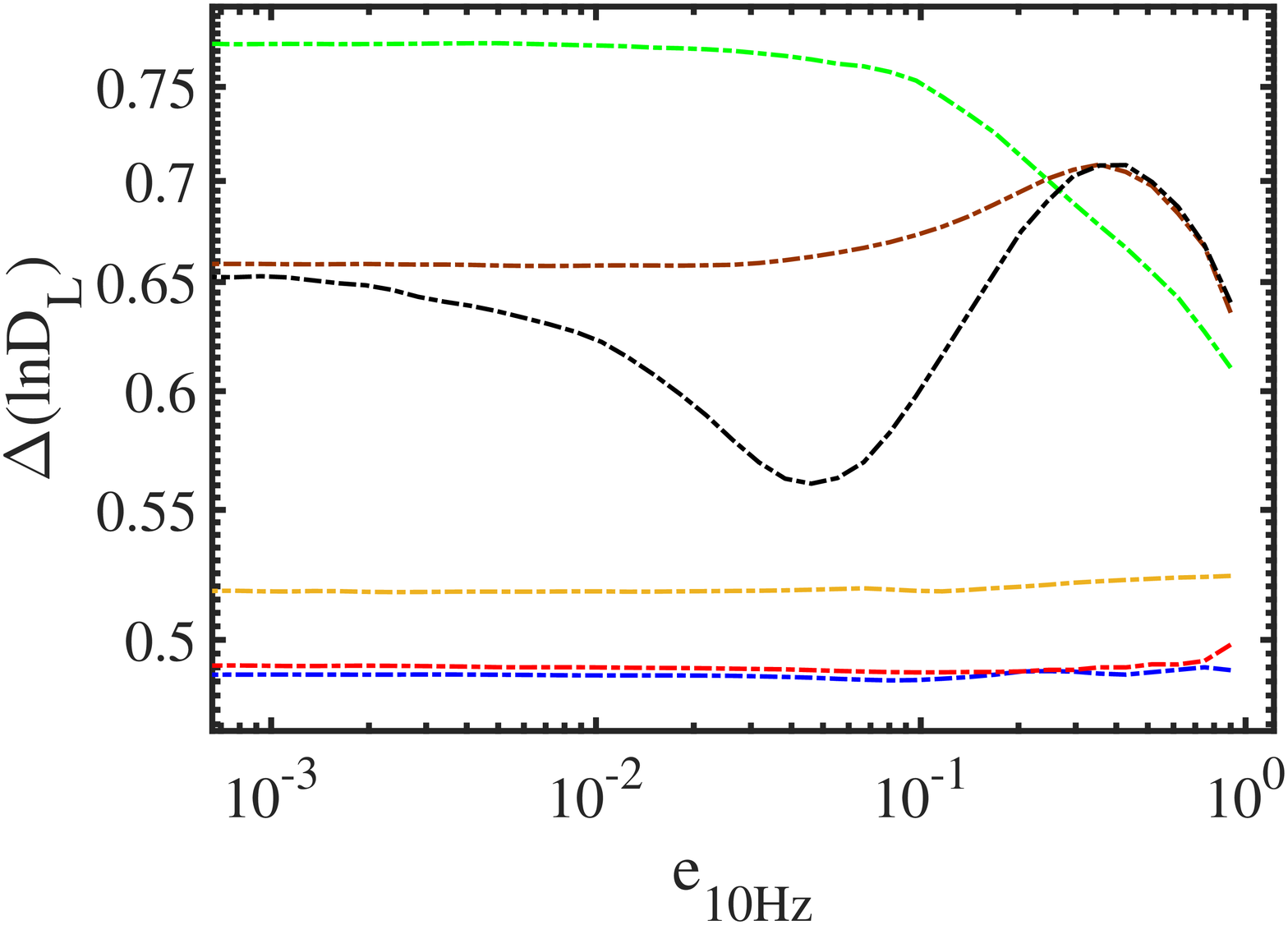}      
\\
    \includegraphics[width=80mm]{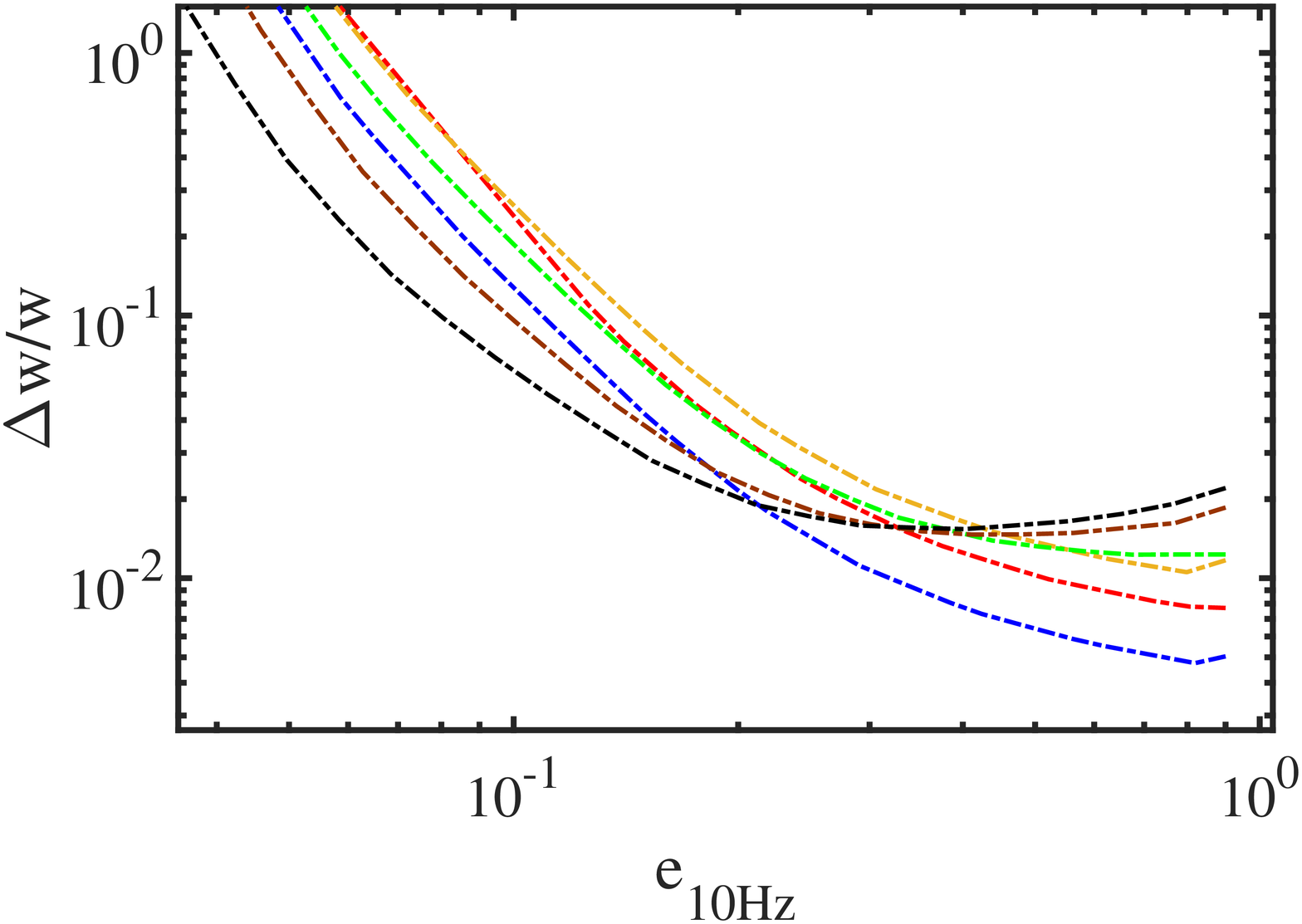} 
  \caption{Same as Figure \ref{fig:ParamEst_BHNS_SNRtot20}, but for precessing, eccentric, equal-mass \mbox{BH--BH} binaries of different masses, showing binaries that form outside of the detectors' frequency band. }  \label{fig:ParamEst_BHBH_SNRtot20}
\end{figure*}

\subsection{Measurement Errors of Binary Parameters as a Function of $e_{10\,\rm Hz}$ for Fixed $\mathrm{S/N} _\mathrm{tot}$} 
\label{subsec:MeasErrFixSNR}

 In Figures \ref{fig:ParamEst_BHNS_SNRtot20} and \ref{fig:ParamEst_BHBH_SNRtot20}, we scale the $e_{10\,\rm Hz}$ dependent measurement errors of binary parameters $(e_0, \rho_{{\rm p}0}, \ln \mathcal{M}_z, \ln D_L, \ln w, e_{\rm 10 Hz}, e_{\rm LSO})$ to \mbox{$\mathrm{S/N} _\mathrm{tot} = 20$}, and show results for binaries that form outside of the detectors' frequency band. Fixing the \mbox{$\mathrm{S/N}_\mathrm{tot}$} eliminates the observational selection bias from the detection likelihood. For a homogeneous isotropic source population in Eucledian space, the number of events with a given \mbox{$\mathrm{S/N}_\mathrm{tot}$} value is approximately proportional to $\mbox{$(\mathrm{S/N}_\mathrm{tot})^{-4}$}$. Thus, the parameter dependence of the measurement precision shown in Figures \ref{fig:ParamEst_BHNS_SNRtot20} and \ref{fig:ParamEst_BHBH_SNRtot20} for fixed \mbox{$\mathrm{S/N}_\mathrm{tot}$} represents the relative number of sources with different measurement accuracies in a magnitude-limited sample.
 
 To obtain these figures, note that in Appendix \ref{sec:DL_Dep} we show that measurement errors and $\mathrm{S/N} _\mathrm{tot}$ scale roughly as $\propto D_L$ and $\propto 1/ D_L$, respectively, implying that $\Delta \lambda \propto (\mathrm{S/N} _\mathrm{tot}) ^{-1}$. Thus, the measurement errors can be given for $\mathrm{S/N} _\mathrm{tot} = 20$ by the product $\Delta \lambda (e_{\rm 10 Hz}) \times [\mathrm{S/N} _\mathrm{tot} (e_{\rm 10 Hz}) / 20]$, where both $\Delta \lambda (e_{\rm 10 Hz})$ and $\mathrm{S/N} _\mathrm{tot}$ are calculated for $D_L = 100 \, \Mpc$. Further, we note that measurement errors can be scaled to other $\mathrm{S/N} _\mathrm{tot}$ values by multiplying the displayed results by a factor of $20 \times (\mathrm{S/N} _\mathrm{tot})^{-1}$. Results for an $\mathrm{S/N} _\mathrm{tot}$ of $20$ are displayed in Figures \ref{fig:ParamEst_BHNS_SNRtot20} and \ref{fig:ParamEst_BHBH_SNRtot20}. 

 For fixed $\mathrm{S/N}_\mathrm{tot}$, the mass and $e_{\rm 10 Hz}$ dependent trends of measurement errors vary relative to that for fixed $D_L$ as follows: For precessing eccentric \mbox{NS--BH} and \mbox{BH--BH} binaries with BH masses $\lesssim 30 \Msun$, the $\mathrm{S/N} _\mathrm{tot}$ is nearly constant (increases weakly) as a function of $e_{\rm 10 Hz}$ (Figure \ref{fig:SNRtot_rho}), consequently, the $e_{\rm 10 Hz}$ dependent trends of measurement errors are similar for fixed $\mathrm{S/N} _\mathrm{tot}$ or for fixed  $D_L$. However, for precessing eccentric \mbox{NS--BH} and \mbox{BH--BH} binaries with BH masses $\lesssim 30 \Msun$, these trends  change significantly depending on the type of binary and the parameter; see errors for $\ln D_L$ for \mbox{BH--BH} binaries in Figure \ref{fig:ParamEst_BHBH_SNRtot20} as an example.
 
 We identify the following trends. Note that many of these statements are nearly identical to those made earlier for fixed $D_{\mathrm{L}}$.
 \begin{itemize}
 
 \item The measurement error of $e_{10\,\rm Hz}$ is in the range $\Delta e_{10\,\rm Hz} \sim (10^{-3}-10^{-2}) \times 20 \times (\mathrm{S/N} _\mathrm{tot})^{-1}$ for precessing eccentric \mbox{NS--NS}, \mbox{NS--BH}, and \mbox{BH--BH} binaries if $e_{10\,\rm Hz} > 0.1$. 
 
 \item $\Delta e_{\rm LSO}$ is at the level of $\Delta e_{\rm LSO} \sim (10^{-3}-10^{-5}) \times 20 \times (\mathrm{S/N} _\mathrm{tot})^{-1}$ for precessing eccentric \mbox{NS--BH} and \mbox{BH--BH} binaries if \mbox{$e_{10\,\rm Hz} > 0.1$}, while measurement errors for precessing eccentric \mbox{NS--NS} binaries are in the range $\Delta e_{\rm LSO} \sim (10^{-6}-10^{-5}) \times 20 \times (\mathrm{S/N} _\mathrm{tot})^{-1}$. 
 
 \item The measurement error of $e_0$ is at the level of $\Delta e_0 \sim (10^{-3} - 10^{-1}) \times 20 \times (\mathrm{S/N} _\mathrm{tot})^{-1}$ for precessing eccentric \mbox{NS--NS}, \mbox{NS--BH}, and \mbox{BH--BH} binaries if $e_{10\,\rm Hz} \gtrsim 0.1$, where errors systematically decrease with $e_{10\,\rm Hz}$ for \mbox{NS--NS} and \mbox{NS--BH} binaries and for \mbox{BH--BH} binaries with BH masses $\lesssim 30 \Msun$. 
 
 \item $\rho_{{\rm p}0}$ can be measured at the level of $\Delta \rho_{{\rm p}0} \sim (10^{-3} - 10^{-1}) \times 20 \times (\mathrm{S/N} _\mathrm{tot})^{-1}$ for precessing eccentric \mbox{NS--NS} and \mbox{NS--BH} binaries if $e_{10\,\rm Hz} \gtrsim 0.1$, where the errors decrease with $e_{10\,\rm Hz}$. $\Delta \rho_{{\rm p}0}$ is in the range $10^{-1} \times 20 \times (\mathrm{S/N} _\mathrm{tot})^{-1}$ and $(10^{-1} - 10) \times 20 \times (\mathrm{S/N} _\mathrm{tot})^{-1} $ for precessing eccentric \mbox{BH--BH} binaries with BH masses \mbox{$\lesssim 90 \Msun$} and \mbox{$90 \Msun \lesssim$}, respectively.
 
 \item In comparision to circular binaries, the measurement error of source distance, sky location, and binary orientation slightly improves/deteriorates for precessing eccentric \mbox{NS--NS}, \mbox{NS--BH}, and \mbox{BH--BH} binaries.
 
 \item The measurement accuracy of $\mathcal{M}_z$ can improve by up to a factor of $\sim 20$ and \mbox{$\sim 30-100$} for precessing eccentric \mbox{NS--NS} and \mbox{NS--BH} binaries with BH companion masses \mbox{$\lesssim 40 \, \Msun$}, in particular, compared to similar binaries in the circular limit. Even higher precision may be achieved for BH companion masses above $\sim 40 \, \Msun$, in particular a factor of $\sim 200$ may be achieved for \mbox{NS--BH} binaries with BH masses of $\sim 100 \, \Msun$. In comparison to circular binaries, the measurement accuracy improves by up to a factor of $\sim 10-10^2$ and $100 - 150$ for precessing eccentric \mbox{BH--BH} binaries with BH masses below and above $\sim 40 \, \Msun$, respectively.
 
 \item The measurement accuracy of the characteristic relative velocity is in the range $\Delta w /w \sim (10^{-3} - 10^{-1}) \times 20 \times (\mathrm{S/N} _\mathrm{tot})^{-1}$ for precessing eccentric \mbox{NS--NS}, \mbox{NS--BH}, and \mbox{BH--BH} binaries if $e_{10\,\rm Hz} \gtrsim 0.1$, where higher precision corresponds to higher $e_{10\,\rm Hz}$ values. Further, errors systematically decrease with $e_{10\,\rm Hz}$ for \mbox{NS--NS} and \mbox{NS--BH} binaries and for \mbox{BH--BH} binaries with BH masses $\lesssim 30 \Msun$. 
 
 \end{itemize}
 
 Finally, we compare our results with those of \citet{AjithBose2009} for the $M_{\rm tot}$ dependence of $\Delta (\ln D_L)$ and $\Delta (\ln \mathcal{M}_z)$ (i.e., for the mutual parameters) for fixed $\mathrm{S/N} _\mathrm{tot}$ in the circular limit. We find for \mbox{NS--NS} and \mbox{NS--BH} binaries that $\Delta (\ln D_L)$ slightly increases below a certain $M_{\rm tot}$ value, but rapidly increases above it and $\Delta (\ln \mathcal{M}_z)$ systematically increases with $M_{\rm tot}$ (Figures \ref{fig:ParamEst_BHNS_SNRtot20} and \ref{fig:ParamEst_BHBH_SNRtot20}), which qualitatively agrees with results presented for the considered waveform model in the inspiral phase in Figures 6-8 in \citet{AjithBose2009}. We find similar trends for \mbox{BH--BH} binaries with $M_{\rm tot} \lesssim 130 \Msun$, but these trends break for $ M_{\rm tot} \gtrsim 130 \Msun$ because the $M_{\rm tot}$ dependence of the $\mathrm{S/N} _\mathrm{tot}$  decreases more rapidly above $M_{\rm tot} \sim 130 \Msun$ for the considered waveform model (Section \ref{subsec:TimeDomWaveform}) in the circular limit than in the inspiral phase in \citet{AjithBose2009}.

\section{Discussion and Conclusion}
\label{sec:DiscAndConcl}
 
 In this paper, we have carried out a Fisher matrix type analysis to estimate the expected measurement errors of the physical parameters characterizing non-spinning precessing eccentric binary inspirals with the aLIGO--AdV--KAGRA GW detector network. We carried out calculations for compact object masses that correspond to precessing eccentric \mbox{NS--NS}, \mbox{NS--BH}, and \mbox{BH--BH} binaries with BH masses in the range \mbox{$5\,\Msun-110\,\Msun$} and a wide range of dimensionless initial pericenter distances $\rho_{{\rm p}0}$ to cover binaries that form in the advanced GW detectors' frequency band with high eccentricities to those that enter it at $10 \, \Hz$ with eccentricities between \mbox{$8 \times 10^{-4} \leqslant e_{10\,\rm Hz} \leqslant 0.9$}. 
 
 The results show that the source distance, sky location, and binary orientation parameter measurement precision (the so-called slow parameters; \citealt{Kocsisetal2007}) are roughly independent of $e_{10\,\rm Hz}$ for precessing eccentric \mbox{NS--NS} and \mbox{NS--BH} binaries unless the BH companion mass is higher than $\sim 50\, \Msun$. The distance, source sky location, and binary orientation measurement precision improve significantly by a factor of $2-20$ for $e_{10\,\rm Hz} \gtrsim 0.1$ for precessing eccentric \mbox{NS--BH} and \mbox{BH--BH} binaries with BH masses in the range $50 \, \Msun - 110 \, \Msun$ compared to similar binaries in the circular limit, where higher accuracies correspond to higher $e_{10\,\rm Hz}$ values and binary masses. The measurement precision further improves to $2-50$ when these binaries form in the advanced GW detectors' frequency band.
 
 The measurement accuracy of the characteristic relative velocity is in the range $\Delta w /w \sim (10^{-3} - 10^{-1}) \times (D_L / 100\,\rm Mpc)$ for $0.04 \lesssim e_{10\,\rm Hz}$, where lower errors correspond to higher $e_{10\,\rm Hz}$ values, and results weakly depend on the binary mass. The relative velocity is expected to be a proxy for the velocity dispersion of the host environment or it may estimate the initial orbital velocity of the binary in a single-binary resonant scattering encounter.  
 
 The measurement accuracy of other parameters such as $\mathcal{M}_z$, $\rho_{{\rm p}0}$, $e_{0}$, $e_{10\,\rm Hz}$, and $e_{\rm LSO}$ are sensitive to the eccentricity for all component masses. 
 
 $\mathcal{M}_z$ is expected to be measured to a much higher measurement accuracy for more eccentric precessing \mbox{NS--NS}, \mbox{NS--BH}, and \mbox{BH--BH} sources with $e_{10\,\rm Hz} \gtrsim 0.1$ than for similar binaries in the circular limit. The improvement is by a factor of $\sim 10-10^3$, where higher accuracies correspond to higher masses and $e_{10\,\rm Hz}$ values.
 
 The eccentricity measurement errors at a GW frequency of $10 \, {\rm Hz}$ and at the LSO are in the range $\Delta e_{10\,\rm Hz} \sim (10^{-4}-10^{-3}) \times (D_L / 100\,\rm Mpc)$ and $\Delta e_{\rm LSO} \sim (10^{-6}-10^{-3}) \times (D_L / 100\,\rm Mpc)$, respectively, for precessing eccentric \mbox{NS--NS}, \mbox{NS--BH}, and \mbox{BH--BH} binaries if \mbox{$e_{10\,\rm Hz} \gtrsim 0.1$}, where lower errors correspond to lower masses and higher $e_{10\,\rm Hz}$ values. Further, we showed that $e_{10\,\rm Hz}$ can be measured effectively for the GW transients analogous to GW150914, GW151226, GW170104, GW170608, GW170729, GW170809, GW170814, GW170817, GW170818, and GW170823 if $e_{10 \,\rm Hz} \gtrsim \{0.051, \, 0.046, \, 0.058, \, 0.043, \, 0.081, \, 0.063, \, 0.053, \, 0.023,$ $\, 0.062, \, 0.072\}$, respectively.
 
 The initial\footnote{I.e., when GWs start to drive the inspiral of the binary and dominate the evolution, such as after a binary-forming close encounter of single--single objects due to GW losses, or after a binary--single or binary--binary encounter that places the binary on an inspiraling trajectory} eccentricity and pericenter distance can be measured with the aLIGO--AdV--KAGRA detector network either if the binary forms in the detectors' frequency band or if it forms not far outside of it such that $e_{10\,\rm Hz} \gtrsim 0.1$. The errors are in the range \mbox{$\Delta e_0 \sim (10^{-3}-10^{-1}) \times (D_L / 100\,\rm Mpc)$} and \mbox{$\Delta \rho_{{\rm p}0} \sim (0.01-10) \times (D_L / 100\,\rm Mpc)$}, with the smallest errors for binaries that form in the GW detectors' frequency band near the low-frequency cutoff of the detectors or at \mbox{$e_{10 \, \rm Hz} \sim e_0$}, where errors systematically decrease with $e_{10\,\rm Hz}$ and weakly depend on the binary mass.
 
 These results are encouraging as they suggest that the upcoming second-generation GW detector network \citep{Abbottetal2018} at design sensitivity has the capability to distinguish between eccentric waveforms and circular waveforms, implying that the eccentricity may be used to constrain the astrophysical origin of GW source populations. These may include binary--single or binary--binary encounters in dense stellar systems such as in globular clusters or in active galactic nucleus gaseous disks, single--single GW capture encounters, Kozai--Lidov effect in field triples, or Kozai--Lidov effect caused by a supermassive black hole for binaries in galactic nuclei (see Introduction). However, further work is needed to verify if these conclusions hold when including higher post-Newtonian order waveforms or waveforms of numerical relativity simulations, spin effects, perturbations related to tidal deformability, posterior parameter distributions with Monte Carlo Markov chain methods and Bayesian inference, and theoretical errors due to inaccuracies in the waveform model are included. In particular, the errors may be much worse than in our estimates for precessing eccentric \mbox{NS--BH} and \mbox{BH--BH} binaries with high BH masses or in cases where binaries form at small $\rho_{{\rm p}0}$ because in this case, the detected signal starts at a separation where these corrections are already significant. Indeed, note that the PN expansion parameter approximately satisfies $v^2/c^2 \propto \rho_{\rm p}^{-1}$, so that $\rho_{\rm p}\gg 10$ is required for higher-order PN effects to be negligible. Our conclusions may be accurate in cases where these corrections are negligible, when the binary waveform enters the GW detectors' frequency band at relatively high pericenter distance $\rho_{\rm p}$. For instance, for initially highly eccentric \mbox{NS--NS} binaries, the binary forms with $f_{\rm GW} < 10\,\rm Hz$ at $\rho_{{\rm p}0} \gtrsim 150$. Similarly, for initially highly eccentric \mbox{NS--BH} binaries with relatively low BH masses below $\sim 20 \, \Msun$, the initial pericenter is $\rho_{{\rm p}0} \gtrsim 40$ if the binary forms with $f_{\rm GW} < 10 \, \rm Hz$. In these cases, the 1PN results may be expected to be relatively accurate. 
 
 We conclude that further development of GW data analysis algorithms to detect eccentric inspiraling GW sources and to measure their physical parameters may offer high rewards in GW astronomy for interpreting the astrophysical origin of GW sources.

\acknowledgments
 We thank the anonymous referee for constructive comments that helped improve the quality of the paper. This work received funding from the European Research Council (ERC) under the European Union's Horizon 2020 Programme for Research and Innovation ERC-2014-STG under grant agreement No. 638435 (GalNUC), and from the Hungarian National Research, Development, and Innovation Office under grant NKFIH KH-125675.

\appendix

\section{Residual Eccentricity as a Function of Waveform Parameters}
\label{sec:ParamestErrore10Hz}

 In this section, we express $e_{\rm 10 Hz}$ as a function of $e_{\rm LSO}$ and $M_{\rm tot}$. Based on \citet{Peters1964}, we showed in \citet{Gondanetal2018} that the product $\nu (e) H(e)$ is conserved during the evolution of the binary, where $H(e)$ is defined as
\begin{equation}  \label{eq:He}
  H(e) = e^{18/19}(1-e^2)^{-3/2} \left(1+\frac{121}{304} e^2 \right)^{\frac{1305}{2299}} \, ,
\end{equation}
 and $\nu(e)$ is the redshifted Keplerian mean orbital frequency:
\begin{equation}  \label{eq:nuKepler} 
 \nu(e,\rho_\mathrm{p}) = \frac{ (1-e)^{3/2} }{2 \pi M_\mathrm{tot} \rho_\mathrm{p}^{3/2}(e) } \, . 
\end{equation}
 Here, we use the leading-order orbital evolution equation \citep{Peters1964}
\begin{equation}  \label{eq:rhoe}
   \rho_\mathrm{p} (e) = \frac{ c_0 }{ M_{\mathrm{tot}} } \frac{ e^{12/19}}{(1+e)}
   \left( 1+ \frac{ 121 }{ 304 } e^2 \right)^{ \frac{870}{2299} } \, ,
\end{equation}
 where $c_0/M_{\mathrm{tot}}$ may be expressed with $e_0$ and $\rho_\mathrm{p0}$ by solving Equation (\ref{eq:rhoe}) for $\rho_\mathrm{p} = \rho_\mathrm{p0}$ and $e=e_0$. Here, \mbox{$\nu (e) H(e) \equiv {\rm const}$} for $e_\mathrm{LSO} \leqslant e \leqslant e_0$. Thus, $e_{\rm 10 Hz}$ may be expressed with either $e_0$ or $e_\mathrm{LSO}$. 

 Here, we write out the expressions for $e_\mathrm{LSO}$. By setting 
\begin{equation}  \label{eq:e10HzEQUIVeLSO}
 \nu (e_{\rm 10 Hz}) H(e_{\rm 10 Hz}) = \nu (e_\mathrm{LSO}) H(e_\mathrm{LSO}) \, 
\end{equation} 
 and substituting $\nu (e)$ with Equation (\ref{eq:nuKepler}) in Equation (\ref{eq:e10HzEQUIVeLSO}). This leads to the following implicit equation for $e_{\rm 10 Hz}$:
\begin{equation}  \label{eq:Implicit_First}
  \frac{ (1-e_{\rm 10 Hz})^{3/2} H( e_{\rm 10 Hz} ) }{ \rho_{\rm 10 Hz}^{3/2} } = \frac{ (1-e_\mathrm{LSO} )^{3/2} H( e_\mathrm{LSO} ) }{ \rho_\mathrm{p,LSO}^{3/2} } \, .
\end{equation} 
 Here, $\rho_{\rm 10 Hz}$ is defined by rewriting Equation (\ref{eq:rhoDet_Gen}) at $e=e_{\rm 10 Hz}$ as
\begin{equation}  \label{eq:rho10Hz}
  \rho_{\rm 10 Hz} =  \left( (1+e_{\rm 10 Hz})^{0.3046} \pi M_{\mathrm{tot},z} f_{\rm det} \right)^{-2/3} \, ,
\end{equation} 
where $f_{\rm det} = 10\,\rm Hz $ (Section \ref{subsec:NumSims_2}). In Equation (\ref{eq:Implicit_First}), we define $\rho_{\rm LSO}$ for eccentric \mbox{BH--BH} binaries as \mbox{$\rho_\mathrm{p,LSO} = (6 + 2e_\mathrm{LSO})/(1 + e_\mathrm{LSO})$}, and for eccentric \mbox{NS--NS} and \mbox{NS--BH} binaries as $8.5$ and $7.5$, respectively; see Section \ref{subsec:PhaseEvolution} for details. After combining Equation (\ref{eq:Implicit_First}) with Equation (\ref{eq:rho10Hz}), we obtain a result of the form \mbox{$e_{\rm 10 Hz} \equiv e_{\rm 10 Hz}(e_\mathrm{LSO}, M_{\mathrm{tot},z})$}. A similar derivation gives \mbox{$e_{\rm 10 Hz} \equiv e_{\rm 10 Hz}(e_0, M_{\mathrm{tot},z})$} when $e_0$ is set instead of $e_\mathrm{LSO}$ in Equation (\ref{eq:e10HzEQUIVeLSO}). We tested the algorithms numerically to ensure that both $e_{\rm 10 Hz}$ and $\Delta e_{\rm 10 Hz}$ are correctly calculated.

\section{The Total Signal-to-noise Ratio and Parameter Measurement Errors as a Function of $D_L$}
\label{sec:DL_Dep} 
 
 In this section, we determine numerically how the $\mathrm{S/N} _\mathrm{tot}$ and the parameter measurement errors scale with $D_L$ in order to scale our $D_L = 100 \, {\rm Mpc}$ results to higher $D_L$ values. 
 
 The $D_L$ dependence of the $\mathrm{S/N} _\mathrm{tot}$ and parameter measurement errors may be given by their respective definition.\footnote{The $(\mathrm{S/N} _\mathrm{tot})^2$ is the scalar product of the waveform and parameter measurement errors are the root mean squares of the diagonal elements of the covariance matrix; see Sections 4 and 5 in \citet{Gondanetal2018} for details.} The signal-to-noise ratio and the Fisher matrix elements depend on the spectral GW amplitude, which scales as $1/D_L$. Furthermore, there is a cosmological redshift dependence of mass parameters that varies with distance, i.e., \mbox{$m_z = (1 + z) m$}, where $z = z(D_L)$ according to Equation (\ref{eq:CovDLz}). Here, $z(D_L)$ may be obtained numerically using the inverse function  of Equation (\ref{eq:CovDLz}). 

 If $D_L$ is relatively small such that $z \ll 1$, the $\mathrm{S/N} _\mathrm{tot}$ and parameter measurement errors scale as $\propto 1 / D_L$ and $\propto D_L$, respectively. However, for higher $D_L$, the $z$ dependence becomes significant, which modifies the distance dependence. The $\mathrm{S/N} _\mathrm{tot}$ and the components of the Fisher matrix have a non-monotonic $M_{{\rm tot},z}$ dependence for the considered waveform model\footnote{See Appendix D and related sections in \citet{Gondanetal2018} for expressions defining the $\mathrm{S/N} _\mathrm{tot}$ and components of the Fisher matrix for precessing eccentric binaries.}. 
 
 We determine $\mathrm{S/N} _\mathrm{tot}$ and parameter measurement errors numerically as in Section \ref{subsec:NumSetup}, but now vary $D_L$ instead of $\rho _{\mathrm{p}0}$. We present scaling relations for these quantities between $100 \, \Mpc$ and $1 \, {\rm Gpc}$, where the numerical algorithms have been validated. 
 
 We find that the $\mathrm{S/N} _\mathrm{tot}$ is best fit with a power-law profile $\propto D_L^{-\gamma}$, where \mbox{$\gamma \approx 1.03 - 1.07$} depending on the component masses and $\rho_\mathrm{p0}$. Similarly, measurement errors of binary parameters can also be approximated as $\propto D_L^{\kappa}$, where \mbox{$\kappa \approx 1.02-1.1$} depending on the component masses, $\rho_\mathrm{p0}$, and the parameter. Using $\gamma=1$ leads to inaccuracies at the level of $5$--$12 \%$ and $7$--$17 \%$ for \mbox{$D_L = 500 \, \Mpc$} and $1 \, {\rm Gpc}$, respectively, for the $\mathrm{S/N} _\mathrm{tot}$. Similarly for the measurement errors of binary parameters, these inaccuracies are at the level of $3$--$15 \%$ and $5$--$21 \%$ for the same $D_L$ values when $\kappa = 1$ is used.

\bibliographystyle{yahapj}
\bibliography{refs}

\end{document}